\documentclass[]{article}
\usepackage{epsf}
\usepackage{latexsym}
\begin{document}
\newcommand {\bea}{\vspace{-.25in}\begin{eqnarray}}   
\newcommand {\eea}{\vspace{-.25in}\end{eqnarray}}
\newcommand {\p}{{\cal P}}
\newcommand {\jj}{{\cal J}}
\newcommand {\OR}{{\cal O}}
\newcommand {\V}{V(\la)}
\newcommand {\Vp}{V(\la')}
\newcommand {\gla}{g_{_{\hspace{-.01in}\la}}}
\newcommand {\glap}{g_{_{\hspace{-.02in}\la'}}}
\newcommand {\D}[2]{\, \delta_{#1,#2} \, }
\newcommand {\dV}{\delta V}
\newcommand {\CD}{\mathrm{C\hspace{.2mm}D}}
\newcommand {\CI}{\mathrm{C\hspace{.2mm}I}}
\newcommand {\ths}{{\hspace{.1mm}\mathrm{th}}}
\newcommand {\Mf}{{\cal M}_\mathrm{free}^2}
\newcommand {\n}{\nonumber\\}
\newcommand {\noi}{\noindent}
\newcommand {\la}{\Lambda}
\newcommand {\M}{{\mathcal{M}}^2(\la)}
\newcommand {\Mp}{{\cal M}^2(\la')}
\newcommand {\lll}[1]{\label{eq:#1}}
\newcommand {\reff}[1]{Eq. (\ref{eq:#1})}
\newcommand {\bt}{\begin{tabbing}}
\newcommand {\et}{\end{tabbing}}
\newcommand {\lbrak}{<\!\!\!\!\!|\;}
\newcommand {\rbrak}{\hspace{-.04in}>\hspace{-.112in}|\;}
\newcommand {\MI}{{\cal M}^2_\mathrm{int}(\la)}
\newcommand {\bx}[2]{\Delta^{#1}_{#2}}
\newcommand {\Bx}[2]{\Delta^{#1}_{#2}}
\newcommand {\dVme}{\left< F \right| \delta V \left| I \right>}
\newcommand {\SP}{16 \pi^3}
\newcommand {\dVo}[1]{\delta V^{(#1)}}
\newcommand {\dVome}[1]{\left< F \right| \delta V^{(#1)} \left| I \right>}
\newcommand {\lz}{{\la \; \mathrm{terms}}}
\newcommand {\lzb}{\:\rule[-4mm]{.1mm}{8mm}_{\; \la \; \mathrm{terms}}}
\newcommand {\state}{\left| \Psi^{jn}(P) \right>}
\newcommand {\wavep}{\Phi^{jn}_{s_1 s_2}(p_1,p_2)}
\newcommand {\tim}{\n &\times&}
\newcommand {\kp}{\vec k_\perp}
\newcommand {\kpp}{\vec k_\perp^{\, \prime}}
\newcommand {\Pp}{\vec P_\perp}
\newcommand {\Ppsq}{\vec P_\perp^{\, 2}}
\newcommand {\kpsq}{k^{\prime \, 2}}
\newcommand {\xpsq}{x^{\prime \, 2}}
\newcommand {\wavex}{\Phi^{jn}_{s_1 s_2}(x,\vec k_\perp)}
\newcommand {\R}{R\put(-5.5,8.5){\line(1,0){5}}}
\newcommand {\IN}{\mathrm{I\hspace{.2mm}N}}
\newcommand {\SE}{\mathrm{S\hspace{.2mm}E}}
\newcommand {\EX}{\mathrm{E\hspace{.2mm}X}}
\newcommand {\CON}{\mathrm{C\hspace{.2mm}O\hspace{.2mm}N}}
\newcommand {\KE}{\mathrm{K\hspace{.2mm}E}}
\newcommand {\INA}{_\IN^\mathrm{A}}
\newcommand {\INB}{_\IN^\mathrm{B}}
\newcommand {\INBF}{_\IN^\mathrm{B,F}}
\newcommand {\SEF}{_\SE^\mathrm{F}}
\newcommand {\SED}{_\SE^\mathrm{D}}
\newcommand {\EXF}{_\EX^\mathrm{F}}
\newcommand {\EXD}{_\EX^\mathrm{D}}
\newcommand {\INX}{\mathrm{IN+EX}}
\newcommand {\ME}{\left< q',l',t',j \right| \M \left| q,l,t,j \right>}
\newcommand {\BIT}{\hspace{.5mm}}
\newcommand {\eval}[1]{\:\rule[-4mm]{.1mm}{8mm}_{\,#1}}
\newcommand {\Qp}{\vec Q_\perp}
\newcommand {\Np}{\vec N_\perp}
\newcommand {\rp}{\vec r_\perp}
\newcommand {\Wp}{\vec w_\perp}
\newcommand {\ala}{\alpha_{_\Lambda}}
\newcommand {\Nt}{N_{\mathrm{t}}}
\newcommand {\Nl}{N_{\mathrm{l}}}
\newcommand {\T}{T\put(-6.2,8.5){\line(1,0){6}}}
\def\bbox#1{%
\relax\ifmmode
\mathchoice
{{\hbox{\boldmath$\displaystyle#1$}}}%
{{\hbox{\boldmath$\textstyle#1$}}}%
{{\hbox{\boldmath$\scriptstyle#1$}}}%
{{\hbox{\boldmath$\scriptscriptstyle#1$}}}%
\else
\mbox{#1}%
\fi
}

\pagestyle{empty}

{\Large \bf \centerline{Glueballs in a Hamiltonian Light-Front Approach}}
{\Large \bf \centerline{to Pure-Glue QCD}}

\vspace{.2in}

\centerline{Brent H. Allen\footnote{E-mail: allen@mps.ohio-state.edu} and Robert J. Perry\footnote{E-mail:
perry@mps.ohio-state.edu}}

\vspace{.1in}

\centerline{\it Department of Physics, Ohio State University, Columbus, Ohio 43210}

\centerline{(August 1999)}
\vspace{.5in}

\centerline{\bf Abstract}

\vspace{-.1in}

\begin{quotation}
We calculate a renormalized Hamiltonian for pure-glue QCD and diagonalize it.  The renormalization
procedure is designed to produce a Hamiltonian that will yield physical states that rapidly converge in an expansion in free-particle
Fock-space sectors.  To make this possible, we use light-front field theory to isolate vacuum effects, and we place a smooth
cutoff on the Hamiltonian to force its free-state matrix elements to quickly decrease as the difference 
of the free masses of the states increases.  The cutoff violates a number of physical principles of 
light-front pure-glue QCD, including Lorentz covariance and gauge
covariance.  This means that the operators in the Hamiltonian are not required to respect these physical principles.  
However, by requiring the Hamiltonian to produce cutoff-independent 
physical quantities and by requiring it to respect the unviolated physical principles of pure-glue QCD, we are able to
derive recursion relations that define the Hamiltonian to all orders in perturbation theory
in terms of the running coupling.  We approximate all physical states as two-gluon states (thus they are relatively simple single-glueball states),
and use our recursion relations to calculate to second order the part of the Hamiltonian that is required to compute the spectrum.
We diagonalize the Hamiltonian using basis-function expansions for the gluons' color, spin, and momentum degrees 
of freedom.  We examine the sensitivity of our results to the cutoff and use them
to analyze the nonperturbative scale dependence of the coupling.  We investigate the effect of the
dynamical rotational symmetry of light-front field theory on the rotational degeneracies of the
spectrum and compare the spectrum to recent lattice results.  Finally, we examine our wave functions and analyze the various sources of
error in our calculation.
\end{quotation}

\newpage
\pagestyle{plain}

\setcounter{footnote}{0}

\section{Introduction}

A solution to a quantum field theory that is close to our physical intuition is possible if
we can develop a formalism in which the physical states of the theory rapidly converge in an 
expansion in free-particle Fock-space sectors (free sectors).
Such an expansion is unlikely to be possible in an equal-time approach to many of the more interesting theories, such as QCD.  This is
because in equal-time approaches to these theories, the physical states must be built on top of a complicated vacuum
unless the volume of space is severely limited.  For this reason, we work in light-front field theory (LFFT).
In LFFT, it is possible to force the vacuum to be empty by removing from the theory all particles 
that have zero longitudinal momentum\footnote{This is because there are no negative longitudinal 
momenta and momentum conservation requires the three-momenta of the constituents of the vacuum to  sum to
zero.}.  Any physical effects of these
particles must be incorporated into the operators of the theory in order to obtain correct physical
quantities.

In LFFT, the Hamiltonian is trivially related to the invariant-mass operator\footnote{The invariant-mass operator is given by the
square of the momentum operator: ${\cal P}^\mu {\cal P}_\mu={\cal M}^2$.  See Appendix A for more details.} (IMO),
and it is more natural to work with the IMO because it is manifestly boost-invariant.
If the IMO satisfies three conditions in the basis of free-particle Fock-space states (free states),
then its eigenstates will rapidly converge in an expansion in free sectors.  
First, the diagonal matrix elements of the IMO must be dominated by the free part of the IMO.  
Second, the off-diagonal
matrix elements of the IMO must quickly decrease as the difference of the free masses of the states increases.
If the IMO satisfies these first two conditions, then each  of its eigenstates
will be dominated by free-state components with free masses that are close to the mass of the eigenstate.  
The third condition on the IMO is that the free mass of a free state must quickly increase as the number of particles in
the state increases.  If the IMO satisfies all three conditions, 
then the number of particles in a free-state component that dominates an eigenstate will be limited from above.
This means that the IMO's eigenstates will rapidly converge in an expansion in free sectors\footnote{There 
are three subtleties here.  The first subtlety is that the first and
third conditions on the IMO will not be satisfied for those free states in which many of the particles have negligible center-of-mass transverse
momentum and little or no mass.  However, the contributions of these free states to the physical states in which we are interested are typically 
suppressed.  
For example, in QCD these free states have very large widths in transverse position space and are thus highly suppressed by confinement.  In QED,
the particles with negligible center-of-mass transverse momentum and no mass are long-transverse-wavelength photons.  These photons decouple from
the physical states in which we are typically interested, e.g. charge-singlet states like hydrogen and positronium.  Thus the contributions to
these physical states from the free states containing these photons are suppressed.

The second subtlety is that exactly how quickly the
IMO's off-diagonal matrix elements must decrease and the free mass of a free state must increase are not
known.  We assume that the rates that we are able to achieve are sufficient. This can be verified by
diagonalizing the IMO and examining the rate of convergence of the free-sector expansion of its eigenstates.

The third subtlety is that the coefficients of the expansion for highly excited eigenstates may grow for a number of free sectors and then peak
before diminishing and becoming rapidly convergent.}.

To satisfy the first condition on the IMO, we assume that we can derive the IMO in perturbation theory.  If this is true, then
the couplings are small and the diagonal matrix elements of the IMO are dominated by the
free part of the IMO.  To satisfy the second condition, we place a smooth cutoff on the IMO to force its matrix elements
to quickly decrease as the difference of the free masses of the states increases.
Once we have removed the particles with zero longitudinal momentum from the theory,
it is reasonable to expect that the third and final condition on the IMO will be satisfied automatically.  This is because the
free-particle dispersion relation of LFFT should force 
the free mass of a free state to quickly increase as the number of particles in the state increases (see Appendix A of Ref. \cite{brent}).

By suppressing the matrix elements of the IMO that have large changes in free mass, the cutoff regulates the 
ultraviolet divergences of the theory.  Unfortunately, it also
violates a number of physical principles of LFFT, including Lorentz covariance and gauge
covariance.  This means that the operators in the IMO are not required to respect these physical principles, 
and renormalization is {\bf not} simply a matter of adjusting a few canonical
parameters.  The simplest way to systematically determine the IMO in this case
is in perturbation theory.  In order for a perturbative computation of the IMO to be strictly valid,
the theory must be asymptotically free\footnote{Our method may work even if the theory is not asymptotically free.  For example, it works in QED
because the scale at which the electron charge is large is astronomical.}.  If this is the case, then 
by requiring the IMO to produce cutoff-independent physical quantities and by requiring 
it to respect the unviolated physical principles of the theory\footnote{\label{broken}Some of the physical principles, such as 
cluster decomposition, are violated in a very specific manner and can still be used to restrict the form of the IMO.  However, the
restriction in this situation is always weaker than it would have been had there been no violation of
the principle.}, we can derive recursion relations that define the IMO to all orders in perturbation theory
in terms of the fundamental parameters of the field theory.  If our cutoff is large enough, then the couplings will be small and the 
perturbative approximation to the IMO may work well.

The physical principles that we use to determine the IMO form a subset of the full set of physical principles of light-front field theory.
This raises the question of how the remaining principles, which are violated by our cutoff, are restored in physical quantities.
Since the IMO is uniquely determined by the principles that we use, the remaining principles
must be automatically respected by physical quantities derived from our IMO, at least perturbatively.  
If they are not, then they contradict the principles that we use and no
consistent theory can be built upon the complete set of principles.  The reason that this process is possible is that there are 
redundancies among the various physical principles.  

It is possible to compute operators other than the IMO in our approach.  Although we compute operators
perturbatively, we can use these operators to compute nonperturbative quantities.  For example, the spectrum
can be computed by diagonalizing the IMO (see Section 6).  However, there are drawbacks to computing operators perturbatively.  It is 
possible that there are
intrinsically nonperturbative effects in the theory that require nonperturbative renormalization.  Any such effects 
are neglected in this approach.  Another problem is that perturbative renormalization makes nonperturbative physical
quantities somewhat cutoff-dependent.

In general, field theories have an infinite number of degrees of freedom. However, since our IMO will
cause the physical states of the theory to rapidly converge in an expansion in free sectors, we can truncate this expansion. 
This means that approximate computations of physical quantities will require only a finite number of 
finite-body matrix elements of operators.  In addition, since we assume
that we can compute these matrix elements perturbatively, we do not implement particle creation and annihilation
nonperturbatively in these calculations.  This allows us to always work with a finite number of degrees of
freedom.

When we perturbatively calculate matrix elements of operators, we do not truncate the space of intermediate states that
can appear; so our approach does not use a
Tamm-Dancoff truncation\footnote{The only degrees of freedom that we remove from the full theory are the particles 
with zero longitudinal momentum.  We should be able to replace the physical effects of these particles with 
interactions without compromising the validity of the
theory.  This is because these particles are vacuum effects or have infinite kinetic
energies (or both), and thus are not observable in the laboratory as particles.} \cite{tamm,dancoff,robert td}.  
We also do not completely eliminate any interactions, such as those
that change particle number.  These strengths of our approach allow us to better describe physical theories.
However, the truncation of the free-sector expansion of physical states has drawbacks that are similar to those of perturbative renormalization.  
It neglects any physical effects that require an infinite number of particles and contributes to the cutoff dependence
of nonperturbative physical quantities.

The accuracy of our results and the 
strength of the cutoff dependence of our nonperturbative physical quantities are determined by the order in perturbation theory
to which we calculate the operators of the theory and the number of free sectors that we keep in the expansion of physical states.
If we use a cutoff that is too small, then the couplings of the theory will be large, and it will be necessary for us to keep many terms
in the expansion of the operators.  If we use a cutoff that is too large, then the free-sector expansion of the
physical states will converge slowly, and it will be necessary for us 
to keep many sectors in the states.  We assume that if the order of perturbation 
theory and the number of free sectors are manageable, then there is a range of cutoff values for which the approximations
work well and physical quantities are relatively accurate and cutoff-independent.

As we mentioned, we remove from the theory all particles with zero longitudinal momentum.  We should replace their physical effects with interactions.
However, due to the limitations of our method, we can reproduce only those effects of these particles that can be derived with perturbative
renormalization and require only a small number of particles.  

There are a number of approaches that are similar to ours \cite{glazek and wilson,long paper,brazil,billy,stan,elana,walhout,zhang}, and 
some of these methods have been used to calculate the physical states of QCD \cite{zhang,martina}.  
These calculations are based on nonrelativistic 
approximations and use sharp step-function cutoffs.  Nonrelativistic approximations drastically simplify the diagonalization of the IMO,
but are insufficient for states containing light quarks or gluons.  Sharp cutoffs prevent the complete cancellation of
the infrared divergences that appear in light-front gauge theories\footnote{These appear due to the exchange of massless gauge
particles with arbitrarily small longitudinal momentum.}.

Our approach is completely relativistic and uses smooth cutoffs to ensure the complete cancellation of
the light-front infrared divergences.  It is largely based on the
renormalization methods of Perry \cite{cc1}, Perry and Wilson \cite{cc2},
Wilson \cite{wilson}, and G{\l}azek and Wilson \cite{glazek and wilson}, as well as the Hamiltonian-diagonalization methods of Wegner 
\cite{wegner}.  In Ref. \cite{brent}, we developed and tested our 
method in massless $\phi^3$ theory in six dimensions.  In Ref. \cite{roger}, Kylin, Allen, and Perry extended this method to include particle masses.
In this paper, we extend our method to pure-glue QCD.  This theory is simpler than full QCD due to the reduced number of vertices and the 
absence of quark masses.  For this application,
we derive the recursion relations that determine the IMO to all orders in perturbation theory
in the running coupling.  We approximate all physical states as two-gluon states and use the recursion relations to compute 
to second order the part of the IMO
that is required to compute the spectrum.  We diagonalize the IMO in a basis-state expansion and analyze the results.

In the remainder of the Introduction, we outline the rest of the paper.  Appendix A contains our conventions for light-front pure-glue QCD; so the
reader may wish to examine it first.  In Section 2 we present our method for computing the
free-state matrix elements of the IMO.  
After defining the action of our cutoff, we place a number of restrictions on the IMO.  First we force the IMO at a given cutoff 
to be unitarily equivalent to itself 
at a higher cutoff.  This implies that the IMO is unitarily equivalent to itself at an infinite cutoff, and will therefore yield cutoff-independent
physical quantities.  From the statement of unitary equivalence, we develop
a perturbative series that relates the interactions at two different cutoffs.  We then proceed to use physical principles to restrict the
form of the IMO.  We require it to conserve momentum and to be invariant under boosts and rotations about the 3-axis.  Although our 
cutoff violates exact transverse locality, we are able to require the IMO
to respect an approximate transverse locality.  In practice this means that the IMO's matrix elements are analytic functions of
transverse momenta.  The cutoff also violates cluster decomposition, but we show that 
the implications of this violation are simple enough that we can still use this
principle to restrict the form of the IMO.  

In order to represent pure-glue QCD, the IMO must become the free IMO in the noninteracting limit, and it must be a function only of the cutoff
and the coupling.  The final physical restriction on the IMO is that it must reproduce the perturbative
scattering amplitudes of pure-glue QCD.  This restriction specifies the form of the first-order interaction and part of the second-order interaction.

We use the restrictions from physical principles and the perturbative series that relates the interactions 
at two different cutoffs to derive the recursion relations that determine the IMO.  A
crucial step in this process is the removal of the coupling from the perturbative
series.  This allows us to separate the cutoff dependences of the operators in the interaction from the cutoff dependences of their
couplings.  Most of the details of this derivation are relegated to Appendix B.

In Section 3 we define a basis for the expansion of physical states.  
We construct the physical states to be simultaneous eigenstates of the
IMO, the three-momentum operator, and the part of the generator of rotations about the 3-axis that generates rotations of the internal
degrees of freedom of states (as opposed to their centers of mass).  We approximate all physical states as two-gluon states, which means
that they are relatively simple single-glueball states.  We analytically calculate the color parts of the states by requiring the 
states to be color
singlets, and then we expand the states' spin and momentum degrees of freedom in a complete set of orthonormal basis functions.  
We conclude Section 3 by deriving the eigenvalue equation for the IMO in our basis.

In Section 4 we use the recursion relations for the IMO to compute its two-gluon to two-gluon matrix element to second order in the running
coupling.  This is the only free-state matrix element that we need to solve the eigenvalue equation.  

In Section 5 we compute the matrix elements of the IMO in the basis that we have defined, 
in terms of integrals that must be evaluated numerically.
To avoid roundoff error, we evaluate the integrals for the kinetic energy and the two-point interaction (the self-energy) 
by writing them as sums of gamma functions.
The remaining integrals are five-dimensional, and we compute them with Monte Carlo methods.  However, before we can do this,
it is necessary for us to make manifest the cancellation of the infrared divergences from exchanged gluons with infinitesimal 
longitudinal momentum.  This is nontrivial, and even after the divergences are gone, we must manipulate the integrals quite a bit
to get them into a form that is amenable to Monte Carlo integration.  We present some of these details and other technical issues
in Appendix C.

In Section 6 we derive our results and analyze the sources of error in the calculation.  We begin by discussing how we assign quantum
numbers to our numerical results for glueball states and proceed with a discussion of the procedure that we use to compute these results.  We derive the
nonperturbative cutoff dependence of the coupling and discuss the cutoff dependence of our glueball masses.  
We use this analysis to choose the value of the cutoff that minimizes our errors.  We then present the spectrum that
we find with this optimal cutoff and compare it to recent lattice results.  The last results that we present are the probability densities for
our five lightest glueballs.  We conclude Section 6 by discussing the sources of error in our calculation and estimating the sizes of 
these errors.

Finally, in Section 7 we conclude with a summary and a discussion of the direction of future work.

\section{The Method for Computing Free-State Matrix Elements of the Invariant-Mass Operator}

This section summarizes the extension of the results of Sections 2-4 and Appendix D of Ref. \cite{brent} to the case
of pure-glue QCD.

\subsection{The Cutoff}

Our goal is to derive recursion relations that uniquely determine the IMO to all orders in perturbation theory
in the running coupling.  The IMO is a function of the cutoff, and can be split into the canonical free IMO and 
an interaction (see Appendix A for our light-front pure-glue QCD conventions):

\bea
\M = \Mf + \MI .
\lll{split}
\eea

\noi
The cutoff is implemented on the matrix elements of the IMO:

\bea
\left< F \right| \M \left| I \right> &=& \left< F \right| \Mf \left| I \right> + 
\left< F \right| \MI \left| I \right> \n
&=& M_F^2 \left< F \right| \!I \left. \!\right> + 
e^{-\frac{\bx{2}{FI}}{\la^4}} \left<F \right| \V \left| I \right> ,
\lll{cutoff}
\eea

\noi
where $\left| F \right>$ and $\left| I \right>$ are eigenstates of the free IMO
with eigenvalues $M_F^2$ and $M_I^2$, and $\Bx{}{FI}$ is the difference of these eigenvalues:

\bea
\Bx{}{FI} = M_F^2 - M_I^2 .
\eea

\noi
$\V$ is the interaction with the Gaussian cutoff factor removed, and we refer to it as the ``reduced interaction."
To determine the IMO, we must determine the reduced interaction.

We will see that $\left<F \right| \V \left| I \right>$ 
does not grow exponentially as $\Bx{2}{FI}$ gets large; so the
exponential in \reff{cutoff} forces the off-diagonal matrix elements of the IMO to rapidly diminish as
$\Bx{2}{FI}$ grows.  This satisfies the second of our three conditions on the IMO and regulates it.

\subsection{The Restriction to Produce Cutoff-Independent Physical Quantities}

Our cutoff violates a number of physical principles of LFFT, including Lorentz covariance and gauge
covariance\footnote{Our regulator breaks these symmetries because the mass of a free state is neither
gauge-invariant nor rotationally invariant (except for rotations about the 3-axis).}.  
This means that the operators in the IMO are not required to respect these physical principles.
In addition, since there is no locality in the longitudinal direction in Hamiltonian
LFFT\footnote{That there is no
longitudinal locality in Hamiltonian LFFT is evident from the fact that the longitudinal momentum of a free particle appears
in the denominator of its dispersion relation, $(\vec p_\perp^{\,2} + m^2)/p^+$.}, these operators can contain
arbitrary functions of longitudinal momenta.  To uniquely determine the IMO in this case,
we have to place some restrictions on it. The first restriction is that
it must produce cutoff-independent physical quantities.  To enforce this, we require $\M$ to satisfy

\bea
\M = U(\la,\la') \: {\cal M}^2(\la') \: U^\dagger(\la,\la') ,
\lll{unitary equiv}
\eea

\noi
where $U$ is a unitary transformation that changes the IMO's cutoff and 

\bea
\la < \la' < 2 \la .
\lll{lap}
\eea

\noi
We have placed an upper limit on $\la'$ because \reff{unitary equiv} is perturbatively valid only if $\la'$ is not too
much larger than $\la$ \cite{ken}.  Note that we are considering $\M$ to be a function
of its argument; i.e. ${\cal M}^2(\la')$ has the same functional dependence on $\la'$ that $\M$ has on
$\la$.  In Ref. \cite{brent}, we proved that \reff{unitary equiv} forces $\M$ to produce cutoff-independent physical quantities.

The unitary transformation that we use is designed to alter the cutoff implemented in \reff{cutoff}, and is a simplified 
version of a transformation introduced by Wegner \cite{wegner}, modified for implementation with the IMO.  It
is uniquely defined by a linear first-order differential equation:

\bea
\frac{d U(\la,\la')}{d (\la^{-4})} = T(\la) U(\la,\la') ,
\lll{diff eq}
\eea

\noi
with one boundary condition:

\bea
U(\la,\la) = {\bf 1} .
\eea

\noi
$U(\la,\la')$ is unitary as long as $T(\la)$ is anti-Hermitian and linear \cite{brent}.  We define 

\bea
T(\la) = \left[ \Mf, \M \right] ,
\lll{T def}
\eea

\noi
which is anti-Hermitian and linear.

To solve for $\M$ perturbatively, we need to turn \reff{unitary equiv} into a perturbative restriction on
the reduced interaction, $\V$.  We outlined how to do this in Ref. \cite{brent}, and here we simply state the
results that we need.  The perturbative version of \reff{unitary equiv} in terms of the reduced interaction is

\bea
\V - \Vp &=& \dV ,
\lll{unitary const}
\eea

\noi
where $\dV$ is the change to the reduced interaction and is a function of both $\la$ and $\la'$:

\bea
\dVme &=& \frac{1}{2} \sum_K
\left<F \right| \Vp \left| K \right> \left<K \right| \Vp \left| I \right> T_2^{(\la,\la')}(F,K,I) \n
&+& \frac{1}{4} \sum_{K,L}  \left<F \right| \Vp \left| K \right> \left<K \right| \Vp \left| L \right>
\left<L \right| \Vp \left| I \right> T_3^{(\la,\la')}(F,K,L,I) \n
&+& \OR\left(\left[\Vp \right]^4\right) .
\lll{RFI}
\eea

\noi
In this equation, the sums are over complete sets of free states and the cutoff functions are defined by

\bea
T_2^{(\la,\la')}(F,K,I) &=& \left( \frac{1}{\Bx{}{FK}} - \frac{1}{\Bx{}{KI}} \right) \left( 
e^{2 \la'^{-4} \bx{}{FK} \bx{}{KI}} - e^{2 \la^{-4} \bx{}{FK} \bx{}{KI}} \right)
\eea

\noi
and

\bea
&&T_3^{(\la,\la')}(F,K,L,I) = \n
&&\left( \frac{1}{\Bx{}{KL}} - \frac{1}{\Bx{}{LI}} \right) \left( \frac{1}{\Bx{}{KI}} -
\frac{1}{\Bx{}{FK}} \right) e^{2 \la'^{-4} \bx{}{KL} \bx{}{LI}} \left( e^{2 \la^{-4} \bx{}{FK} \bx{}{KI}} -
e^{2 \la'^{-4} \bx{}{FK} \bx{}{KI}} \right) \n
&+& \left( \frac{1}{\Bx{}{KL}} - \frac{1}{\Bx{}{LI}} \right) 
\frac{\Bx{}{FK} + \Bx{}{IK}}{\Bx{}{KL} \Bx{}{LI} + \Bx{}{FK} \Bx{}{KI}} \left(  e^{2 \la'^{-4} (\bx{}{FK}
\bx{}{KI}
+ \bx{}{KL} \bx{}{LI} )} - 
e^{2 \la^{-4} (\bx{}{FK} \bx{}{KI} + \bx{}{KL} \bx{}{LI} )} \right) \n
&+& \left( \frac{1}{\Bx{}{FK}} -
\frac{1}{\Bx{}{KL}} \right) \left( \frac{1}{\Bx{}{LI}} -
\frac{1}{\Bx{}{FL}} \right) e^{2 \la'^{-4} \bx{}{FK} \bx{}{KL}} \left( e^{2 \la^{-4} \bx{}{FL} \bx{}{LI}} -
e^{2 \la'^{-4} \bx{}{FL} \bx{}{LI}} \right) \n
&+& \left( \frac{1}{\Bx{}{FK}} - \frac{1}{\Bx{}{KL}} \right) 
\frac{\Bx{}{FL} + \Bx{}{IL}}{\Bx{}{FK} \Bx{}{KL} + \Bx{}{FL} \Bx{}{LI}}  \left(  e^{2 \la'^{-4} (\bx{}{FK}
\bx{}{KL} + \bx{}{FL} \bx{}{LI} )} - 
e^{2 \la^{-4} (\bx{}{FK} \bx{}{KL} + \bx{}{FL} \bx{}{LI} )}  \right) . \n
\eea

\noi
The above definitions for the cutoff functions assume that none of the $\Bx{}{}$'s that appear
in the denominators is zero.  In the event one of them is zero, the appropriate cutoff function is defined by

\bea
T^{(\la,\la')}_i(\bx{}{}=0) &=& \lim_{\bx{}{} \rightarrow 0} T^{(\la,\la')}_i(\bx{}{}) .
\eea

\subsection{Restrictions from Physical Principles}

\reff{unitary const} is the first restriction on the IMO.
To uniquely determine the IMO, we need to place additional restrictions on it, and we do this using 
the physical principles of the theory that are not violated by the cutoff. (See Footnote \ref{broken}.)

\subsubsection{Symmetry Principles}

Any LFFT should exhibit manifest momentum conservation, boost covariance, and covariance under rotations about the 3-axis.  
Our cutoff does not violate any
of these principles; so we restrict the IMO to conserve momentum and to be invariant under boosts
and rotations about the 3-axis.

\subsubsection{Transverse Locality}

Ideally, the IMO should be local in the transverse directions, and thus each of its matrix elements should be
expressible as a finite series of powers of transverse momenta with expansion coefficients that are functions of
longitudinal momenta.  In our case, the cutoff suppresses
interactions that have large transverse-momentum transfers and replaces them with interactions that have smaller
transverse-momentum transfers.  This is equivalent to suppressing interactions that occur over small transverse
separations and replacing them with interactions that occur over larger transverse separations; so we do not
expect our interactions to be perfectly transverse-local.  Nonetheless, we expect that interactions in $\M$
should appear local relative to transverse separations larger than $\la^{-1}$ or, equivalently, to transverse
momenta less than $\la$.  This means that for transverse momenta less than $\la$ we should be able to
approximate each matrix element of $\M$ as a finite power series in $\vec p_\perp/\la$. We enforce this
by assuming that transverse locality is violated in the weakest manner possible, i.e. that any matrix
element of the IMO can be expressed as an {\it infinite} series of powers of transverse momenta with an
infinite radius of convergence.  In other words, we assume that the matrix elements of the IMO are analytic functions
of transverse momenta.

\subsubsection{Cluster Decomposition}

Since the matrix elements of the IMO conserve momentum, they can
be written as a sum of terms, with each term containing a unique product of momentum-conserving delta functions \cite{weinberg}.
We require the IMO to satisfy approximate cluster decomposition \cite{brent}; i.e. when any of its matrix
elements is written as an expansion in the possible products of momentum-conserving delta functions, 
the coefficient of any term in the expansion is restricted as follows.  (If there
is more than one possible set of spectators for a given product of momentum-conserving delta functions, then
the coefficient has to be broken into a distinct part for each possible set, and these restrictions hold for each part
separately.) 
It can depend on the cutoff, the quantum numbers of the interacting particles, and the total longitudinal momentum.
It must be proportional to a quantum-number-conserving Kronecker delta for each discrete quantum number for each spectator.
It can have no other dependence on the quantum numbers of spectators, and it cannot contain delta functions that fix momenta.

The reason that the IMO does not respect exact cluster decomposition and that the coefficients in the
delta-function expansion can depend on the momenta of spectators (through a dependence on the total longitudinal momentum)
is that our cutoff on free-mass differences violates cluster decomposition.  To see this, note that the change in free
mass for some process is given by $P^+ ( \sum_j p_j^{\prime \, -} - \sum_{i} p_i^- )$, where
the $p_i$'s are the momenta of the particles in the initial state, the
$p_j'$'s are the momenta of the particles in the final state, and $P^+$ is the
total longitudinal momentum of each state.  The minus momenta of any spectators cancel in this difference, but 
their longitudinal momenta still contribute to the overall factor of $P^+$.

\subsubsection{Representation of the Theory of Interest}

\label{subsubsec: rep of toi}

The preceding restrictions on $\M$ are valid for any LFFT in more than two dimensions.
In order to represent a particular theory, we must place additional restrictions on $\M$.  

We assume that we can compute $\M$
perturbatively, which means that we can expand $\V$ in powers of the coupling  at
the scale $\la$.  Our cutoff has no effect in the noninteracting limit; so our IMO must
reproduce free pure-glue QCD in this limit.  According to \reff{cutoff}, this means that 
$\V$ vanishes in the noninteracting limit.

In pure-glue QCD, the
only fundamental parameter is the coupling; so we require the IMO to depend only on it and
the scale.  (For an example of the application of our method to a theory with more than one parameter, see Ref. \cite{roger}.)
In this case, the expansion of $\V$ takes the form

\bea
\V = \sum_{r=1}^\infty \gla^r V^{(r)}(\la) ,
\lll{order exp}
\eea

\noi
where $\gla$ is the coupling at the scale $\la$.  We refer to $V^{(r)}(\la)$ as the $\OR(\gla^r)$ reduced
interaction, although for convenience the coupling is factored out.

$\gla$ is the correct fundamental parameter for pure-glue QCD if and only if
its definition is consistent with the canonical definition of the coupling.  The canonical definition 
is

\bea
g &=& \left[ 16 \pi^3 p_1^+ \delta^{(3)}(p_1 - p_2 - p_3)
\lbrak g_2 g_3 |
v | g_1 \rbrak  \right]^{-1} \left< g_2 g_3 \right| {\cal
M}^2 \left| g_1 \right>,
\lll{can g}
\eea

\noi
where $v$ is the canonical interaction and $\lbrak j | v | i \rbrak$ denotes a modified matrix element of $v$.  (See Appendix A 
for our light-front pure-glue QCD conventions.)  
The denominator removes all dependence on momentum, spin, and color in the canonical matrix element for gluon emission, and thus
isolates the coupling.  We define the coupling $\gla$ by

\bea
\gla &=& \bigg\{ \left[ 16 \pi^3 p_1^+ \delta^{(3)}(p_1 - p_2 - p_3)
\lbrak g_2 g_3 | v | g_1 \rbrak  \right]^{-1}
\left< g_2 g_3 \right| \M \left| g_1 \right> \bigg\}^{s_n=1; \,
c_n=n}_{\vec p_{2 \perp}
=\vec p_{3 \perp} ; \; p_2^+ = p_3^+; \, \epsilon=0} \n
&=& \bigg\{ \left[ 16 \pi^3 p_1^+ \delta^{(3)}(p_1 - p_2 - p_3)
\lbrak g_2 g_3 | v | g_1 \rbrak   \right]^{-1}
\left< g_2 g_3 \right| \V \left| g_1 \right> \bigg\}^{s_n=1; \,
c_n=n}_{\vec p_{2 \perp}
=\vec p_{3 \perp} ; \; p_2^+ = p_3^+; \, \epsilon=0}
\hspace{-.1in} ,
\lll{g def}
\eea

\noi
where $n=1,2,3$, and $\epsilon$ is a longitudinal-momentum cutoff that we define below.
Momentum conservation and boost invariance imply that the matrix element that defines the coupling 
can depend only on the transverse momentum of
particle 2 in the center-of-mass frame and the ratio $p_2^+/p_1^+$.  The restrictions $\vec p_{2 \perp} = \vec p_{3 \perp}$ and $p_2^+ = p_3^+$ 
fix these quantities to be 0 and 1/2, respectively.  

Our definition of the coupling is consistent with the canonical definition
because the conditions on the matrix elements in \reff{g def} have no effect on the right-hand side (RHS) of
\reff{can g} and do not force $\left< g_2 g_3 \right| {\cal
M}^2 \left| g_1 \right>$ to vanish.  According to \reff{order exp}, the IMO is {\it coupling coherent} \cite{cc1,cc2,zimm} 
because the couplings of its noncanonical operators are functions only of the fundamental parameters 
of the theory and they vanish in the noninteracting limit.

We have assumed that the IMO obeys approximate transverse locality, which means that we can expand
any matrix element $\left<F \right| V^{(r)}(\la) \left| I \right>$ in powers of transverse momenta.
Each term in this expansion is either cutoff-dependent or
cutoff-independent.  We define
$V^{(r)}_\CD(\la)$ to be the cutoff-dependent part of $V^{(r)}(\la)$, i.e. 
the part that produces the cutoff-dependent terms in transverse-momentum
expansions of matrix elements of $V^{(r)}(\la)$.  We define
$V^{(r)}_\CI$ to be the cutoff-independent part of $V^{(r)}(\la)$, i.e. 
the part that produces the cutoff-independent terms in transverse-momentum
expansions of matrix elements of $V^{(r)}(\la)$.  Then

\bea
V^{(r)}(\la) = V^{(r)}_\CD(\la) + V^{(r)}_\CI .
\eea

\noi
This separation is necessary because the procedures for computing $V^{(r)}_\CD(\la)$ and $V^{(r)}_\CI$ differ.

We believe that if $\M$ is to reproduce the perturbative scattering amplitudes of
pure-glue QCD, then it is necessary and sufficient for the reduced interaction to
satisfy the following conditions\footnote{Any longitudinal regulator that is consistent
with the physical principles that we use to restrict the IMO is sufficient.  It is not necessary to
use our $\epsilon$ cutoff in order to reproduce the perturbative scattering amplitudes of
pure-glue QCD.}.
We prohibit $V^{(r)}(\la)$ from having a three-point interaction unless $r$ is odd, and
we prohibit $V^{(r)}(\la)$ from having a four-point interaction unless $r$ is even.  We require

\bea
V^{(1)} &=& \p^+ \frac{16 \pi^3}{2!} \int D_1 D_2 D_3 \, a_2^\dagger
a_3^\dagger a_1 \, \delta^{(3)}(p_1 - p_2 - p_3)
\lbrak g_2 g_3 | v | g_1 \rbrak \n
&+& \p^+ \frac{16 \pi^3}{2!} \int D_1 D_2 D_3 \, a_3^\dagger a_1 a_2 \,
\delta^{(3)}(p_1 + p_2 - p_3)
\lbrak g_3 | v | g_1 g_2 \rbrak
\lll{O1}
\eea

\noi
and

\bea
V^{(2)}_\CI &=& \p^+ \frac{16 \pi^3}{2! \, 2!} \int D_1 D_2 D_3 D_4 \,
a_3^\dagger a_4^\dagger a_1 a_2 \,
\delta^{(3)}(p_1 + p_2 - p_3 - p_4)  \sum_{i=1}^4 \theta_{1,2;3,4}^{(i)}
\lbrak g_3 g_4 |
v | g_1 g_2 \rbrak_{\hspace{.005in} i} \n
&+& \p^+ \frac{16 \pi^3}{3!} \int D_1 D_2 D_3 D_4 \, a_2^\dagger
a_3^\dagger a_4^\dagger a_1 \,
\delta^{(3)}(p_1 - p_2 - p_3 - p_4)
\sum_{i=1}^4 \theta_{1;2,3,4}^{(i)} \lbrak g_2 g_3 g_4 | v | g_1
\rbrak_{\hspace{.005in} i} \n
&+& \p^+ \frac{16 \pi^3}{3!} \int D_1 D_2 D_3 D_4 \, a_4^\dagger a_1 a_2
a_3 \, \delta^{(3)}(p_1 + p_2 + p_3 - p_4)
\sum_{i=1}^4 \theta_{1,2,3;4}^{(i)} \lbrak g_4 | v | g_1 g_2 g_3
\rbrak_{\hspace{.005in} i} ,
\lll{O2}
\eea

\noi
where

\bea
\theta_{1,2;3,4}^{(1)} &=& 1 , \n
\theta_{1,2;3,4}^{(2)} &=& \theta(|p_1^+ - p_3^+| - \epsilon \p^+) , \n
\theta_{1,2;3,4}^{(3)} &=& \theta(|p_1^+ - p_4^+| - \epsilon \p^+) , \n
\theta_{1,2;3,4}^{(4)} &=& 1,
\eea

\noi
and

\bea
\theta_{1;2,3,4}^{(1)} &=& 1 , \n
\theta_{1;2,3,4}^{(2)} &=& \theta(|p_1^+ - p_2^+| - \epsilon \p^+), \n
\theta_{1;2,3,4}^{(3)} &=& \theta(|p_1^+ - p_3^+| - \epsilon \p^+), \n
\theta_{1;2,3,4}^{(4)} &=& \theta(|p_1^+ - p_4^+| - \epsilon \p^+),
\eea

\noi
and

\bea
\theta_{1,2,3;4}^{(1)} &=& 1 , \n
\theta_{1,2,3;4}^{(2)} &=& \theta(|p_1^+ - p_4^+| - \epsilon \p^+), \n
\theta_{1,2,3;4}^{(3)} &=& \theta(|p_2^+ - p_4^+| - \epsilon \p^+), \n
\theta_{1,2,3;4}^{(4)} &=& \theta(|p_3^+ - p_4^+| - \epsilon \p^+).
\eea

\noi
The presence of $\epsilon$ in these step-function cutoffs ensures that we will avoid
divergences from exchanged gluons (either instantaneous or real) with infinitesimal longitudinal
momentum.  In Section \ref{sec: bs me} we show how we can take $\epsilon$ to zero before we diagonalize 
$\M$.

We have not yet proved that the above conditions on the reduced interaction are necessary and sufficient
to reproduce the perturbative scattering amplitudes of pure-glue QCD.  
We can, however, show in $\phi^3$ theory \cite{brent} that analogous conditions are necessary and sufficient to reproduce all
second-order scattering amplitudes.

\subsection{The Recursion Relations for the Invariant-Mass Operator}

\label{subsec: recursion}

The restrictions that we have placed on the IMO are sufficient to uniquely determine it order-by-order in
perturbation theory.  In this subsection, we present the recursion relations that define $\M$ in terms
of $V^{(1)}$ and $V^{(2)}_\CI$, which we have defined above.  To begin, we consider the restriction that
forces the IMO to produce cutoff-independent physical quantities:

\bea
\V - \Vp &=& \dV .
\lll{main}
\eea

\noi
This restriction is in terms of the reduced interaction and the change to the reduced interaction.

$\dV$ is defined in \reff{RFI}, which makes it clear that since $\Vp$ can be
expanded in powers of $\glap$, so can $\dV$:

\bea
\dV = \sum_{t=2}^\infty \glap^t \dVo{t} .
\lll{RFI exp}
\eea

\noi
We refer to $\dVo{t}$ as the $\OR(\glap^t)$ change to the reduced interaction, although for convenience 
the coupling
is factored out. Note that $\dVo{t}$ is a function of $\la$ and $\la'$.

Now \reff{main} can be expanded in powers of $\gla$ and $\glap$:

\bea
\sum_{t=1}^\infty \gla^t V^{(t)}(\la) - 
\sum_{t=1}^\infty \glap^t V^{(t)}(\la') = \sum_{t=2}^\infty \glap^t 
\dVo{t} .
\lll{u expand}
\eea

\noi
This equation is a bit tricky to use because it involves the couplings at two different scales.
To see how they are related, consider the matrix element of \reff{main} for $g_1 \rightarrow g_2 g_3$:

\bea
&&\left<g_2 g_3 \right| \V \left| g_1 \right> - \left<g_2 g_3 \right| \Vp\left| g_1 \right>  
= \left<g_2 g_3 \right| \dV \left| g_1 \right> .
\lll{gen 3p me}
\eea

\noi
According to the definition of the coupling, this equation implies that

\bea
&&\gla - \glap  = \bigg\{ \left[ 16 \pi^3 p_1^+ \delta^{(3)}(p_1 - p_2 - p_3)
\lbrak g_2 g_3 | v | g_1 \rbrak   \right]^{-1}
\left< g_2 g_3 \right| \dV \left| g_1 \right> \bigg\}^{s_n=1; \,
c_n=n}_{\vec p_{2 \perp}
=\vec p_{3 \perp} ; \; p_2^+ = p_3^+; \epsilon=0}
\hspace{-.1in} .
\lll{running 1}
\eea

\noi
Since $V^{(1)}$ changes particle number by 1, inspection of \reff{RFI} reveals that 
$\left<g_2 g_3 \right| \dV \left| g_1 \right>$ is $\OR(\glap^3)$; so

\bea
\gla = \glap + \OR(\glap^3) .
\eea

\noi
This implies that

\bea
\gla = \glap + \sum_{s=3}^\infty \glap^s C_s(\la,\la') ,
\lll{scale dep}
\eea

\noi
where the $C_s$'s are functions of $\la$ and $\la'$.  For an integer $t \ge 1$, \reff{scale dep} implies that

\bea
\gla^t = \glap^t + \sum_{s=2}^\infty \glap^{t+s} B_{t,s}(\la,\la') ,
\lll{scale dep 2}
\eea

\noi
where the $B_{t,s}$'s are functions of $\la$ and $\la'$, and can be calculated in terms of the $C_s$'s by
raising \reff{scale dep} to the $t^\ths$ power.  

We substitute \reff{scale dep 2} into \reff{u expand} and demand that it hold order-by-order in $\glap$.
At $\OR(\glap^r)$ ($r \ge 1$), this implies that

\bea
&&V^{(r)}(\la) - 
V^{(r)}(\la') = \delta V^{(r)} - \sum_{s=2}^{r-1}  B_{r-s,s} V^{(r-s)}(\la),
\lll{coupled}
\eea

\noi
where $\delta V^{(1)} = 0$, and we define any sum to be zero if its upper limit is less than its lower limit.
The cutoff-independent parts of $V^{(r)}(\la)$ and $V^{(r)}(\la')$ cancel on the left-hand side (LHS), leaving

\bea
&&V^{(r)}_{\CD}(\la) - 
V^{(r)}_{\CD}(\la') = \delta V^{(r)} - \sum_{s=2}^{r-1}  B_{r-s,s}
V^{(r-s)}(\la) .
\lll{CD coupled}
\eea

\noi
This equation can be used to derive the desired recursion relations.

In the remainder of this section, we summarize the results of Appendix B, which contains a derivation of
the recursion relations that define the IMO.  Appendix B is an extension of
Appendix D of Ref. \cite{brent} to the case of pure-glue QCD.

Recall that momentum conservation implies that any matrix element 
$\left<F \right| V(\la) \left| I \right>$ can be written as an expansion in unique products of
momentum-conserving delta functions.  This means that an arbitrary matrix element of 
\reff{CD coupled} can be expanded in products of
delta functions and thus is equivalent to a set of equations, one for each possible product of delta
functions.  Given approximate transverse locality, each of the resulting equations can be expanded
in powers of transverse momenta.  Matching the coefficients of the powers of transverse momenta on either
side of these equations allows us to rigorously derive the following results (see Appendix B for details).

First, the cutoff-dependent part of the 
$\OR(\gla^r)$ reduced interaction is given in terms of lower-order reduced interactions by

\bea
\left<F \right| V^{(r)}_{\CD}(\la) \left| I \right> = \left[ \dVome{r} - \sum_{s=2}^{r-1}  B_{r-s,s}
\left<F \right| V^{(r-s)}(\la) \left| I \right>\right]_\lz ,
\lll{cc soln}
\eea

\noi
where ``$\la \; \mathrm{terms}$" means that the RHS is to be 
expanded in powers of transverse momenta and only the terms in the expansion that are proportional to powers or inverse powers of
$\la$ contribute.  Recall that $\dV^{(r)}$ is defined by Eqs. (\ref{eq:RFI}) and (\ref{eq:RFI exp}).

Second, the cutoff-independent part of $V^{(r)}(\la)$ has two contributions: a four-point interaction with no 
transverse-momentum dependence, and a three-point interaction that is linear in transverse momenta.  If there are no such 
contributions to $V^{(r)}(\la)$, then it
is completely determined by \reff{cc soln}.  Third, the coupling runs at odd orders; i.e. $C_s$ is zero if $s$ is even
[see \reff{scale dep}].  Fourth, there is no wave-function renormalization at any order in perturbation theory
in our approach because this would violate the restrictions that we have placed on the IMO.  

The fifth and final result from Appendix B is that the cutoff-independent parts of the 
$\OR(\gla^r)$ and $\OR(\gla^{r+1})$ reduced interactions for odd $r \ge 3$ 
are determined by the coupled integral 
equations\footnote{It is very difficult to prove that integral equations of this type have a unique
solution; so we simply assume that it is true in this case.}

\bea
\left<F \right| V^{(r)}_\CI \left| I \right> &=& \frac{1}{B_{r,2}} \left[ \dVome{r+2} -
\sum_{s=3}^{r+1}  B_{r+2-s,s}
\left<F \right| V^{(r+2-s)}(\la) \left| I \right>\right]^\mathrm{3-point}_{\vec p_\perp^{\, 1} \; \mathrm{term}},
\lll{CI cc soln 1}
\\
\left<F \right| V^{(r+1)}_\CI \left| I \right> &=& \frac{1}{B_{r+1,2}} \left[ \dVome{r+3} -
\sum_{s=3}^{r+2}  B_{r+3-s,s}
\left<F \right| V^{(r+3-s)}(\la) \left| I \right>\right]^\mathrm{4-point}_{\vec p_\perp^{\, 0} \; \mathrm{term}} .
\lll{CI cc soln 2}
\eea

\noi
To use these equations, the right-hand sides have to be expanded in powers of transverse momenta.  Only 
three-point interactions that are linear in transverse momenta contribute to $V^{(r)}_\CI$, and only 
four-point interactions that are independent of all transverse momenta contribute to $V^{(r+1)}_\CI$.  

These equations are coupled integral equations because both $V^{(r)}_\CI$ and $V^{(r+1)}_\CI$ appear
on the RHS of \reff{CI cc soln 1} inside integrals in $\delta V^{(r+2)}$, and $V^{(r+1)}_\CI$ appears on the
RHS of \reff{CI cc soln 2} inside integrals in $\delta V^{(r+3)}$.  It would seem that $V^{(r+2)}_\CI$ also appears
on the RHS of \reff{CI cc soln 2} inside integrals in $\delta V^{(r+3)}$, but $V^{(r+2)}_\CI$
cannot couple to $V^{(1)}$ to produce a transverse-momentum-independent four-point contribution to $\delta V^{(r+3)}$.  
This is because the cutoff function
$T_2^{(\la,\la')}$ vanishes when the intermediate state is massless and all external transverse momenta are zero.  This means that since we
specified $V^{(1)}$ and $V^{(2)}_\CI$ in Subsection \ref{subsubsec: rep of toi}, we can use Eqs. (\ref{eq:CI cc soln 1}) and
(\ref{eq:CI cc soln 2}) to solve for $V^{(3)}_\CI$ and $V^{(4)}_\CI$ simultaneously, and $V^{(5)}_\CI$ and $V^{(6)}_\CI$ simultaneously,
and so on.  Note that before we can use these equations to solve for $V^{(r)}_\CI$ and $V^{(r+1)}_\CI$ simultaneously, we must
first use \reff{cc soln} both to compute $V^{(r)}_\CD(\la)$ in terms of lower-order interactions and to express $V^{(r+1)}_\CD(\la)$
in terms of lower-order interactions and $V^{(r)}(\la)$.

\section{The Basis for the Expansion of Physical States}

\label{basis section}

\subsection{Preliminaries}

In the remainder of this paper, we use the results of our renormalization procedure to compute the physical states of pure-glue QCD and their 
masses.  The states will be eigenstates of $\M$.  Since our cutoff preserves translational covariance
and covariance under rotations about the 3-axis, we would like the states to also be simultaneous eigenstates of the generators
of these symmetries, but this is impossible
because translations do not commute with rotations.  However, a rotation about the 3-axis separates into a part that
rotates the centers of mass of states and a part that rotates states' internal degrees of freedom, and translations do
commute with these internal rotations.

To be precise, we define $\jj_3$ to be the generator of rotations about the 3-axis, and $\jj_3^\mathrm{R}$ to
be the part of $\jj_3$ governing gluons' momenta in the center-of-mass frame and spin polarizations.  $\M$, $\p^+$, $\vec \p_\perp$,  and
$\jj_3^\mathrm{R}$ are a set of commuting observables; so an eigenstate of $\M$ can be labeled by their eigenvalues.
We choose to write a physical state as $\state$, where $P$ is the three-momentum of the state, $j$ is the
eigenvalue of $\jj_3^\mathrm{R}$ for the state, 
and $n$ labels the mass eigenvalue of the state ($n=1$ has the smallest mass, $n=2$ has the second-smallest mass, etc.).  
Note that because $\state$ will be determined by $\M$, it will implicitly depend on $\la$ and $\gla$.

An examination of the matrix elements of $\M$ leads us to believe that the light-front infrared divergences (see the discussion below)
will not cancel unless the physical states are color singlets, although we do not have a rigorous proof of this.  Therefore,
we assume that the physical states are color singlets.  Using the Fock-space expansion for the identity operator 
(see Appendix A for our light-front pure-glue QCD conventions), we can expand a physical state in terms of the number
of gluons:

\bea
\state = {\bf 1} \state \simeq \frac{1}{2!} \int D_1 D_2 \big< g_1 g_2 \! \state
\left| g_1 g_2 \right> ,
\lll{fock exp}
\eea

\noi
where there is no one-gluon component because there is no color-singlet gluon.  We neglect contributions to the
states with more than two gluons, which is a severe approximation.  This means that we are approximating all physical states as relatively 
simple single-glueball states, and
$j$ is then the projection of the glueball's spin onto the 3-axis.  From now
on, we refer to the approximate physical states simply as glueball states.  

Any eigenstate of $\M$ can be written as a superposition of cross-product states:

\bea
\state &=& \sum_i \left| \Upsilon_{n,i} \right> \otimes \left| \Gamma^{jn}_i(P) \right> ,
\lll{cross sup}
\eea

\noi
where $\left| \Upsilon_{n,i} \right>$ is the color part of the $i^\ths$ contribution to the state and $\left| \Gamma^{jn}_i(P) \right>$ is the
momentum/spin part.  

The free state $\left| g_1 g_2 \right>$ is defined as a two-gluon state
in which one particle has quantum numbers $p_1$, $s_1$, and $c_1$, and the other has
quantum numbers $p_2$, $s_2$, and $c_2$.  $\left| g_1 g_2 \right>$ is symmetrized with respect to particle labels.
Thus we can write

\bea
\left| g_1 g_2 \right> &=& \frac{1}{\sqrt{2}} \Big[ \left| g_1 ; g_2 \right> + \left| g_2 ; g_1 \right> \Big] ,
\eea

\noi
where the semi-colon in $\left| g_1; g_2 \right>$ indicates that particle \#1 has quantum numbers
$p_1$, $s_1$, and $c_1$, and particle \#2 has quantum numbers $p_2$, $s_2$, and $c_2$.  Then we can write the wave function as the
sum of two permutations:

\bea
\left< g_1 g_2 \state \right. &=& \frac{1}{\sqrt{2}} \Big[ \left< g_1; g_2 \right| + \left< g_2; g_1 \right| \Big] \state .
\lll{wave split}
\eea

\noi
This implies that the wave function $\left< g_1 g_2 \state \right.$ is symmetric under exchange of particles 1 and 2.  
Using \reff{cross sup}, we can write the first permutation as

\bea
\left< g_1; g_2 \state \right. &=& \sum_i \left< c_1; c_2 \left| \Upsilon_{n,i} \right> \right.  \left< p_1, s_1; p_2, s_2 \left|
\Gamma^{jn}_i(P) \right> \right. .
\lll{super}
\eea

\noi
$\left< c_1; c_2 \left| \Upsilon_{n,i} \right> \right.$ and $\left< p_1, s_1; p_2, s_2 \left| \Gamma^{jn}_i(P) \right> \right.$ 
are the color
and momentum/spin wave functions for the first permutation of the $i^\ths$ contribution to the total wave function.

\subsection{Calculation of the Color Wave Function}

In our two-gluon approximation, the fact that the glueball states are 
color singlets uniquely determines their color wave functions.  Using the completeness relation
for color states, the color part of the $i^\ths$ contribution to the glueball state can be written

\bea
\left| \Upsilon_{n,i} \right> &=& \sum_{c_1 c_2} \left| c_1 ; c_2 \right> \left< c_1; c_2 \left| \Upsilon_{n,i} \right> \right. .
\eea

\noi
Let $Y$ be the unitary operator that rotates two-gluon states in color space.  Then under a color rotation, we have

\bea
\left| \Upsilon_{n,i} \right> \rightarrow \left| \Upsilon_{n,i}' \right> &=& \sum_{c_1 c_2} Y \left| c_1 ; c_2 \right> 
\left< c_1; c_2 \left| \Upsilon_{n,i} \right> \right. \n
&=& \sum_{c_1 c_2} \left| c_1' ; c_2' \right> 
\left< c_1; c_2 \left| \Upsilon_{n,i} \right> \right. ,
\eea

\noi
where the primes on the basis vectors denote rotated basis states.  Since $\left| \Upsilon_{n,i} \right>$ is a color singlet, it does not change
under a color rotation:

\bea
\left| \Upsilon_{n,i} \right> = \left| \Upsilon_{n,i}' \right> = \sum_{c_1 c_2} \left| c_1' ; c_2' \right> 
\left< c_1; c_2 \left| \Upsilon_{n,i} \right> \right. .
\lll{state rot}
\eea

A gluon color-rotated state is given by \cite{greiner}

\bea
\left| c_1' \right> = U \left| c_1 \right> = \sum_{c_2} \left| c_2
\right> \left< c_2 \right| U \left| c_1 \right>
=  \sum_{c_2} U_{c_2 c_1} \left| c_2 \right> ,
\lll{single rot}
\eea

\noi
where

\bea
U = e^{-i \theta_{c_1} \tilde F^{c_1}}
\eea

\noi
and

\bea
\tilde F^{c_1}_{c_2 c_3} = -i f^{c_1 c_2 c_3} .
\eea

\noi
Here $\theta_c$ is a vector of $N_\mathrm{c}^2-1$ real numbers that parameterize the color rotation.  Since

\bea
\left| c_1' ; c_2' \right> = \left| c_1' \right> \otimes \left| c_2' \right> ,
\eea

\noi
Eqs. (\ref{eq:state rot}) and (\ref{eq:single rot}) imply that

\bea
\left| \Upsilon_{n,i} \right> =
\sum_{c_1 c_2 c_3 c_4} \left< c_1; c_2 \left| \Upsilon_{n,i} \right> \right. U_{c_3 c_1} U_{c_4 c_2}
\left| c_3 ; c_4 \right> .
\lll{more rot}
\eea

As a guess, let us try

\bea
\left< c_1; c_2 \left| \Upsilon_{n,i} \right> \right. = \frac{1}{N} \delta_{c_1, c_2} ,
\eea

\noi
where $N$ is some constant.  Then \reff{more rot} becomes

\bea
\left| \Upsilon_{n,i} \right> = \frac{1}{N} \sum_{c_1 c_3 c_4} U_{c_3 c_1} U_{c_4 c_1} \left| c_3 ; c_4 \right> .
\eea

\noi
Note that

\bea
(U^{-1})_{cc'} = (U^\dagger)_{cc'} = U^*_{c'c} = U_{c'c} .
\eea

\noi
Then

\bea
\left| \Upsilon_{n,i} \right> &=& \frac{1}{N} \sum_{c_3 c_4} \delta_{c_3, c_4} \left| c_3 ; c_4 \right> \n
&=& \sum_{c_1 c_2}  \left< c_1; c_2 \left| \Upsilon_{n,i} \right> \right.\left| c_1 ; c_2 \right> \n
&=& \left| \Upsilon_{n,i} \right> .
\eea

\noi
Since this is true, our guess for a color-singlet wave function was correct.  This solution is unique \cite{cheng}; so we can
drop the subscript $i$ in \reff{super}.  $N$ is determined by normalization to be $\sqrt{N_\mathrm{c}^2-1}$.

Now \reff{wave split} becomes

\bea
\left< g_1 g_2 \state \right. &=& \frac{1}{\sqrt{2 (N_\mathrm{c}^2 - 1)}} \delta_{c_1, c_2}
\Bigg[ \left< p_1, s_1; p_2, s_2 \left|  \Gamma^{jn}(P) \right> \right. + \left< p_2, s_2; p_1, s_1 \left|
\Gamma^{jn}(P) \right> \right. \Bigg] .
\eea

\noi
Since the IMO conserves momentum, the factor in brackets must be proportional to a
momentum-conserving delta function.  Thus we can write

\bea
\left< g_1 g_2 \right| \left. \! \!\Psi^{jn}(P) \right> = 2! (16 \pi^3)^\frac{3}{2}
\delta^{(3)}(P - p_1 - p_2) \frac{1}{\sqrt{2 (N_\mathrm{c}^2-1)}} \delta_{c_1, c_2} \sqrt{p_1^+ p_2^+} \Phi^{jn}_{s_1 s_2}(p_1,p_2) ,
\lll{color result}
\eea

\noi
where $\Phi^{jn}_{s_1 s_2}(p_1,p_2)$ is the momentum/spin wave function with the mo\-men\-tum-con\-ser\-ving del\-ta function removed, and all the
extra factors in this equation are present to simplify the normalization of $\Phi^{jn}_{s_1 s_2}(p_1,p_2)$.
We must solve for $\wavep$, and this equation indicates that it is symmetric under exchange of particles $1$ and $2$.  

\subsection{Jacobi Variables}

Using \reff{color result}, the Fock-state expansion for a glueball state in \reff{fock exp} becomes

\bea
\state &=& \frac{1}{\sqrt{16 \pi^3}} \frac{1}{\sqrt{2 (N_\mathrm{c}^2-1)}}  \sum_{s_1 s_2 c_1 c_2} \delta_{c_1, c_2} \int 
\frac{d^2 p_{1 \perp} d p_1^+}{\sqrt{p_1^+ p_2^+}}
\theta(p_1^+ - \epsilon P^+) \theta(p_2^+ - \epsilon P^+) \tim \wavep \left| g_1 g_2 \right> ,
\lll{fock}
\eea

\noi
where momentum conservation implies that $p_2 = P - p_1$.  It is useful to separate the motion of the center of mass of the
state from the internal motions of the gluons.  To do this, we change variables from $p_1$ to the Jacobi variables $x$ and $\kp$:

\bea
p_1 &=& (x P^+, x \vec P_\perp + \vec k_\perp) , \n
p_2 &=& ([1-x] P^+, [1-x] \vec P_\perp - \vec k_\perp) .
\lll{jacobi}
\eea

\noi
Here $x$ is the fraction of the total longitudinal momentum that is carried by particle 1, and $\kp$ is the transverse
momentum of particle 1 in the center-of-mass frame.  We only display the longitudinal and transverse components of the momenta.  (Since the 
glueball state is a superposition of free-particle states, and since the momentum of a free gluon satisfies $p^2=0$, 
the minus components of the momenta of $g_1$ and $g_2$ are constrained to be given by $p_i^- = \vec p_{i \perp}^{\, 2}/p_i^+$.)

In terms of the Jacobi variables, the glueball state is

\bea
\left| \Psi^{jn}(P)\right> &=& \frac{1}{\sqrt{16 \pi^3}} \frac{1}{\sqrt{2 (N_{\mathrm{c}}^2-1)}}  \sum_{s_1 s_2 c_1 c_2} \delta_{c_1, c_2} \int 
\frac{d^2 k_\perp dx}{\sqrt{x(1-x)}} \theta(x - \epsilon) \theta(1-x-\epsilon) \tim \Phi^{jn}_{s_1 s_2}(x,\vec k_\perp,P) 
\left| g(x,\vec k_\perp,P;s_1,c_1) g(1-x,-\vec k_\perp,P;s_2,c_2) \right> ,
\lll{pre boost}
\eea

\noi
where we explicitly show the dependence of the RHS ket on the Jacobi variables and the total momentum.  We can use boost covariance to show that
$\Phi^{jn}_{s_1 s_2}(x,\vec k_\perp,P)$ is independent of $P$.  To do this, we note that under a longitudinal boost, 
the longitudinal momentum of each particle in pure-glue QCD 
(whether the particle is point-like or composite) transforms according to

\bea
p^+ \rightarrow e^\nu p^+ ,
\eea

\noi
where $\nu$ is a boost parameter \cite{kogut,leutwyler}.  Under a transverse boost, each particle's transverse momentum transforms
according to

\bea
\vec p_\perp \rightarrow \vec p_\perp + p^+ \vec v_\perp ,
\eea

\noi
where $\vec v_\perp$ is a boost parameter.  This means that if we apply a boost operator to both sides of \reff{pre boost}, we find that

\bea
\left| \Psi^{jn}(P')\right> &=& \frac{1}{\sqrt{16 \pi^3}} \frac{1}{\sqrt{2 (N_{\mathrm{c}}^2-1)}}  \sum_{s_1 s_2 c_1 c_2} \delta_{c_1, c_2} \int 
\frac{d^2 k_\perp dx}{\sqrt{x(1-x)}} \theta(x - \epsilon) \theta(1-x-\epsilon) \tim \Phi^{jn}_{s_1 s_2}(x,\vec k_\perp,P) 
\left| g(x,\vec k_\perp,P';s_1,c_1) g(1-x,-\vec k_\perp,P';s_2,c_2) \right> ,
\lll{post boost}
\eea

\noi
where the boost takes the glueball's momentum from $P$ to $P'$.  Note that the boost does not affect the wave function, only
the kets.  Since \reff{pre boost} holds for all $P$, it holds in particular for $P'$:

\bea
\left| \Psi^{jn}(P')\right> &=& \frac{1}{\sqrt{16 \pi^3}} \frac{1}{\sqrt{2 (N_{\mathrm{c}}^2-1)}}  \sum_{s_1 s_2 c_1 c_2} \delta_{c_1, c_2} \int 
\frac{d^2 k_\perp dx}{\sqrt{x(1-x)}} \theta(x - \epsilon) \theta(1-x-\epsilon) \tim \Phi^{jn}_{s_1 s_2}(x,\vec k_\perp,P') 
\left| g(x,\vec k_\perp,P';s_1,c_1) g(1-x,-\vec k_\perp,P';s_2,c_2) \right> .
\lll{alternate}
\eea

\noi
Eqs. (\ref{eq:post boost}) and (\ref{eq:alternate}) contradict each other unless the wave function $\Phi^{jn}_{s_1 s_2}(x,\vec k_\perp,P)$ 
is independent of $P$:

\bea
\Phi^{jn}_{s_1 s_2}(x,\vec k_\perp,P) = \Phi^{jn}_{s_1 s_2}(x,\vec k_\perp) .
\eea

\noi
Thus we can write $\state$ as

\bea
\state = \frac{1}{\sqrt{16 \pi^3}} \frac{1}{\sqrt{2 (N_{\mathrm{c}}^2-1)}} \sum_{s_1 s_2 c_1 c_2} \delta_{c_1, c_2} 
\int d^2 k_\perp dx \frac{1}{\sqrt{x(1-x)}} \theta_\epsilon \wavex \left| g_1 g_2 \right> ,
\eea

\noi
where $\theta_\epsilon=\theta(x-\epsilon) \theta(1-x-\epsilon)$.

\subsection{The Momentum and Spin Wave-Function Bases}

To solve for $\wavex$, we expand it in a complete orthonormal basis for each degree of freedom.  
Since $\wavex$ is symmetric under exchange of particles 1 and 2,

\bea
\wavex = \Phi^{jn}_{s_2 s_1}(1-x,-\vec k_\perp) .
\eea

\noi
To use this, we define

\bea
\wavex &=& \sum_{q=1}^4 \chi_{q}^{s_1 s_2} \Omega^{jn}_{q}(x,\vec k_\perp) ,
\eea

\noi
where the spin wave functions are

\bea
\chi_{1}^{s_1 s_2} &=& \delta_{s_1, 1} \delta_{s_2, 1} ,\n
\chi_{2}^{s_1 s_2} &=& \delta_{\bar s_1, 1} \delta_{\bar s_2, 1} ,\n
\chi_{3}^{s_1 s_2} &=& \frac{1}{\sqrt{2}}\left[ \delta_{s_1, 1} \delta_{\bar s_2, 1} +
\delta_{\bar s_1, 1} \delta_{s_2, 1} \right] ,\n
\chi_{4}^{s_1 s_2} &=& \frac{1}{\sqrt{2}}\left[ \delta_{s_1, 1} \delta_{\bar s_2, 1} -
\delta_{\bar s_1, 1} \delta_{s_2, 1} \right] ,
\eea

\noi
where $\bar s = - s$, and the momentum wave functions satisfy

\bea
\Omega^{jn}_{1}(x,\vec k_\perp) &=& \Omega^{jn}_{1}(1-x,-\vec k_\perp), \n
\Omega^{jn}_{2}(x,\vec k_\perp) &=& \Omega^{jn}_{2}(1-x,-\vec k_\perp), \n
\Omega^{jn}_{3}(x,\vec k_\perp) &=& \Omega^{jn}_{3}(1-x,-\vec k_\perp), \n
\Omega^{jn}_{4}(x,\vec k_\perp) &=& -\Omega^{jn}_{4}(1-x,-\vec k_\perp) .
\eea

\noi
Note that

\bea
\sum_{s_1 s_2} \chi_{q}^{s_1 s_2} \chi_{q'}^{s_1 s_2} = \delta_{q,q'} .
\eea

\noi
We define $k = | \vec k_\perp|$, and we define the angle $\phi$ by

\bea
\vec k_\perp = k \cos \phi \: \hat x + k \sin \phi \: \hat y .
\lll{phi}
\eea

We expand the momentum wave function in complete orthonormal bases:

\bea
\Omega_{q}^{jn}(x,\vec k_\perp) = \sum_{l=0}^\infty \sum_{t=0}^\infty
\sum_{a=-\infty}^\infty \R^{jn}_{q l t a} L^{(e)}_l(x) T^{(d)}_t(k) A_a(\phi) ,
\lll{basis func exp}
\eea

\noi
where $L^{(e)}_l(x)$, $T^{(d)}_t(k)$, and $A_a(\phi)$ are basis functions for the longitudinal,
transverse-magnitude, and transverse-angular degrees of freedom.  $e$ and $d$ are parameters that govern the widths of the longitudinal and
transverse-magnitude basis functions, respectively.  We can adjust these widths to optimize the bases.
Note that if we do not truncate the sums in \reff{basis func exp}, 
then the $\R^{jn}_{q l t a}$'s depend on $e$ and $d$ such that $\Omega_{q}^{jn}(x,\vec k_\perp)$ 
is independent of $e$ and $d$, although we do not indicate this dependence explicitly.

We define the transverse-magnitude basis functions $T^{(d)}_t(k)$ by

\bea
T_t^{(d)}(k) = d \BIT e^{-k^2 d^2} \sum_{s=0}^t \sigma_{t,s} k^s d^s ,
\lll{T_t def}
\eea

\noi
where the $\sigma_{t,s}$'s are constants that we have computed numerically using the Gram-Schmidt orthogonalization procedure and are such that

\bea
\int_0^\infty dk \BIT k \BIT T_t^{(d)}(k) T_{t'}^{(d)}(k) = \delta_{t,t'} .
\eea

\noi
Adjusting $d$ allows us to adjust the width of the Gaussian weight function in \reff{T_t def}.  $d$ cannot be allowed to pass through zero, because
when $d$ is zero, $T_t^{(d)}(k)$ is zero.  We choose $d$ to be positive without loss
of generality.  When doing numerical computations, we work with a dimensionless form of these basis functions:

\bea
\T_t(k d) = \frac{1}{d} T_t^{(d)}(k) .
\eea

\noi
Under exchange of the two particles, $k$ is unaffected; so $T_t^{(d)}(k)$ is unaffected.

We define the longitudinal basis functions $L_l^{(e)}(x)$ by

\bea
L_l^{(e)}(x) = [x(1-x)]^e \sum_{m=0}^l \lambda^{(e)}_{\:l,m} x^m ,
\lll{L_l def}
\eea

\noi
where

\bea
&& \lambda^{(e)}_{l,m} = (-1)^{l-m} \frac{1}{m! (l-m)!} \frac{\Gamma(1+4e+l+m)}{\Gamma(1+2e+m)}
\sqrt{\frac{l!(1+4e+2l)}{\Gamma(1+4e+l)}} .
\eea

\noi
These definitions imply that

\bea
\int_0^1 dx L^{(e)}_l(x) L^{(e)}_{l'}(x) = \delta_{l,l'} ,
\eea

\noi
as long as $e > -1/2$. (If $e \leq -1/2$, then the state is not normalizable.)  We can adjust the width of the weighting function in \reff{L_l def}
by adjusting $e$.  It is often more convenient to work with

\bea
\bar L_l^{(e)}(x) = \frac{1}{\sqrt{x(1-x)}} L_l^{(e)}(x) .
\eea

\noi
Under exchange of the two particles, $x \rightarrow 1-x$, which means that $L_l^{(e)}(x) \rightarrow (-1)^l L_l^{(e)}(x)$.

We define the transverse-angular basis functions $A_a(\phi)$ by

\bea
A_a(\phi) = \frac{1}{\sqrt{2 \pi}} e^{i a \phi}, 
\eea

\noi
and then

\bea
\int_0^{2 \pi} d \phi A_{a'}^*(\phi) A_a(\phi) = \delta_{a, a'} .
\eea

\noi
These basis functions are useful because they are eigenfunctions of ${\cal L}_3^\mathrm{R}$, the part of the generator of rotations about the 3-axis
that governs gluons' momenta in the center-of-mass frame.
Under exchange of the two particles, $\phi \rightarrow \phi+\pi$, which means that $A_a(\phi) \rightarrow (-1)^a A_a(\phi)$.

Using the above definitions of the bases, $\state$ can be written

\bea
\state = \frac{1}{\sqrt{16 \pi^3}} \frac{1}{\sqrt{2 (N_\mathrm{c}^2-1)}} \sum_{s_1 s_2 c_1 c_2 q l t a} \delta_{c_1, c_2} 
\R^{jn}_{qlta} \chi_{q}^{s_1 s_2} \int d^2 k_\perp dx \theta_\epsilon \bar L^{(e)}_l(x) T^{(d)}_t(k) A_a(\phi)
\left| g_1 g_2 \right> .
\eea

\noi
Since the glueball state is symmetric under exchange of the two gluons, the behaviors of the spin and momentum wave functions 
under exchange of the two gluons imply that if $q=4$, then $l+a$ must be odd, and if
$q\neq4$, then $l+a$ must be even (so that the spin and
momentum wave functions have the same symmetry under exchange).

\subsection{Rotations about the 3-Axis}

We want to ensure that $\state$ is an eigenstate of $\jj_3^\mathrm{R}$ with eigenvalue $j$.
The action of $\jj_3$ on the state is given by

\bea
\jj_3 \state &=& \frac{1}{\sqrt{16 \pi^3}} \frac{1}{\sqrt{2 (N_\mathrm{c}^2-1)}} \sum_{s_1 s_2 c_1 c_2 q l t a} \delta_{c_1, c_2} \R^{jn}_{qlta} 
\chi_{q}^{s_1 s_2} \int d^2 k_\perp dx \theta_\epsilon \n
&\times& \left\{ \left[  i p_{1 \perp}^2 \frac{\partial}{\partial p_{1
\perp}^1} - i p_{1 \perp}^1 \frac{\partial}{\partial p_{1 \perp}^2} + i p_{2 \perp}^2
\frac{\partial}{\partial p_{2 \perp}^1} -
i p_{2 \perp}^1 \frac{\partial}{\partial p_{2 \perp}^2} + s_1 + s_2 \right] \bar L^{(e)}_l(x) T^{(d)}_t(k) A_a(\phi) \right\} \tim
\left| g_1 g_2 \right> .
\eea

\noi
Using the definitions of the Jacobi variables, we can separate the center-of-mass and internal degrees of freedom:

\bea
\jj_3 \state &=& \frac{1}{\sqrt{16 \pi^3}} \frac{1}{\sqrt{2 (N_\mathrm{c}^2-1)}} \sum_{s_1 s_2 c_1 c_2 q l t a} \delta_{c_1, c_2} \R^{jn}_{qlta} 
\chi_{q}^{s_1 s_2} \int d^2 k_\perp dx \theta_\epsilon \n
&\times& \left\{ \left[  i P_\perp^2 \frac{\partial}{\partial P_\perp^1} - i P_\perp^1 \frac{\partial}{\partial P_\perp^2} 
-i \frac{\partial}{\partial \phi} + s_1 + s_2 \right] 
\bar L^{(e)}_l(x) T^{(d)}_t(k) A_a(\phi) \right\}
\left| g_1 g_2 \right> .
\eea

\noi
This implies that the action of $\jj_3^\mathrm{R}$ on the glueball state is given by

\bea
\jj_3^\mathrm{R} \state &=& \frac{1}{\sqrt{16 \pi^3}} \frac{1}{\sqrt{2 (N_\mathrm{c}^2-1)}} \sum_{s_1 s_2 c_1 c_2 q l t a} \delta_{c_1, c_2} 
\R^{jn}_{qlta} \chi_{q}^{s_1 s_2} \int d^2 k_\perp dx \theta_\epsilon \n
&\times& \left\{ \left[  -i \frac{\partial}{\partial \phi} + s_1 + s_2 \right] 
\bar L^{(e)}_l(x) T^{(d)}_t(k) A_a(\phi) \right\} \left| g_1 g_2 \right> \n
&=& \frac{1}{\sqrt{16 \pi^3}} \frac{1}{\sqrt{2 (N_\mathrm{c}^2-1)}} \sum_{s_1 s_2 c_1 c_2 q l t a} \delta_{c_1, c_2} 
\R^{jn}_{qlta} \chi_{q}^{s_1 s_2} \int d^2 k_\perp dx \theta_\epsilon \left[  a + s_1 + s_2 \right] 
\bar L^{(e)}_l(x) \tim T^{(d)}_t(k) A_a(\phi) \left| g_1 g_2 \right> .
\eea

\noi
Since $\state$ is an eigenstate of $\jj_3^\mathrm{R}$ with eigenvalue $j$, it must be the case that $a+s_1+s_2=j$
for all values of $a$, $s_1$, and $s_2$ that contribute to the sums in this equation.  This implies that for some
set of coefficients $R^{jn}_{qlt}$,

\bea
\R^{jn}_{qlta} = R^{jn}_{qlt} \left[ \delta_{q,1} \delta_{a,j-2} + \delta_{q,2} \delta_{a,j+2} + \delta_{q,3}
\delta_{a,j} + \delta_{q,4} \delta_{a,j} \right] .
\eea

\noi
This means that $\state$ can be written

\bea
\state &=& \sum_{qlt} R^{jn}_{qlt} \left| q,l,t,j \right>,
\eea

\noi
where

\bea
\left| q,l,t,j \right> = \frac{1}{\sqrt{16 \pi^3}} \frac{1}{\sqrt{2 (N_\mathrm{c}^2-1)}} \sum_{s_1 s_2 c_1 c_2} \int d^2 k_\perp dx
\theta_\epsilon \delta_{c_1, c_2} \chi_{q}^{s_1 s_2} \bar L^{(e)}_l(x) T^{(d)}_t(k) A_{j-s_1-s_2}(\phi)
\left| g_1 g_2 \right> .
\lll{bsb}
\eea

We are going to calculate the matrix elements of the IMO
in the $\left| q,l,t,j\right>$ basis and diagonalize it.  
Since $\M$ commutes with $\jj_3^\mathrm{R}$, we can do this for each value of $j$ separately.
The diagonalization procedure will yield mass eigenvalues and the $R^{jn}_{qlt}$ coefficients.  
As long as the coefficients satisfy

\bea
\delta_{j,j'} \sum_{q l t} R^{j' n' *}_{q l t} R^{j n}_{q l t} = \delta_{j, j'} \delta_{n, n'} ,
\eea

\noi
the glueball state will have a plane-wave normalization:

\bea
&& \big< \Psi^{j'n'}(P') \! \left| \right. \!\! \Psi^{jn}(P) \big> = 16
\pi^3 P^+ \delta^{(3)}(P - P') \delta_{j, j'} \delta_{n, n'} .
\lll{norm}
\eea

Due to the symmetry of the glueball state under the exchange of particles 1 and 2, the basis state
$\left| q,l,t,j \right>$ is zero if $l+j$ is odd and $q \neq 4$, or if $l+j$ is even
and $q=4$.  To take advantage of this, we consider only the subspace in which $l+j$ is even if $q \neq 4$,
and $l+j$ is odd if $q=4$.  

\subsection{The Eigenvalue Equation}

The IMO's eigenvalue equation is

\bea
\M \state = M_n^2 \state ,
\eea

\noi
where the lower-case subscript on $M_n^2$ indicates that it is an eigenvalue of $\M$ rather than an eigenvalue of $\Mf$. (An example
of an eigenvalue of $\Mf$ is $M_K^2$.)

In the basis that we have defined, the eigenvalue equation takes the form

\bea
\M \sum_{qlt} R^{jn}_{qlt} \left| q,l,t,j \right> = M^2_n \sum_{qlt} R^{jn}_{qlt} \left| q,l,t,j \right> .
\eea

\noi
If we project $\left< q',l',t', j \right|$ onto the left of this equation and use the identity

\bea
&&\left< q',l',t',j \right. \! \left| q,l,t,j \right> = 16 \pi^3 P^+ \delta^{(3)}(P - P') \delta_{q,q'} \delta_{l,l'} \delta_{t,t'} ,
\eea

\noi
then we find that the eigenvalue equation in our basis is

\bea
\sum_{qlt} \frac{\left< q',l',t',j \right| \M \left| q,l,t,j \right>}{16 \pi^3 P^+ \delta^{(3)}(P - P')} R^{jn}_{qlt} &=& M^2_n
R^{jn}_{q'l't'} .
\eea

\section{Cal\-cu\-la\-tion of the Free-State Ma\-trix El\-e\-ment of the In\-var\-i\-ant-Mass Op\-er\-a\-tor}

\label{sec:free}

\subsection{Preliminaries}

To solve the eigenvalue equation, we must calculate the matrix elements $\left< q',l',t',j \right| \M \left| q,l,t,j \right>$.
These can be written in terms of the free-state matrix element $\left< g_1' g_2' \right| \M \left| g_1 g_2 \right>$, which is 
specified by our renormalization procedure.  We are going to calculate this matrix element to second order in perturbation theory.

Before continuing, we would like to point out three simplifications that we repeatedly use in this section.  
To make these simplifications clear, we note that the matrix elements $\left< q',l',t',j \right| \M \left| q,l,t,j \right>$
can be written as integrals of wave functions times the free-state matrix element $\left< g_1' g_2' \right| \M \left| g_1 g_2 \right>$.
To make the first simplification, we observe that $\left< g_1' g_2' \right| \M \left| g_1 g_2 \right>$ has step functions on
the particles' longitudinal momenta, and these step functions are redundant because they also appear in the integrals that define 
$\left< q',l',t',j \right| \M \left| q,l,t,j \right>$.  We drop these step functions in the formulas that we present for
$\left< g_1' g_2' \right| \M \left| g_1 g_2 \right>$.  To make the second simplification, we point out that
the matrix elements $\left< q',l',t',j \right| \M \left| q,l,t,j \right>$ are symmetric under exchange of the
two initial-state gluons and also under exchange of the two final-state gluons.  Because of this, 
when we are computing $\left< g_1' g_2' \right| \M \left| g_1 g_2 \right>$, 
we combine terms that are related by exchange of the two initial-state or two final-state gluons.  Finally, due to the color-singlet
nature of the glueball states, certain parts of $\left< g_1' g_2' \right| \M \left| g_1 g_2 \right>$ do not contribute to
$\left< q',l',t',j \right| \M \left| q,l,t,j \right>$, and we drop these terms.

We begin by defining Jacobi variables for the final state of the free matrix element:

\bea
p_1' &=& (x' P^+, x' \vec P_\perp + \kpp) , \n
p_2' &=& ([1-x'] P^+, [1-x'] \vec P_\perp - \kpp) .
\eea

\noi
Then using the definition of the IMO in terms of the reduced interaction in Eqs. (\ref{eq:split}), (\ref{eq:cutoff}), and (\ref{eq:order exp}),

\bea
\left< g_1' g_2' \right| \M \left| g_1 g_2 \right> &=& \left< g_1' g_2' \right| \Mf \left| g_1 g_2 \right> +
\left< g_1' g_2' \right| \MI \left| g_1 g_2 \right> \n
&=& M_I^2 \left< g_1' g_2' \right| \left. \! g_1 g_2 \right> + \gla^2 e^{-\la^{-4} \Delta_{FI}^2}
\left< g_1' g_2' \right| V^{(2)}_\CI \left| g_1 g_2 \right> \n &+& \gla^2 e^{-\la^{-4} \Delta_{FI}^2}
\left< g_1' g_2' \right| V^{(2)}_\CD(\la) \left| g_1 g_2 \right> ,
\lll{split 1}
\eea

\noi
where the free masses of the final and initial states are given by

\bea
M_F^2 &=& \frac{k^{\prime \, 2}}{x' (1-x')} ,\n
M_I^2 &=& \frac{k^2}{x (1-x)} .
\eea

\subsection{The Cutoff-Independent Part of the Reduced Interaction}

Based on the definition of $V^{(2)}_\CI$ in \reff{O2}, we find that

\bea
&& \left< g_1' g_2' \right| V^{(2)}_\CI \left| g_1 g_2 \right> = 16 \pi^3 P^+ \delta^{(3)}(P-P')
\sum_{i=1}^4 \theta_{1,2;1',2'}^{(i)}
\lbrak g_1' g_2' | v | g_1 g_2 \rbrak_{\hspace{.005in} i} .
\lll{sum}
\eea

\noi
We can divide this matrix element into contact (momentum-independent) and instantaneous-exchange interactions:

\bea
\left< g_1' g_2' \right| V^{(2)}_\CI \left| g_1 g_2 \right> =
\left< g_1' g_2' \right| V^{(2)}_\CI \left| g_1 g_2 \right>_\CON + \left< g_1' g_2' \right| V^{(2)}_\CI \left| g_1 g_2
\right>_\IN ,
\lll{split 2}
\eea

\noi
where

\bea
\left< g_1' g_2' \right| V^{(2)}_\CI \left| g_1 g_2 \right>_\CON &=& 16 \pi^3 P^+ \delta^{(3)}(P-P')
\theta_{1,2;1',2'}^{(1)} \lbrak g_1' g_2' | v | g_1 g_2 \rbrak_{\hspace{.005in} 1} \n
&=& 32 \pi^3 P^+ \delta^{(3)}(P-P') f^{c_1 c_1' c} f^{c_2' c_2 c} \left( \delta_{s_2, s_1'}
\delta_{s_1, s_2'} - \delta_{s_1', \bar s_2'} \delta_{s_1, \bar s_2} \right),
\lll{contact}
\eea

\noi
and

\bea
\left< g_1' g_2' \right| V^{(2)}_\CI \left| g_1 g_2 \right>_\IN &=& 16 \pi^3 P^+ \delta^{(3)}(P-P') \sum_{i=2}^3
\theta_{1,2;1',2'}^{(i)}
\lbrak g_1' g_2' | v | g_1 g_2 \rbrak_{\hspace{.005in} i} \n
&=& 32 \pi^3 P^+ \delta^{(3)}(P-P')
\theta(|x-x'| - \epsilon) f^{c_1 c_1'
c} f^{c_2' c_2 c} \delta_{s_1, s_1'} \delta_{s_2, s_2'} \tim
\frac{1}{(x-x')^2} (x+x') (1-x+1-x') .
\lll{instantaneous}
\eea

\noi
(The remaining term in \reff{sum} vanishes because the $\left| q,l,t,j \right>$ states are color singlets.)
The contact interaction is displayed in Figure 1 and the instantaneous-exchange interaction is displayed
in Figure 2.

\begin{figure}
\centerline{\epsffile{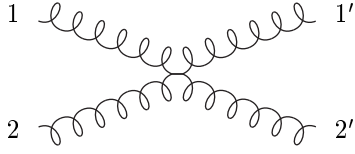}}
\caption{A diagrammatic representation of $\left< g_1' g_2' \right| V^{(2)}_\CI \left| g_1 g_2 \right>_\CON$, 
the two-gluon contact interaction. The numbers label the particles.}
\end{figure}

\begin{figure}
\centerline{\epsffile{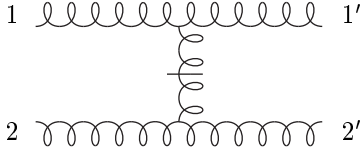}}
\caption{A diagrammatic representation of $\left< g_1' g_2' \right| V^{(2)}_\CI \left| g_1 g_2 \right>_\IN$, 
the two-gluon instantaneous-exchange interaction. The numbers label the particles.}
\end{figure}

\subsection{The Cutoff-Dependent Part of the Reduced Interaction}

\subsubsection{Preliminaries}

According to our recursion relation for the cutoff-dependent part of the reduced interaction, \reff{cc soln},
$\left< g_1' g_2' \right| V^{(2)}_\CD(\la) \left| g_1 g_2 \right>$ is given by

\bea
\left< g_1' g_2' \right| V^{(2)}_\CD(\la) \left| g_1 g_2 \right> &=& \left< g_1' g_2' \right| \delta V^{(2)}
\left| g_1 g_2 \right>\rule[-1.7mm]{.1mm}{4.4mm}_{\; \la \; \mathrm{terms}} .
\eea

\noi
Using the definition of the $\OR(\glap^r)$ change to the reduced interaction in Eqs. (\ref{eq:RFI}) and (\ref{eq:RFI exp}), we find that

\bea
\left< g_1' g_2' \right| V^{(2)}_\CD(\la) \left| g_1 g_2 \right> &=& \frac{1}{2} \sum_K
\left<g_1' g_2' \right| V^{(1)} \left| K \right> \left<K \right| V^{(1)} \left| g_1 g_2 \right>
T_2^{(\la,\la')}(F,K,I) \,\rule[-1.7mm]{.1mm}{4.4mm}_{\; \la \; \mathrm{terms}} .
\eea

\noi
The intermediate state can be either a one-particle or three-particle state.  The contribution to the eigenvalue equation
from the one-particle-intermediate-state part is zero because the $\left| q,l,t,j \right>$ states are color singlets.  This means
that

\bea
\left< g_1' g_2' \right| V^{(2)}_\CD(\la) \left| g_1 g_2 \right> &=& \frac{1}{12} \int D_3 D_4 D_5
\left<g_1' g_2' \right| V^{(1)} \left| g_3 g_4 g_5 \right> \left< g_3 g_4 g_5\right| V^{(1)} \left| g_1 g_2 \right> \tim
T_2^{(\la,\la')}(F,K,I) \,\rule[-1.7mm]{.1mm}{4.4mm}_{\; \la \; \mathrm{terms}} \; ,
\eea

\noi
where $M_K^2$ is the mass of the state $\left| g_3 g_4 g_5 \right>$.  Substituting the definition of $V^{(1)}$ in \reff{O1} into this equation,
and simplifying, we find that

\bea
&&\left< g_1' g_2' \right| V^{(2)}_\CD(\la) \left| g_1 g_2 \right> = 
\left< g_1' g_2' \right| V^{(2)}_\CD(\la) \left| g_1 g_2 \right>_\SE + \left< g_1' g_2' \right| V^{(2)}_\CD(\la) \left|
g_1 g_2 \right>_\EX \; ,
\lll{split 3}
\eea

\noi
where the self-energy interaction is given by

\bea
\left< g_1' g_2' \right| V^{(2)}_\CD(\la) \left| g_1 g_2 \right>_\SE = P^{+ 2} (16 \pi^3)^2 \int D_3 D_4 D_5
T_2^{(\la,\la')}(F,K,I) \D{2}{5} \D{2'}{5} \beta_{134} \beta^*_{1'34} \,\rule[-1.7mm]{.1mm}{4.4mm}_{\; \la \; \mathrm{terms}} \; ,
\eea

\noi
the exchange interaction is given by

\bea
\left< g_1' g_2' \right| V^{(2)}_\CD(\la) \left|
g_1 g_2 \right>_\EX = 2 P^{+ 2} (16 \pi^3)^2 \int D_3 D_4 D_5
T_2^{(\la,\la')}(F,K,I) \D{2}{5} \D{1'}{3} \beta_{134} \beta^*_{2'45} \,\rule[-1.7mm]{.1mm}{4.4mm}_{\; \la \; \mathrm{terms}}  \; ,
\eea

\noi
and we have defined

\bea
\beta_{ijk} = \theta(p_i^+) \theta(p_j^+) \theta(p_k^+) \delta^{(3)}(p_i - p_j - p_k) \lbrak g_j g_k | v | g_i \rbrak 
\eea

\noi
and

\bea
\delta_{i,k} = 16 \pi^3 p_i^+ \delta^{(3)}(p_i - p_k) \delta_{s_i, s_k} \delta_{c_i, c_k} .
\eea

\noi
The self-energy interaction is displayed in Figure 3 and the exchange interaction is displayed in Figure 4.

\begin{figure}
\centerline{\epsffile{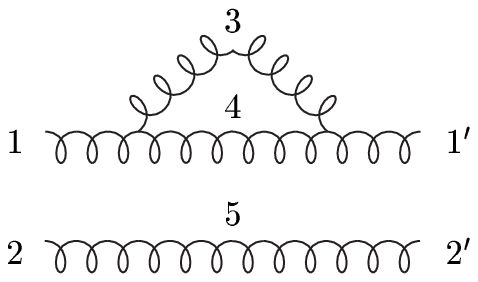}}
\caption{A diagrammatic representation of $\left< g_1' g_2' \right| V^{(2)}_\CD(\la) \left| g_1 g_2 \right>_\SE$, 
the self-energy interaction. The numbers label the particles.}
\end{figure}

\begin{figure}
\centerline{\epsffile{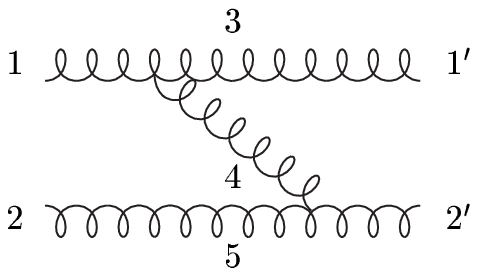}}
\caption{A diagrammatic representation of $\left< g_1' g_2' \right| V^{(2)}_\CD(\la) \left| g_1 g_2 \right>_\EX$, 
the exchange interaction. The numbers label the particles.}
\end{figure}
\subsubsection{The Self-Energy Interaction}

After some simplification, the self-energy interaction takes the form

\bea
\left< g_1' g_2' \right| V^{(2)}_\CD(\la) \left| g_1 g_2 \right>_\SE &=& P^{+ 2} \D{2}{2'} \delta^{(3)}(p_1 - p_1')
\sum_{s_3 s_4 c_3 c_4} \int \frac{d^2 p_{3 \perp} d p_3^+}{p_3^+}
\theta(p_3^+ - \epsilon P^+) \frac{1}{p_4^+} \theta(p_4^+ - \epsilon P^+) \n &\times&
T_2^{(\la,\la')}(F,K,I) \lbrak g_3 g_4 | v | g_1 \rbrak \lbrak g_1' | v | g_3 g_4 \rbrak \, \rule[-1.7mm]{.1mm}{4.4mm}_{\; \la \; \mathrm{terms}} \; ,
\eea

\noi
where $p_4 = p_1 - p_3$.  At this point, it is useful to change variables from $p_3$ to Jacobi variables $y$ and $\vec r_\perp$:

\bea
p_3 = (y p_1^+, y \vec p_{1 \perp} + \vec r_\perp) .
\eea

\noi
Then the free mass of the intermediate state is

\bea
M_K^2 &=& ( p_3^- + p_4^- + p_5^- ) P^+ - \vec P_\perp^{\,2} \n
&=& \frac{k^2}{x (1-x)} + \frac{r^2}{x y (1-y)} ,
\eea

\noi
and after a lot of additional simplification,

\bea
\left< g_1' g_2' \right| V^{(2)}_\CD(\la) \left| g_1 g_2 \right>_\SE = \frac{N_\mathrm{c}}{2 \pi^2} \sqrt{\frac{\pi}{2}} 
\left[ \la^2 - \la^{\prime \, 2} \right] \D{1}{1'} \D{2}{2'} 
\theta(x - 2 \epsilon) \bigg[ \log x - \log \epsilon - \frac{11}{12} \bigg]  \rule[-3.5mm]{.1mm}{8.5mm}_{\; \la \; \mathrm{terms}} .
\eea

\noi
Recalling
that ``$\la$ terms'' means that we are to expand the RHS of this equation in powers of transverse momenta and
keep only the terms that are proportional to powers or inverse powers of $\la$, we find that

\bea
&&\left< g_1' g_2' \right| V^{(2)}_\CD(\la) \left| g_1 g_2 \right>_\SE = \frac{N_\mathrm{c}}{2 \pi^2} \sqrt{\frac{\pi}{2}} 
\la^2 \D{1}{1'} \D{2}{2'} \theta(x - 2 \epsilon) \bigg[ \log x - \log \epsilon - \frac{11}{12} \bigg] .
\lll{sei}
\eea

\noi
This result shows that our cutoff violates cluster decomposition.  The evidence is that 
the self-energy depends on $x=p_1^+/(p_1^+ + p_2^+)$, even though
$p_2$ is the momentum of a spectator (see Figure 3).

\subsubsection{The Exchange Interaction}

For the exchange interaction, the free mass of the intermediate state is

\bea
M_K^2 &=& ( p_3^- + p_4^- + p_5^- ) P^+ - \vec P_\perp^{\,2} \n
&=& \frac{\kpsq}{x'} + \frac{(\kp - \kpp)^2}{x-x'} + \frac{k^2}{1-x} .
\eea

\noi
Then the changes in free mass are

\bea
\Delta_{FK} &=& - \frac{\kpsq [1-x]^2 + k^2 [1-x']^2 - 2 k k' (1-x') (1-x) \cos \gamma}{(1-x') (x-x') (1-x)} , \n
\Delta_{IK} &=& - \frac{k^2 \xpsq + \kpsq x^2 - 2 k k' x x' \cos \gamma}{x x' (x-x')} ,
\eea

\noi
where $\gamma = \phi - \phi'$.

Using the identity

\bea
\vec \varepsilon_{\perp s} \! \cdot \kp &=& \frac{-1}{\sqrt{2}} k \BIT s \BIT e^{i s \phi} ,
\eea

\noi
the exchange interaction becomes

\bea
&& \left< g_1' g_2' \right| V^{(2)}_\CD(\la) \left| g_1 g_2 \right>_\EX = 64 \pi^3 P^+ \delta^{(3)}(P - P')
f^{c_1 c c_1'} f^{c c_2 c_2'}
\frac{\theta(x-x'- \epsilon)}{x-x'} \tim \left( \frac{1}{\Delta_{FK}} +
\frac{1}{\Delta_{IK}} \right) \left( e^{-2 \la^{\prime -4} \Delta_{FK} \Delta_{IK}} - e^{-2 \la^{-4} \Delta_{FK} \Delta_{IK}} \right)
\sum_{i,m=1}^3 Q^{(i,m)} \lzb ,
\lll{ex}
\eea

\noi
where

\bea
Q^{(1,1)} &=& \frac{1}{x-x'} s_1 s_1' \delta_{s_2, s_2'} \left\{\frac{x'}{x} k
e^{i s_1 \phi} - k' e^{i s_1 \phi'}
\right\} \left\{ k' e^{- i s_1' \phi'} [ 1 - x ] - k e^{- i s_1' \phi} [1 - x'] \right\} , \n
Q^{(1,2)} &=& s_1 s_2' \delta_{s_2, s_1'} \left\{\frac{x'}{x} k e^{i s_1 \phi} - k' e^{i s_1 \phi'}
\right\} \left\{ k e^{-i s_2' \phi} - \frac{1-x}{1-x'} k' e^{-i s_2' \phi'} \right\} , \n
Q^{(1,3)} &=& - s_1 s_2 \delta_{\bar s_1', s_2'} \left\{\frac{x'}{x} k e^{i s_1 \phi} - k' e^{i s_1 \phi'}
\right\} \left\{ k' e^{i s_2 \phi'}  - \frac{1-x'}{1-x} k e^{i s_2 \phi} \right\} , \n
Q^{(2,1)} &=& 2\delta_{s_1, s_1'} \delta_{s_2, s_2'} \frac{1}{(x-x')^2}
\left\{ x (1-x) \kpsq + x' (1-x') k^2 - k k' \left[ x' (1-x) + x (1-x') \right] \cos \gamma \right\} , \n
Q^{(2,2)} &=&  \delta_{s_1, s_1'} s_2 s_2' \frac{1}{x-x'} \left\{ x k' e^{i s_2 \phi'} - x' k e^{i s_2 \phi}
\right\} \left\{ k e^{-i s_2' \phi} - \frac{1-x}{1-x'} k' e^{-i s_2' \phi'} \right\} , \n
Q^{(2,3)} &=&  s_2 s_2' \delta_{s_1, s_1'} \frac{1}{x-x'} \left\{ x k' e^{-i s_2' \phi'} - x' k e^{-i s_2' \phi}
\right\} \left\{ k' e^{i s_2 \phi'}  - \frac{1-x'}{1-x} k e^{i s_2 \phi} \right\} , \n
Q^{(3,1)} &=&  s_1 s_1' \delta_{s_2, s_2'} \frac{1}{x-x'} \left\{ \frac{x}{x'} k' e^{-i s_1' \phi'} - k e^{- i s_1' \phi}
\right\} \left\{ k' e^{i s_1 \phi'} [ 1 - x ] - k e^{i s_1 \phi} [1 - x'] \right\} , \n
Q^{(3,2)} &=& - s_1' s_2' \delta_{s_1, \bar s_2} \left\{ \frac{x}{x'} k' e^{-i s_1' \phi'} - k e^{- i s_1' \phi} \right\}
\left\{ k e^{-i s_2' \phi} - \frac{1-x}{1-x'} k' e^{-i s_2' \phi'} \right\} , \n
Q^{(3,3)} &=&  s_2 s_1' \delta_{s_1, s_2'} \left\{ \frac{x}{x'} k' e^{-i s_1' \phi'} - k e^{- i s_1' \phi} \right\}
\left\{ k' e^{i s_2 \phi'}  - \frac{1-x'}{1-x} k e^{i s_2 \phi} \right\} .
\eea

\noi
In \reff{ex}, if we expand everything multiplying the delta function in powers of transverse momenta, the lowest-order terms
from the two exponentials cancel
and leave two types of terms: those proportional to inverse powers of $\la$ and those proportional to inverse powers of 
$\la'$.  We can isolate the
terms that are proportional to inverse powers of $\la$ without altering the cancellation of the lowest term 
by replacing the first exponential with a $1$:

\bea
\left< g_1' g_2' \right| V^{(2)}_\CD(\la) \left| g_1 g_2 \right>_\EX &=& 64 \pi^3 P^+ \delta^{(3)}(P - P')
f^{c_1 c c_1'} f^{c c_2 c_2'} \frac{1}{(x-x')}
\theta(x-x'- \epsilon) \tim \left( \frac{1}{\Delta_{FK}} +
\frac{1}{\Delta_{IK}} \right) \left( 1 - e^{-2 \la^{-4} \Delta_{FK} \Delta_{IK}} \right)
\sum_{i,m=1}^3 Q^{(i,m)} .
\lll{exc}
\eea

\subsection{Combining the Interactions}

In order to get the infrared ($\epsilon \rightarrow 0$) divergences to cancel, it is useful to combine the
interactions in a particular manner.  From Eqs. (\ref{eq:split 1}), (\ref{eq:split 2}), and (\ref{eq:split 3}),

\bea
\left< g_1' g_2' \right| \M \left| g_1 g_2 \right> &=& \left< g_1' g_2' \right| \M \left| g_1 g_2 \right>_\KE
+ \left< g_1' g_2' \right| \M \left| g_1 g_2 \right>_\CON + \left< g_1' g_2' \right| \M \left| g_1 g_2 \right>_\IN \n
&+& \left< g_1' g_2' \right| \M \left| g_1 g_2 \right>_\SE + \left< g_1' g_2' \right| \M \left| g_1 g_2 \right>_\EX ,
\eea

\noi
where

\bea
\left< g_1' g_2' \right| \M \left| g_1 g_2 \right>_\KE &=& M_I^2 \left< g_1' g_2' \right| \left. \! g_1 g_2 \right> ,\n
\left< g_1' g_2' \right| \M \left| g_1 g_2 \right>_\CON &=& \gla^2 e^{-\la^{-4} \Delta_{FI}^2} 
\left< g_1' g_2' \right| V^{(2)}_\CI \left| g_1 g_2 \right>_\CON ,\n
\left< g_1' g_2' \right| \M \left| g_1 g_2 \right>_\IN &=& \gla^2 e^{-\la^{-4} \Delta_{FI}^2} 
\left< g_1' g_2' \right| V^{(2)}_\CI \left| g_1 g_2 \right>_\IN ,\n
\left< g_1' g_2' \right| \M \left| g_1 g_2 \right>_\SE &=& \gla^2 e^{-\la^{-4} \Delta_{FI}^2} 
\left< g_1' g_2' \right| V^{(2)}_\CD(\la) \left| g_1 g_2 \right>_\SE ,\n
\left< g_1' g_2' \right| \M \left| g_1 g_2 \right>_\EX &=& \gla^2 e^{-\la^{-4} \Delta_{FI}^2} 
\left< g_1' g_2' \right| V^{(2)}_\CD(\la) \left| g_1 g_2 \right>_\EX .
\lll{bsb results}
\eea

We break the instantaneous interaction into two parts: a part that is ``above'' the cutoff,
i.e. a part that would vanish if we took $\la \rightarrow \infty$, and a part that is ``below'' the cutoff,
i.e. a part that would survive if we took $\la \rightarrow \infty$:

\bea
\left< g_1' g_2' \right| \M \left| g_1 g_2 \right>_\IN = \left< g_1' g_2' \right| \M \left| g_1 g_2 \right>\INA + 
\left< g_1' g_2' \right| \M \left| g_1 g_2 \right>\INB,
\eea

\noi
where

\bea
\left< g_1' g_2' \right| \M \left| g_1 g_2 \right>\INA &=& \left( 1 - e^{-2 \la^{-4} \Delta_{FK} \Delta_{IK}} \right)
\left< g_1' g_2' \right| \M \left| g_1 g_2 \right>_\IN , \n
\left< g_1' g_2' \right| \M \left| g_1 g_2 \right>\INB &=& e^{-2 \la^{-4} \Delta_{FK} \Delta_{IK}}
\left< g_1' g_2' \right| \M \left| g_1 g_2 \right>_\IN .
\eea

\noi
Next, we break the self-energy and exchange interactions into finite and divergent parts:

\bea
\left< g_1' g_2' \right| \M \left| g_1 g_2 \right>_\SE = \left< g_1' g_2' \right| \M \left| g_1 g_2 \right>\SEF + 
\left< g_1' g_2' \right| \M \left| g_1 g_2 \right>\SED , \n
\left< g_1' g_2' \right| \M \left| g_1 g_2 \right>_\EX = \left< g_1' g_2' \right| \M \left| g_1 g_2 \right>\EXF + 
\left< g_1' g_2' \right| \M \left| g_1 g_2 \right>\EXD ,
\eea

\noi
where the divergent part of the self-energy interaction consists solely of the term containing the $\log \epsilon$, 
and the divergent part of the
exchange interaction
consists solely of the term containing $Q^{(2,1)}$.  Finally, we define an interaction that is a combination of the instantaneous
interaction ``above'' the cutoff and the divergent part of the exchange interaction:

\bea
\left< g_1' g_2' \right| \M \left| g_1 g_2 \right>_{\mathrm{IN+EX}} = \left< g_1' g_2' \right| \M \left| g_1 g_2 \right>\INA +
\left< g_1' g_2' \right| \M \left| g_1 g_2 \right>\EXD .
\eea

\noi
Then

\bea
\left< g_1' g_2' \right| \M \left| g_1 g_2 \right> &=& \left< g_1' g_2' \right| \M \left| g_1 g_2 \right>_\KE +
\left< g_1' g_2' \right| \M \left| g_1 g_2 \right>\SEF + 
\left< g_1' g_2' \right| \M \left| g_1 g_2 \right>_\CON \n &+&
\left< g_1' g_2' \right| \M \left| g_1 g_2 \right>\EXF +
\left< g_1' g_2' \right| \M \left| g_1 g_2 \right>_{\mathrm{IN+EX}}
+ \left< g_1' g_2' \right| \M \left| g_1 g_2 \right>\INB \n &+&
\left< g_1' g_2' \right| \M \left| g_1 g_2 \right>\SED .
\lll{free set}
\eea

Perry showed that with a suitable definition of long-range interactions, a renormalization 
method that is similar to ours yields a logarithmically confining potential
for quark-antiquark bound states at $\OR(\gla^2)$ \cite{brazil}.  His calculation uses sharp step-function cutoffs
and is based on an analysis of the part of the two-body interaction that is most singular in the
limit in which the exchanged gluon has infinitesimal longitudinal momentum.  The corresponding part of our interaction, which is contained
in $\left< g_1' g_2' \right| \M \left| g_1 g_2 \right>_{\mathrm{IN+EX}}$ and $\left< g_1' g_2' \right| \M \left| g_1 g_2 \right>\INB$,
is similar to what Perry found in the quark-antiquark case.  However, to determine whether or not our interaction is truly confining,
we would have to do a careful analysis of the complete two-body potential, not just the most singular part.  This analysis would be
complicated by the smooth cutoff that we employ, and we leave it for future consideration.

\section{Cal\-cu\-la\-tion of the Ma\-trix El\-e\-ments of the In\-var\-i\-ant-Mass Op\-er\-a\-tor in the Basis for Physical States}
\label{sec: bs me}

The matrix elements $\ME$, which appear in the eigenvalue equation, can be divided into contributions 
corresponding to the different terms in \reff{free set}:

\bea
\ME &=& \ME_\KE + \ME\SEF \n
&+& \ME_\CON + \ME\EXF \n
&+& \ME_\INX + \ME\INB \n
&+& \ME\SED .
\lll{bound set}
\eea

\noi
In this section we express these different contributions in terms
of integrals that can be computed numerically.  These integrals fall into two classes: two-dimensional and
five-dimensional.  We treat each class of integral separately.  

Each of the terms in \reff{bound set} is proportional to the
plane-wave normalization factor $16 \pi^3 P^+ \delta^{(3)}(P-P')$.  To make the remaining equations that we present simpler,
we suppress this factor.  We also take $\epsilon \rightarrow 0$ in any contribution to $\ME$ that is finite in this limit.

\subsection{The Two-Dimensional Integrals}

\subsubsection{The Kinetic Energy}

Using the definition of our basis in \reff{bsb} and the expression for the free-state matrix element
of the kinetic energy in \reff{bsb results}, we find that

\bea
&& \ME_\KE = \delta_{q, q'} \frac{1}{d^2} \int_0^1 dx \bar L^{(e)}_{l'}(x) \bar L^{(e)}_l(x) \int_0^\infty dr r^3
\T_{t'}(r) \T_t(r) .
\eea

\noi
Note that the
kinetic energy is infinite unless $e>0$, whereas normalizability requires only $e > -1/2$.

These integrals can be computed with standard numerical integration routines, but the results can have large errors for large
values of the function indices $l$, $l'$, $t$, and $t'$, due to the oscillating nature of the basis functions.  
It is better to rewrite the integrals as
sums that can be computed numerically with Mathematica \cite{math} to any desired precision.  Using the definitions
of the basis functions,

\bea
&& \ME_\KE = \delta_{q, q'} \frac{\Gamma(2 e)}{d^2} \Bigg[ \sum_{m=0}^l \lambda_{l,m}^{(e)} \sum_{m'=0}^{l'} \lambda_{l',m'}^{(e)}
\frac{\Gamma(2e+m+m')}{\Gamma(4e+m+m')} \Bigg] \tim \Bigg[ \sum_{s=0}^t \sigma_{t,s} \sum_{s'=0}^{t'} \sigma_{t',s'} 2^{-3-\frac{s+s'}{2}} 
\Gamma\left(2+\frac{s+s'}{2}\right) \Bigg] .
\eea

\subsubsection{The Finite Part of the Self-Energy Interaction}

The self-energy interaction conserves each particle's momentum, thus $M_I^2=M_F^2$ for this contribution.  This means
that the Gaussian cutoff factor in \reff{bsb results} has no effect on the self-energy.  For the finite part of the self-energy, we use the
same method for evaluating integrals that we use for the kinetic energy.  This yields

\bea
&&\ME\SEF = \delta_{q,q'} \delta_{t,t'} \frac{N_\mathrm{c} \gla^2}{4 \pi^2} \sqrt{\frac{\pi}{2}} \la^2 \int_0^1 dx \bar L^{(e)}_{l'}(x) 
\bar L^{(e)}_l(x) x (1-x) \tim \left[ \log x - \frac{11}{12} \right]
\lll{sed}\\
&=& -\delta_{q, q'} \delta_{t,t'} \frac{N_\mathrm{c} \gla^2}{4 \pi^2} \sqrt{\frac{\pi}{2}} \la^2 \bigg( \delta_{l,l'} \frac{11}{12} - 
\Gamma(1+2e) \sum_{m=0}^l \lambda_{l,m}^{(e)} \sum_{m'=0}^{l'} \lambda_{l',m'}^{(e)}
\frac{\Gamma(1+2e+m+m')}{\Gamma(2+4e+m+m')} \tim \left[ \psi(1+2e+m+m') - \psi(2+4e+m+m') \right] \bigg) ,
\eea

\noi
where the digamma function $\psi(z)$ is given by

\bea
\psi(z) = \frac{\frac{d \Gamma(z)}{dz}}{\Gamma(z)} .
\eea

\subsection{The Five-Dimensional Integrals}

\subsubsection{The Contact Interaction}

Using the definition of our basis in \reff{bsb} and the expression for the free-state matrix element
of the contact interaction in Eqs. (\ref{eq:contact}) and (\ref{eq:bsb results}), we find that

\bea
&& \ME_\CON = -\frac{N_\mathrm{c} \gla^2}{8 \pi^2} 
\Bigg[ \delta_{j,-2} \delta_{q',2} \delta_{q,2} + \delta_{j, 2}
\delta_{q', 1} \delta_{q, 1} - \delta_{j,0} \delta_{q',3} \delta_{q,3} \Bigg]
\int dk \BIT dk' \BIT k \BIT k' \tim \theta(k) \theta(k') T^{(d)}_{t'}(k') T^{(d)}_t(k) \int dx \BIT dx' \BIT \theta(x) \theta(1-x) \theta(x') 
\theta(1-x') \bar L^{(e)}_{l'}(x') \bar L^{(e)}_l(x) e^{-\la^{-4} \Delta_{FI}^2} .
\lll{contact me}
\eea

\noi
The reader may have noticed that this integral is not five-dimensional as we have implied, but rather four-dimensional.
However, when we numerically compute the integrals that have more than two dimensions, it is most efficient if we combine them
into one integral; so we want their integration variables and their ranges
of integration to be identical.  Thus we increase the number of dimensions of this integral by one by introducing an extra integral over
$\gamma = \phi - \phi'$ using the identity

\bea
1 = \frac{1}{2 \pi} \int d\gamma \BIT \theta(\gamma) \BIT \theta(2 \pi - \gamma) .
\eea

\noi
This will help us to combine this integral with others that contain integrals over $\gamma$ that cannot be done analytically.

Since the integration domain of the exchange interaction is restricted so that $x > x'$ when $\epsilon \rightarrow 0$ [note the step function in
\reff{exc}], we would like to enforce this restriction in the other contributions.  Using the identity

\bea
1 = \theta(x - x') + \theta(x' - x) ,
\eea

\noi
we break the longitudinal integral in \reff{contact me} into two parts:

\bea
\int dx \BIT dx' \BIT \theta(x) \theta(1-x) \theta(x') 
\theta(1-x') \bar L^{(e)}_{l'}(x') \bar L^{(e)}_l(x) e^{-\la^{-4} \Delta_{FI}^2} [\theta(x - x') + \theta(x' - x) ] .
\eea

\noi
In the second term, we let $x \rightarrow 1-x$ and $x' \rightarrow 1-x'$, and then the longitudinal integral becomes

\bea
\int dx \BIT dx' \BIT \theta(x) \theta(1-x) \theta(x') 
\theta(1-x') \bar L^{(e)}_{l'}(x') \bar L^{(e)}_l(x) e^{-\la^{-4} \Delta_{FI}^2} \theta(x - x') \left[ 1 + (-1)^{l+l'} \right].
\lll{contact split}
\eea

\noi
Recall that we are restricting ourselves to the subspace of the states $\left| q,l,t,j \right>$ in which $l+j$ is even if $q \neq 4$,
and $l+j$ is odd if $q=4$.  Then since $q=q'$ for the contact interaction, $l$ and $l'$ must both be even or both be odd for this
interaction.
This means that the two terms in \reff{contact split} are equal.  Thus we can write the contact interaction contribution as follows:

\bea
\ME_\CON &=& -\frac{N_\mathrm{c} \gla^2}{8 \pi^3} 
\Bigg[ \delta_{j,-2} \delta_{q',2} \delta_{q,2} + \delta_{j, 2}
\delta_{q', 1} \delta_{q, 1} - \delta_{j,0} \delta_{q',3} \delta_{q,3} \Bigg] \tim
\int D \BIT \xi \BIT e^{-\la^{-4} \Delta_{FI}^2} ,
\eea

\noi
where

\bea
D = dx \BIT dx' \BIT dk \BIT dk' \BIT d\gamma \BIT k \BIT k' \BIT \theta(x) \BIT \theta(1-x) \BIT \theta(x') \BIT \theta(x-x') \BIT \theta(k) 
\BIT \theta(k') \BIT \theta(\gamma) \BIT \theta(2 \pi - \gamma) ,
\eea

\noi
and

\bea
\xi = \bar L^{(e)}_{l'}(x') \bar L^{(e)}_l(x) T^{(d)}_{t'}(k') T^{(d)}_t(k) .
\eea

\subsubsection{The Finite Part of the Exchange Interaction}

In order to simplify the contribution to $\ME$ from the finite part of the exchange interaction, we wish to change
variables from $\phi$ to $\gamma=\phi-\phi'$:

\bea
d^2 k_\perp d^2 k_\perp' &=& dk \BIT dk' \BIT d \phi \BIT d \phi' \BIT k \BIT k' \BIT \theta(k) \BIT \theta(k') \BIT \theta(\phi) \BIT 
\theta(2 \pi - \phi) \BIT \theta(\phi') 
\BIT \theta( 2 \pi - \phi') \n
&=& dk \BIT dk' \BIT d \gamma \BIT d \phi' \BIT k \BIT k' \BIT \theta(k) \BIT \theta(k') \BIT \theta(\gamma + \phi') \BIT 
\theta(2 \pi - \gamma - \phi') \BIT \theta(\phi') \BIT \theta( 2 \pi - \phi') .
\eea

\noi
All the contributions to $\ME$ depend on $\gamma$ only through dependence on $\cos \gamma$ and $\sin \gamma$.  This
means that we can use the identity

\bea
\int_{-\phi'}^{2 \pi - \phi'} d \gamma f(\cos \gamma, \sin \gamma) = \int_{0}^{2 \pi} d \gamma f(\cos \gamma, \sin \gamma)
\eea

\noi
to write

\bea
d^2 k_\perp d^2 k_\perp' = dk \BIT dk' \BIT d \gamma \BIT d \phi' \BIT k \BIT k' \BIT \theta(k) \BIT \theta(k') \BIT \theta(\gamma) \BIT 
\theta(2 \pi - \gamma) \BIT \theta(\phi') \BIT \theta( 2 \pi - \phi') .
\eea

Inspection of \reff{ex} implies that the integrals in $\ME\EXF$ depend on complex exponentials of $\phi'$ and $\gamma$.  
However, using the identity

\bea
\int_{0}^{2 \pi} d \gamma f(\cos \gamma) \sin a \gamma &=& 0,
\eea

\noi
where $a$ is an integer, it is possible to trivially do the $\phi'$ integral in $\ME\EXF$ and write the remainder as a real quantity 
with integrals that depend
on $\gamma$ only through $\cos \gamma$ and $\sin \gamma$.  The result is

\bea
&& \ME\EXF = -\frac{N_\mathrm{c} \gla^2}{8 \pi^3} \int D \BIT \xi \BIT e^{-\la^{-4} \Delta_{FI}^2} \frac{1}{x-x'}  \left( \frac{1}{\Delta_{FK}} +
\frac{1}{\Delta_{IK}} \right) \tim \left( 1 - e^{-2 \la^{-4} \Delta_{FK} \Delta_{IK}} \right)
\left[ M_I^2 S_{q,q'}^{(1)} + M_F^2 S_{q,q'}^{(2)} + \frac{k k'}{x (1-x) x' (1-x')} S_{q,q'}^{(3)} \right] ,
\eea

\noi
where some of the $S^{(1)}_{q,q'}$'s and $S^{(3)}_{q,q'}$'s are given by

\bea
S_{1,1}^{(1)} &=& \cos ([j-2] \gamma) ,\n
S_{1,1}^{(3)} &=& -\cos ([j-1]\gamma) [x (1-x') + x' (1-x)] ,\n
S_{1,3}^{(1)} &=& \frac{-1}{\sqrt{2}} \cos (j \gamma) [\xpsq + (1-x')^2]  ,     \n
S_{1,3}^{(3)} &=& \frac{1}{\sqrt{2}} \cos ([j-1] \gamma) (x [1-x'] + x' [1-x]) (\xpsq + [1-x']^2) , \n
S_{1,4}^{(1)} &=& \frac{1}{\sqrt{2}} \cos (j \gamma) (1 - 2 x')   ,    \n
S_{1,4}^{(3)} &=& \frac{-1}{\sqrt{2}} \cos ([j-1] \gamma) (1-2x') (x [1-x'] + x' [1-x]) ,\n
S_{3,1}^{(1)} &=& \frac{-1}{\sqrt{2}} \cos ([j-2] \gamma) [x^2 + (1-x)^2] ,       \n
S_{3,3}^{(1)} &=& \cos (j \gamma) [x^2 + (1-x)^2 - 2 x' (1-x')] ,\n
S_{3,3}^{(3)} &=& -\cos (j \gamma) \cos \gamma (x[1-x']+x'[1-x])(1-2x[1-x]-2x'[1-x'])   ,     \n
S_{3,4}^{(1)} &=& 0     ,   \n
S_{3,4}^{(3)} &=& \sin \gamma \sin(j \gamma) (2x^3-2x^2[1+x']+x'[1-2x']+x[1-2x'+4\xpsq])  ,     \n
S_{4,1}^{(1)} &=& \frac{1}{\sqrt{2}} \cos ([j-2] \gamma) (1 - 2 x)      ,  \n
S_{4,3}^{(1)} &=& 0    ,    \n
S_{4,4}^{(1)} &=& \cos (j \gamma) (1 - 2 x - 2 x' + 4 x x')  ,   \n
S_{4,4}^{(3)} &=& -\cos \gamma \cos (j \gamma) (1-2x) (1-2x') (x+x'-2xx') ,
\eea

\noi
and

\bea
S^{(3)}_{q,q'} = S^{(3)}_{q',q} \, \rule[-2.1mm]{.1mm}{5.4mm}_{\; x \leftrightarrow x'}.
\eea

\noi
The rest of the $S^{(1)}_{q,q'}$'s and $S^{(3)}_{q,q'}$'s are given by

\bea
S_{1,2}^{(i)} &=& S_{2,1}^{(i)} = 0,\n
S_{2,2}^{(i)} &=& S_{1,1}^{(i)} \,\rule[-2.1mm]{.1mm}{5.4mm}_{\: j \rightarrow -j}, \n
S_{2,3}^{(i)} &=& S_{1,3}^{(i)} \,\rule[-2.1mm]{.1mm}{5.4mm}_{\: j \rightarrow -j}, \n
S_{2,4}^{(i)} &=& -S_{1,4}^{(i)} \,\rule[-2.1mm]{.1mm}{5.4mm}_{\: j \rightarrow -j}, \n
S_{3,2}^{(i)} &=& S_{3,1}^{(i)} \,\rule[-2.1mm]{.1mm}{5.4mm}_{\: j \rightarrow -j}, \n
S_{4,2}^{(i)} &=& -S_{4,1}^{(i)} \,\rule[-2.1mm]{.1mm}{5.4mm}_{\: j \rightarrow -j} ,
\eea

\noi
where $i=1,3$.  The $S^{(2)}_{q,q'}$'s are given by

\bea
S^{(2)}_{q,q'} = S^{(1)}_{q',q} \:\rule[-2.5mm]{.1mm}{6mm}_{\; x \leftrightarrow x'}.
\eea

\subsubsection{The Instantaneous and Exchange Interactions Combination}

Using similar methods for the contribution to $\ME$ from the combination of the instantaneous interaction 
above the cutoff and the divergent part of the
exchange interaction, we find that

\bea
&& \ME_\INX = -\frac{N_\mathrm{c} \gla^2}{8 \pi^3} \int D \BIT \xi \BIT W_{q,q'}
e^{-\la^{-4} \Delta_{FI}^2} \frac{1}{(x-x')^2} \left( 1 - e^{-2 \la^{-4} \Delta_{FK} \Delta_{IK}}
\right) \tim \Bigg[ (x+x') (1-x+1-x') + \frac{2}{x-x'}
\bigg( \frac{1}{\Delta_{FK}} + \frac{1}{\Delta_{IK}} \bigg)
\bigg( x (1-x) \kpsq + x' (1-x') k^2 \n
&-& k k' \left[ x' (1-x) + x (1-x') \right] \cos \gamma \bigg) \Bigg] ,
\lll{combo}
\eea

\noi
where

\bea
W_{q,q'} = \delta_{q,1} \delta_{q',1} \cos (\gamma [j-2]) + \delta_{q,2} \delta_{q',2} 
\cos (\gamma [j+2]) + \delta_{q,3} \delta_{q',3} \cos (\gamma j) + \delta_{q,4} \delta_{q',4} \cos
(\gamma j) .
\eea

\noi
Note that the divergences from the two interactions that comprise $\ME_\INX$ cancel, allowing us to
take $\epsilon \rightarrow 0$ in this contribution.

\subsubsection{The Instantaneous Interaction Below the Cutoff}

The contribution to $\ME$ from the instantaneous interaction below the cutoff is divergent.  In this subsection,
we extract the divergence, show that it cancels the divergent part of the self-energy, and compute the remainder
of $\ME\INB$.

After simplification, the complete contribution to $\ME$ from the instantaneous interaction below the cutoff is

\bea
&& \ME\INB = - \frac{N_\mathrm{c} \gla^2}{16 \pi^4} \int_{2 \epsilon}^{1-\epsilon} dx \int_\epsilon^{x-\epsilon} dx' \frac{1}{x-x'} F(x,x') ,
\lll{instant me}
\eea

\noi
where

\bea
F(x,x') = \int d^2 k_\perp d^2 k_\perp' \frac{\xi}{x-x'} W_{q,q'} e^{-\la^{-4} \Delta_{FI}^2} 
e^{-2 \la^{-4} \Delta_{FK} \Delta_{IK}} (x+x') (1-x+1-x') .
\eea

\noi
To extract the divergence, we integrate \reff{instant me} by parts with respect to $x'$:

\bea
&& \ME\INB = - \frac{N_\mathrm{c} \gla^2}{16 \pi^4} \int_{2 \epsilon}^{1-\epsilon} dx \Bigg[
- \log(x-x') F(x,x') \, \rule[-2.1mm]{.1mm}{4.7mm}_{\; x'=x-\epsilon} \n
&+& \log(x-x') F(x,x') \, \rule[-2.1mm]{.1mm}{4.7mm}_{\; x'=\epsilon} + \int_\epsilon^{x-\epsilon} dx' \log(x-x') \frac{dF(x,x')}{dx'} \Bigg] \n
&\equiv& B_1 + B_2 + B_3 .
\eea

The first contribution to $\ME\INB$ is

\bea
B_1 &=& \frac{N_\mathrm{c} \gla^2}{16 \pi^4} \frac{\log\epsilon}{\epsilon} \int_{2 \epsilon}^{1-\epsilon} dx 
\int d^2 k_\perp d^2 k_\perp' \xi \BIT W_{q,q'} e^{-\la^{-4} \Delta_{FI}^2} 
e^{-2 \la^{-4} \Delta_{FK} \Delta_{IK}} (x+x') \tim (1-x+1-x') \, \rule[-2.1mm]{.1mm}{4.7mm}_{\; x'=x-\epsilon} .
\eea

\noi
To simplify this, we change variables from $x$ to $y$:

\bea
x = y (1-3\epsilon) + 2 \epsilon ,
\eea

\noi
and from $\kp$ and $\kpp$ to $\Qp$ and $\Np$:

\bea
\kp &=& \frac{\Qp + \sqrt{\epsilon} \Np}{2} , \n
\kpp &=& \frac{\Qp - \sqrt{\epsilon} \Np}{2} .
\eea

\noi
Then

\bea
B_1 &=& \frac{N_\mathrm{c} \gla^2}{64 \pi^4} (1-3\epsilon) \log\epsilon \int_0^1 dy 
\int d^2 Q_\perp d^2 N_\perp \xi \BIT W_{q,q'} e^{-\la^{-4} \Delta_{FI}^2} 
e^{-2 \la^{-4} \Delta_{FK} \Delta_{IK}} (2y[1-3\epsilon]+3\epsilon) \tim (2-2y[1-3\epsilon]-3\epsilon) \, \rule[-2.1mm]{.1mm}{4.7mm}_{\; x'=x-\epsilon} .
\eea

\noi
As $\epsilon \rightarrow 0$, the only contribution that survives is

\bea
B_1 &=& \frac{N_\mathrm{c} \gla^2}{16 \pi^4} \delta_{q,q'} \log\epsilon \int_0^1 dy \BIT y (1-y) \bar L^{(e)}_{l'}(y) \bar L^{(e)}_l(y) 
\int d^2 Q_\perp T^{(d)}_{t'}(Q/2) T^{(d)}_t(Q/2) \int d^2 N_\perp e^{-2 \la^{-4} N^4} \n
&=& \delta_{q,q'} \delta_{t,t'} \frac{N_\mathrm{c} \gla^2}{4 \pi^2} \sqrt{\frac{\pi}{2}} \la^2 
\int_0^1 dx \BIT \bar L^{(e)}_{l'}(x) \bar L^{(e)}_l(x) x (1-x) \log\epsilon .
\eea

\noi
Since $\left<g_1'g_2'\right|\M\left|g_1 g_2\right>\SED$ is the part of $\left<g_1'g_2'\right|\M\left|g_1 g_2\right>_\SE$ with the $\log \epsilon$
[see \reff{sei}], from \reff{sed} we see that $\ME\SED$ is just $\ME\SEF$ with the $[\log x - 11/12]$ factor replaced with $-\log \epsilon$.  
This means that

\bea
B_1 = - \ME\SED.
\eea

\noi
Thus the divergence in $\ME\INB$ cancels $\ME\SED$.

The second contribution to $\ME\INB$ is

\bea
B_2 &=& - \frac{N_\mathrm{c} \gla^2}{8 \pi^3} \int_{2 \epsilon}^{1-\epsilon} dx \log(x-x') \int dk \BIT dk' \BIT d \gamma \BIT k \BIT k' \BIT 
\theta(k) \BIT \theta(k') \BIT \theta(\gamma) \BIT 
\theta(2 \pi - \gamma) \frac{\xi}{x-x'} W_{q,q'} 
e^{-\la^{-4} \Delta_{FI}^2} \tim e^{-2 \la^{-4} \Delta_{FK} \Delta_{IK}} (x+x') (1-x+1-x') \, \rule[-2.1mm]{.1mm}{4.7mm}_{\; x'=\epsilon} .
\eea

\noi
To evaluate this, we change variables from $k'$ to $s=k'/\sqrt{\epsilon}$.  Then the leading term as $\epsilon \rightarrow 0$ is

\bea
B_2 &=& - \epsilon^{e+\frac{1}{2}} \frac{N_\mathrm{c} \gla^2}{8 \pi^3} \BIT d \BIT \sigma_{t',0} \lambda_{l,0}^{(e)} 
\int_0^1 dx (2-x) \log x \int dk \BIT ds \BIT d \gamma \BIT k \BIT s \BIT 
\theta(k) \BIT \theta(s) \BIT \theta(\gamma) \BIT \theta(2 \pi - \gamma) \bar L_l^{(e)}(x) \tim T_t^{(d)}(k) W_{q,q'} 
e^{-\la^{-4} \left[ s^4 + \frac{k^4}{x^2 (1-x)^2} \right]} \n
&=& 0,
\eea

\noi
since $e>-1/2$.

To simplify the third contribution to $\ME\INB$, we take the derivative of $F(x,x')$ and take $\epsilon \rightarrow 0$.  Then

\bea
B_3 = -\frac{N_\mathrm{c} \gla^2}{8 \pi^3} \int D \log(x-x') W_{q,q'} T^{(d)}_{t'}(k') \bar L^{(e)}_l(x) 
T^{(d)}_t(k) e^{-\la^{-4} (\Delta_{FK}^2 + \Delta_{IK}^2)} \sum_{i=1}^5 E'_i 
\prod_{m=1; \; m \neq i}^5 E_m ,
\eea

\noi
where

\bea
E_1 &=& \frac{1}{x-x'} ,\n
E_2 &=& 1 ,\n
E_3 &=& \bar L^{(e)}_{l'}(x') , \n
E_4 &=& x+x' ,\n
E_5 &=& 1-x + 1-x' ,
\eea

\noi
and

\bea
E'_1 &=& \frac{1}{(x-x')^2} ,\n
E'_2 &=& - 2 \la^{-4} (\Delta_{FK} \Delta_{FK}' + \Delta_{IK} \Delta_{IK}' ) ,\n 
E'_3 &=& \bar L^{\prime (e)}_{l'}(x') ,\n 
E'_4 &=& 1 ,\n
E'_5 &=& -1,
\eea

\noi
($E'_i=dE_i/dx'$, except for $i=2$) and

\bea
\bar L^{\prime (e)}_{l'}(x') &=& \frac{d \bar L^{(e)}_{l'}(x')}{dx'} \n
&=& \left( e - \frac{1}{2} \right)  [x'(1-x')]^{e-\frac{1}{2}-1} [1-2x'] \sum_{m'=0}^{l'} \lambda^{(e)}_{\,l',m'} x^{\prime \, m'} + 
[x'(1-x')]^{e-\frac{1}{2}} \tim \sum_{m'=1}^{l'} m' \lambda^{(e)}_{\,l',m'} x^{\prime \; m'-1} \!\!,
\eea

\noi
and

\bea
\Delta'_{FK} &=& \frac{d \Delta_{FK}}{dx'} = \frac{\kpsq}{(1-x')^2} - \frac{(\kp - \kpp)^2}{(x-x')^2} , \n
\Delta'_{IK} &=& \frac{d \Delta_{IK}}{dx'} = \frac{\kpsq}{\xpsq} - \frac{(\kp - \kpp)^2}{(x-x')^2} .
\eea

\noi
This means that we can write the contribution to $\ME$ from the instantaneous interaction below the cutoff as a divergent
part and a finite part:

\bea
\ME\INB = -\ME\SED + \ME\INBF ,
\eea

\noi
where 

\bea
\ME\INBF &=& -\frac{N_\mathrm{c} \gla^2}{8 \pi^3} \int D \log(x-x') W_{q,q'} T^{(d)}_{t'}(k') \bar L^{(e)}_l(x) 
T^{(d)}_t(k) e^{-\la^{-4} (\Delta_{FK}^2 + \Delta_{IK}^2)} \tim \sum_{i=1}^5 E'_i 
\prod_{m=1; \; m \neq i}^5 E_m .
\lll{inbf}
\eea

Using the results of this section, the expression in \reff{bound set} for the matrix elements of the IMO becomes

\bea
&& \ME = \ME_\KE + \ME\SEF \n
&+& \ME_\CON + \ME\EXF \n
&+& \ME_\INX + \ME\INBF .
\eea

\noi
Each of these terms is finite and we have taken $\epsilon \rightarrow 0$ everywhere.
We have written the first two terms as sums that can be computed numerically, and the four remaining terms as five-dimensional integrals
that can be grouped
into one integral suitable for numerical calculation.  (See Appendix C for a discussion of some
of the technical issues involved in the numerical calculation of these matrix elements.)  Once we have computed these matrix elements, we
can diagonalize the matrix to obtain glueball states and masses.  

\section{Results and Error Analysis}

In this section we diagonalize the IMO matrix, obtaining glueball states and masses.  We then discuss the sources of 
error in the calculation.  We begin by discussing 
the procedure that we use to calculate the results.

\subsection{The Procedure for Calculating Results}

We represent a state using the notation
$J^{PC}_{j}$, where $J$ is our best guess for the spin of the state, $P$ is our best guess for the parity of the state, $C$ represents 
the charge-conjugation
eigenvalue of the state (it is always $+$ because we have an even number of gluons), and $j$ is the projection of the state's spin onto the 3-axis.  
We need to
distinguish states with identical $J$'s and $P$'s and different $j$'s because we do not have manifest rotational symmetry.  If $J=0$, we
omit the subscript $j$ in the state notation.  We use an asterisk in the state notation next to the value of $C$ to denote 
an excited state with the given quantum numbers.  We will base our
guesses for $J$ and $P$ on the numerical degeneracies of the states that have identical $J$'s and $P$'s and 
different $j$'s, and on the ordering of the states 
according to lattice data (see the discussion below).  
We consider only the five lightest glueballs (not counting as distinct those states that differ only in their value of $j$).  
The five lightest glueballs have spins $J \le 2$.  This means that we need to consider only $|j| \le 2$.  
For a given $|j|$, the states with $j=|j|$ and $j=-|j|$ are degenerate and simply related (see Appendix C); so we explicitly consider only $j=0,1,2$.  
We consider nine values of the coupling: $\ala=\gla^2/(4 \pi)=0.1,0.2,0.3,\ldots,0.9$.  To calculate our results, we implement
the following four-step procedure.

We execute the first step for all pairs $(j,\ala)$.  In this step, 
we define $\la=1$ and diagonalize $\ME$ with all 4 spin basis functions ($q=1,2,3,4$), 
but with only the lowest transverse-magnitude basis function ($t=0$) and the two lowest longitudinal basis functions ($l=0,1$).  
(It is necessary
to use an even number of longitudinal basis functions so that symmetry and antisymmetry of the wave function under $x \rightarrow 1-x$ 
are
equally represented.)  We perform this diagonalization as a function of the basis-function parameters $d$ and $e$ that determine the
widths of the transverse-magnitude and longitudinal wave functions, respectively, and we find the values of $d$ and $e$ that 
minimize the ground-state mass.  This yields what we consider to be the optimal wave-function widths for each pair $(j,\ala)$.

We also execute the second step of the procedure for all pairs $(j,\ala)$.  In this step, 
we fix $d$ and $e$ to be their computed optimal values and again define $\la=1$.  We diagonalize the matrix with all 4
spin basis functions, $\Nt$ transverse-magnitude 
basis functions, and $\Nl=2 \Nt$ longitudinal basis functions, for a total of $8 \Nt^2$ basis functions, with $\Nt=1,2,3,\ldots,10$.
We use twice as many longitudinal functions as transverse-magnitude functions because $\left|q,l,t,j\right>$ is zero if $l+j$ is even and
$q=4$, or if $l+j$ is odd and $q \neq 4$.  We want to use as many basis functions as possible, but we find for all pairs $(j,\ala)$ that
when $\Nt > 7$, the statistical errors from the Monte Carlo integrations of the matrix elements become overwhelming and
the spectrum and wave functions become unreliable.  The evidence of the breakdown is sudden contamination of the low-lying wave functions
with high-order components.  (See Appendix C for a more complete discussion of this topic.)  Thus in the remainder of our procedure, we analyze
the results that we find in this step when $\Nt=7$, which corresponds to 392 basis functions.

In the third step of the procedure, we use the mass of the $0^{-+}$ state (our most numerically reliable state) to determine the value of 
$\la$ for each $\ala$.  To do this, we note that $\la$ is the
only mass scale in the problem.  This means that the mass of the $0^{-+}$ state, $M_{0^{-+}}$, can be written

\bea
M_{0^{-+}}=b(\ala) \la ,
\lll{mass cutoff}
\eea

\noi
where $b$ is a dimensionless function of $\ala$.  
Since we defined $\la=1$ in the second step of our procedure, the diagonalization of the IMO as a function of $\ala$ yielded $b(\ala)$.  
In this third step, we consider $\la$ to
be a parameter and define $M_{0^{-+}}$ to be a constant.  Then for a given coupling, we can use the results of the second step of our procedure 
to write the cutoff in units of $M_{0^{-+}}$:

\bea
\la/M_{0^{-+}}=\frac{1}{b(\ala)} .
\lll{mass cutoff 2}
\eea

Figure 5 shows the result for the third step of our procedure: a plot of the coupling as a function of the cutoff.  When $\ala > 0.7$
it is not a single-valued function of the cutoff.  This is an indication that the coupling is too large.  For this reason, we consider 
only $\ala \le 0.7$ in the remainder of our procedure.  When $\ala \le 0.7$, the coupling decreases as the
cutoff increases, as expected.  However, $\ala$ depends on $\la$ more strongly than one may expect.
We expect that perturbative pure-glue QCD would indicate that $\ala \sim 1/\ln \la$, but the result that we get is much closer to 
$\ala \sim \exp(-a \la)$, where $a$
is a constant.  The reason for this is that the truncation of the perturbative series for $\M$ and the truncation of the free-sector expansion of the 
states introduce spurious cutoff dependence in our results for physical quantities.  $M_{0^{-+}}$ is one such physical quantity.
The spurious cutoff dependence of $M_{0^{-+}}$ is manifested through incorrect dependence of $b(\ala)$
on $\ala$.  This means that $\la$ has to compensate by depending on $\ala$ incorrectly in order to keep $M_{0^{-+}}$ a constant
function of $\ala$.  This
results in the strong dependence of $\ala$ on $\la$ that is shown in Figure 5.  

Using the recent anisotropic Euclidean lattice results of Morningstar and Peardon \cite{morningstar}, 
we can make a rough estimate of the range over which our cutoff is varying in Figure 5.  
They found that the mass of the $0^{-+}$ state is 
$M_{0^{-+}} = 2.590 \pm 0.040 \pm 0.130 $ GeV.  This means that our cutoff is varying from about 3.1 -- 6.0 GeV in Figure 5.

The fourth step of our procedure is to determine the optimal value of the cutoff, or equivalently, the optimal value of the coupling.
We use two criteria to determine this.  First we determine the value of the cutoff for which the computed masses are most
independent of the cutoff.  Figures 6, 7, and 8 show the masses of our states with $j=0,1,2$, respectively, as functions of the cutoff.
The masses and the cutoff are displayed in units of $M_{0^{-+}}$. (Recall that $M_{0^{-+}}$ was defined 
to be independent of the cutoff in the process of defining the cutoff as a function of the coupling.)  
The seven values of the cutoff at the points that we display correspond, from right to left, to 
$\ala=0.1,0.2,0.3,\ldots,0.7$.  It is difficult to tell from these plots where the cutoff dependence is weakest (more points are needed), 
but we see that 
the dependence is relatively weak from $\ala=0.5$ to $\ala=0.7$, which corresponds to $\la/M_{0^{-+}}=1.33$ to $\la/M_{0^{-+}}=1.20$.

The second criterion that we use to determine the optimal cutoff is the degree to which the states with a given $J$ and $P$ and 
different $j$'s are degenerate.
This determines the cutoff that minimizes the violation of rotational symmetry.   We find that these degeneracies are best when 
$\ala=0.5$.
Given this and the fact that Figures 6 -- 8 indicate that the cutoff dependence of the masses is weak when $\ala=0.5$ to $\ala=0.7$, 
we determine that $\ala=0.5$ is the optimal coupling, and thus the optimal cutoff is $\la/M_{0^{-+}}=1.33$.  
Using the result of Morningstar and Peardon for the mass of the $0^{-+}$ state, we estimate that this cutoff is about 3.4 GeV.

\subsection{Results}

Now we present our main results.  Our glueball masses for $\ala=0.5$ 
are summarized in Table 1, in units of the mass of the ground state (the $0^{++}$ state).  
Table 1 also shows an average of
lattice results from a number of different calculations for the sake of comparison \cite{teper}.  
The uncertainties in our results that we report in Table 1 are only the statistical uncertainties
associated with the Monte Carlo evaluation of the matrix elements of $\M$.  The full errors are much larger 
(see the discussion of sources of error below).  
We list three values of the masses for the $2^{++}$ and $2^{++*}$ states for our calculation, corresponding to $j=0,1,2$.  
In each case the three masses would be degenerate if our
calculation were exact.  Our results agree with the lattice results quite well, perhaps better than we should have expected.

We display our spectrum graphically in Figure 9.  The masses are plotted in units of the mass of the $0^{++}$ state and 
the vertical widths of the levels represent the statistical uncertainties in the masses.
The black lines connect the states that we believe should be degenerate.  We see that the $2^{++}_0$ and $2^{++*}_0$ glueballs are 
relatively
degenerate with their $2^{++}_1$ and $2^{++*}_1$ counterparts, and the $2^{++*}_2$ is not too bad, but the $2^{++}_2$ glueball is much 
too light.
Our labeling of the states and subsequent assignment of the expected degeneracies are based on the ordering of the lattice states 
(see Table 1) and 
the apparent degeneracies of the $2^{++}_0$ and $2^{++*}_0$ states with the $2^{++}_1$ and $2^{++*}_1$ states, respectively.

We want to show some of the features of the glueball wave functions.  Rather than presenting the spin-dependent wave functions themselves, 
we present more illuminating spin-independent probability
densities.  A glueball state has the plane-wave normalization shown in \reff{norm} as long as the
wave function $\Phi^{jn}_{s_1 s_2}(x,\kp)$ satisfies

\bea
\int d^2 k_\perp dx \theta(x) \theta(1-x) \sum_{s_1 s_2} \big| \Phi^{jn}_{s_1 s_2}(x,\kp) \big|^2 = 1.
\eea

\noi
This implies that

\bea
\int_0^\infty d\left(\frac{k}{\la}\right) \int_0^1 dx \BIT \Pi(x,k/\la) = 1,
\eea

\noi
where we define the dimensionless probability density $\Pi(x,k/\la)$ by

\bea
\Pi(x,k/\la) = 2 \pi \la k \sum_{s_1 s_2} \big| \Phi^{jn}_{s_1 s_2}(x,\kp) \big|^2 .
\eea

We show the probability densities for some of our glueballs in Figures 10 -- 14. The masses
of the states tend to increase as the probability densities move away from the region $x \sim 1/2$ and more towards the edges.  
This is what we expect based on the form of the kinetic energy of a free state.
Notice that the probability density for the $2^{++}_2$ glueball is peaked around the region $x \sim 1/2$ and looks similar to the
probability density for the $0^{++}$ glueball.  This is consistent with its small mass.

\subsection{Error Analysis}

We now turn to a discussion of the sources of error in our calculation.  The sources of error are:
\bt
\hspace{5mm} $\bullet$ truncation of the renormalized IMO at $\OR(\gla^2)$;\\
\hspace{5mm} $\bullet$ truncation of the free-sector expansion of physical states at two gluons;\\
\hspace{5mm} $\bullet$ truncation of the basis-function expansion of wave functions;\\
\hspace{5mm} $\bullet$ numerical approximation of the matrix elements of $\M$.
\et

We do not know how to estimate the size of any physical effects that require nonperturbative renormalization.  However, we can naively 
estimate the size of the
effects of higher-order perturbative renormalization.  We have calculated the matrix elements of $\M$
through $\OR(\ala)$; so the corrections to
these matrix elements should be $\OR(\ala^2)$.  
This translates to corrections to the mass spectrum of $\OR(\ala^2/2)$.  Since we have used
$\ala=0.5$, we estimate that the uncertainty in the mass spectrum from the effects of higher-order perturbative renormalization is about 13\%.

We do not know how to estimate the size of any physical effects that require an infinite number of particles.  In fact, until we include at
least two free sectors, it is impossible to directly estimate the size of corrections from higher free sectors.  However, we can use the 
lack of degeneracy of the $2^{++}_2$ state with the $2^{++}_0$ and $2^{++}_1$ states to estimate these corrections.  According to 
Table 1,
the discrepancy in the various $2^{++}$ states is about 33\%, if we believe the quoted lattice result.  Since the uncertainty in the 
mass spectrum from effects of higher-order perturbative renormalization is around 13\%, an uncertainty of 30\% due to the truncation of the
free-sector expansion is necessary to explain the lack of rotational symmetry in the spectrum (neglecting the other sources of error, 
which we expect to be small).

As we mentioned, when we increase the number of basis functions that we use to represent the wave functions, we find that we reach a point where the
statistical errors from the Monte Carlo evaluations of the matrix elements are overwhelming and cause our results to become completely
unreliable (see Appendix C).  For this reason, we have to truncate our basis-function expansion for the wave functions at $\Nt=7$
transverse-magnitude functions. ($\Nl=2 \Nt=14$, and there are 4 spin basis functions, for a total of 392 basis functions.)
This truncation results in additional errors in our results.  In Figures 15, 16, and 17, we show the convergence of the masses of the
states with $j=0,1,2$ respectively, as functions of $\Nt$.  The masses do not decrease as rapidly as functions of the number of states 
as one might expect.  This is primarily
because we have already optimized the states quite a bit by determining the widths of the transverse-magnitude and longitudinal
basis functions that minimize the mass of the $0^{++}$ state (using $\Nt=1$).  Our best guess for the uncertainty that we introduce into the
spectrum when we truncate the basis-function expansion, based on Figures 15 -- 17, is a few percent\footnote{Technically, this is not
an uncertainty because improving the states can only reduce their masses, according to the variational principle.  However, our discussion of errors
is not meant to be rigorous.}.

We can estimate the uncertainty in our results associated with the Monte Carlo evaluation of our matrix elements.  To do this, we
compute our results with $\ala=0.5$ four times, obtaining statistically independent results, 
and we compute the standard deviations of the masses that we obtain.  This leads us to estimate that the uncertainty in the spectrum from
the Monte Carlo routine is 1 -- 2\%.  This is the uncertainty that we report in Table 1 and Figure 9.
Combining this uncertainty with the others leads us to estimate that the total uncertainty in our results is about 33\%.

\clearpage

\begin{figure}
\centerline{\epsffile{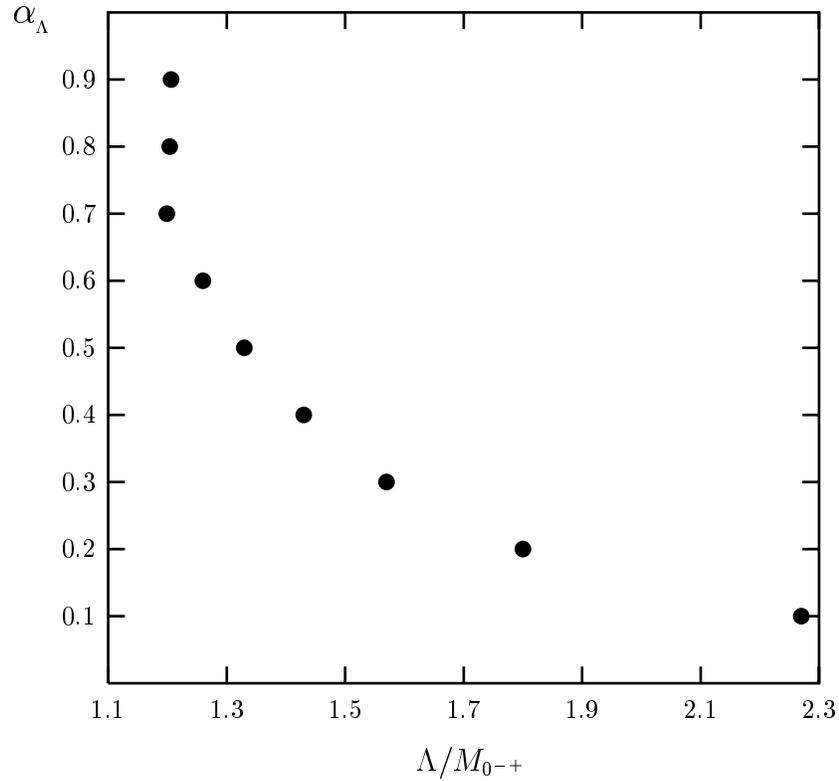}}
\caption{The coupling as a function of the cutoff.  We show the cutoff in units of the mass of the $0^{-+}$ state, and we 
use 14 longitudinal basis functions, 7 transverse-magnitude basis functions, and 4 spin
basis functions, for a total of 392 basis functions.
Using the recent anisotropic Euclidean lattice result of Morningstar and Peardon for the mass of the $0^{-+}$ state
\cite{morningstar}, we estimate that the cutoff is roughly varying from about 3.1 -- 6.0 GeV in this figure.}
\end{figure}

\clearpage

\begin{figure}
\centerline{\epsffile{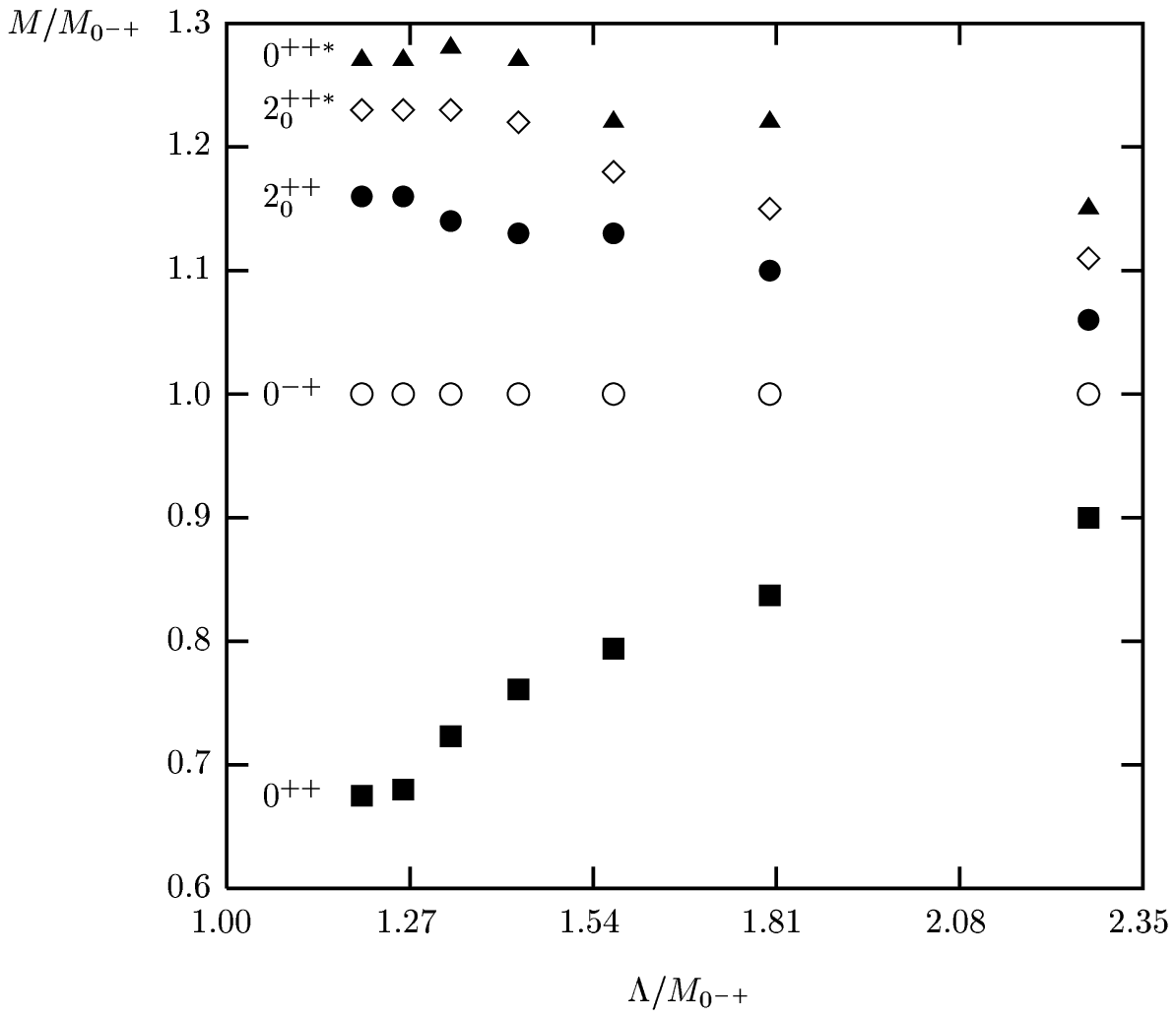}}
\caption{The masses of the five lightest glueballs with $j=0$, as functions of the cutoff.  
The masses and the cutoff are displayed in units of the mass of the $0^{-+}$ state. 
The seven values of the cutoff at the points that we display correspond, from right to left, to 
$\ala=0.1,0.2,0.3,\ldots,0.7$.  We use 14 longitudinal basis functions, 7 transverse-magnitude basis functions, and 4 spin
basis functions, for a total of 392 basis functions.}
\end{figure}

\clearpage

\begin{figure}
\centerline{\epsffile{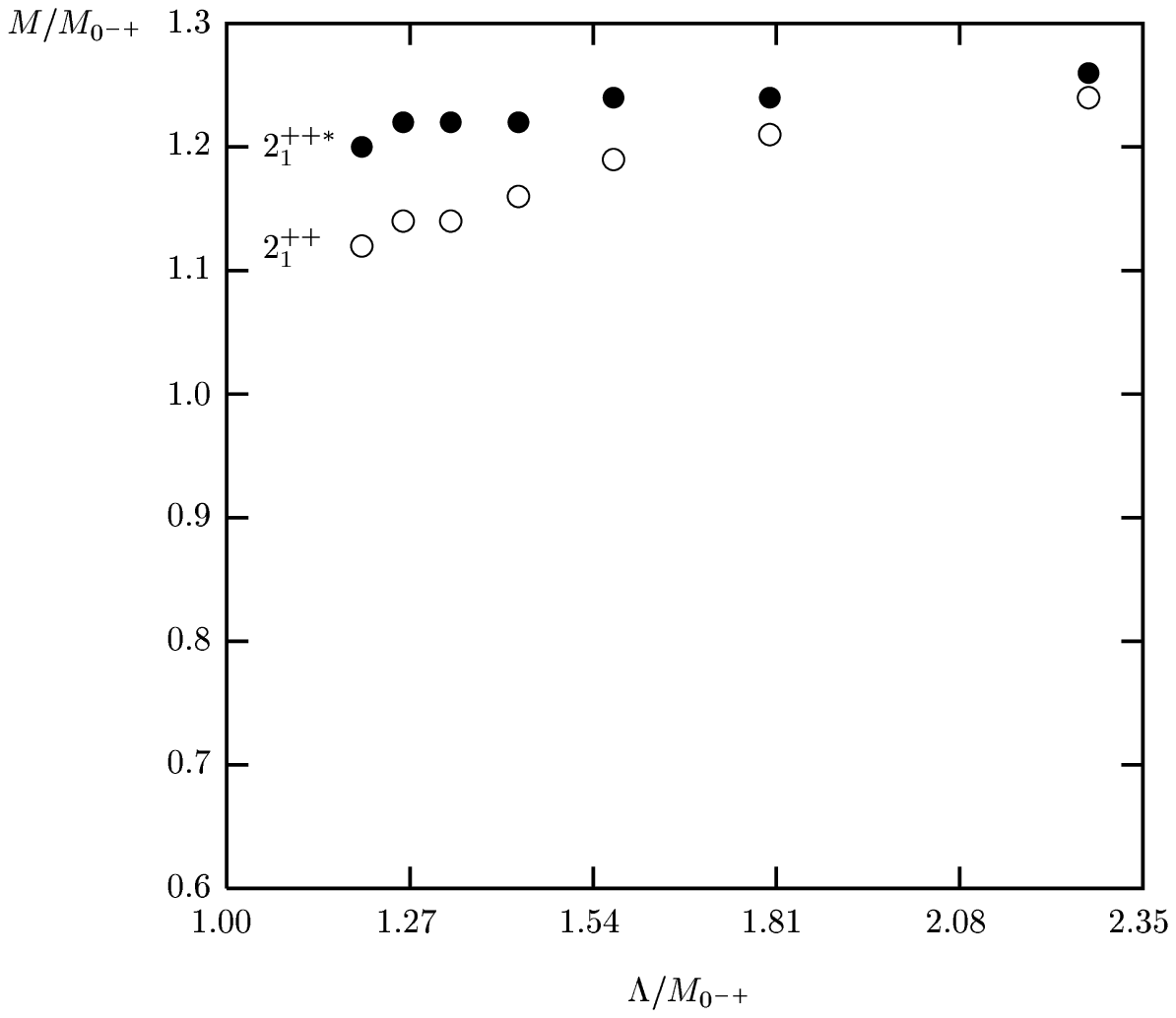}}
\caption{The masses of the two lightest glueballs with $j=1$, as functions of the cutoff.  
The masses and the cutoff are displayed in units of the mass of the $0^{-+}$ state. 
The seven values of the cutoff at the points that we display correspond, from right to left, to 
$\ala=0.1,0.2,0.3,\ldots,0.7$.  We use 14 longitudinal basis functions, 7 transverse-magnitude basis functions, and 4 spin
basis functions, for a total of 392 basis functions.}
\end{figure}

\clearpage

\begin{figure}
\centerline{\epsffile{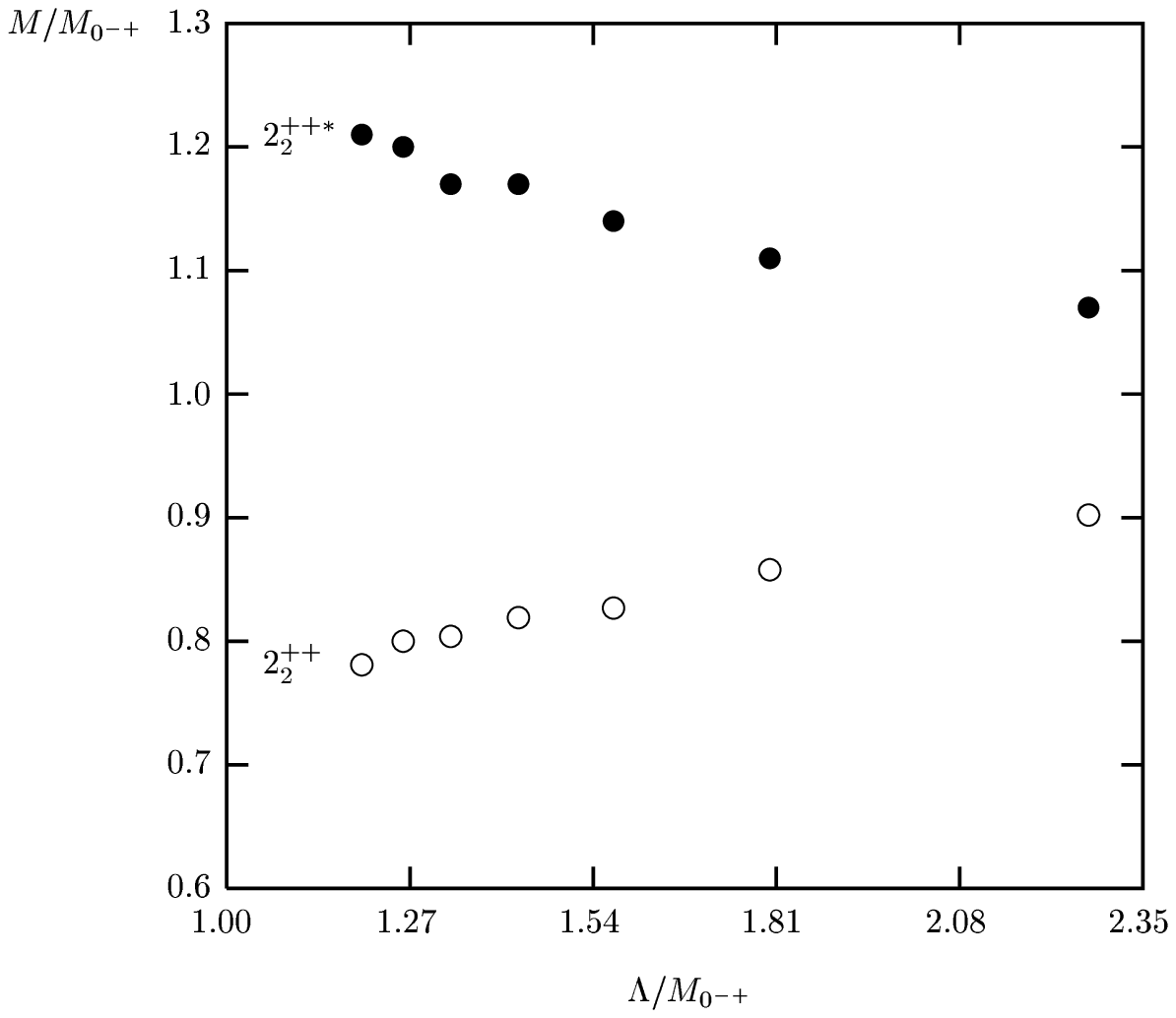}}
\caption{The masses of the two lightest glueballs with $j=2$, as functions of the cutoff.  
The masses and the cutoff are displayed in units of the mass of the $0^{-+}$ state. 
The seven values of the cutoff at the points that we display correspond, from right to left, to 
$\ala=0.1,0.2,0.3,\ldots,0.7$.  We use 14 longitudinal basis functions, 7 transverse-magnitude basis functions, and 4 spin
basis functions, for a total of 392 basis functions.}
\end{figure}

\clearpage

\begin{table}
\centerline{\begin{tabular}{|c|c|c|} \hline
State & $M/M_{0^{++}}$ & Lattice \cite{teper}\\
\hline\hline
$0^{-+}$ & $1.38 \pm 0.02$ & $1.34 \pm 0.18$\\ \hline
& $1.58 \pm 0.01$ & \\
$2^{++}$ & $1.58 \pm 0.02$ & $1.42 \pm 0.06$ \\
& $1.11 \pm 0.01$ & \\
\hline
& $1.70 \pm 0.01$ & \\
$2^{++*}$ & $1.68 \pm 0.02$ & $1.85 \pm 0.20$ \\
& $1.62 \pm 0.02$ & \\ \hline
$0^{++*}$ & $1.77 \pm 0.02$ & $1.78 \pm 0.12$ \\ \hline
\end{tabular}}
\caption{The glueball masses from our calculation compared to an average of
lattice results from a number of different calculations \cite{teper}.  
We display the masses in units of the mass of the $0^{++}$ state.  The 
uncertainties for our results are only the statistical uncertainties
associated with the Monte Carlo evaluation of the matrix elements of $\M$.  The three values
of the masses for the $2^{++}$ and $2^{++*}$ states for our calculation correspond to $j=0,1,2$.  We use the optimal coupling $\ala=0.5$, with 
14 longitudinal basis functions, 7 transverse-magnitude basis functions, and 4 spin
basis functions, for a total of 392 basis functions.}
\end{table}

\clearpage

\begin{figure}
\centerline{\epsffile{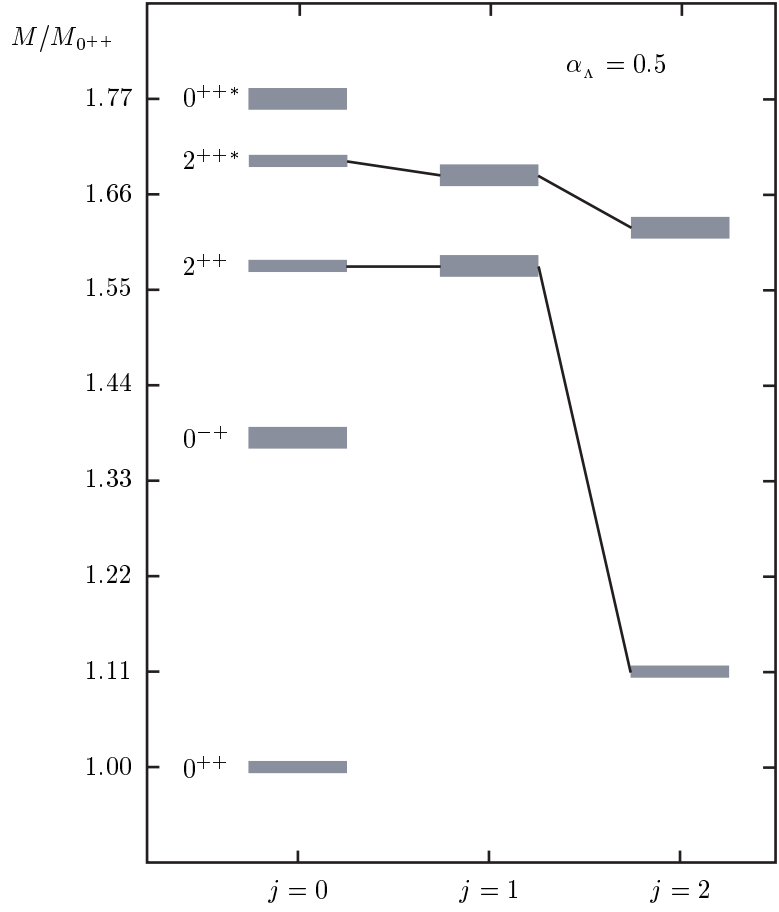}}
\caption{Our glueball spectrum.  The masses are plotted in units of the mass of the $0^{++}$ state and 
the vertical widths of the levels represent the statistical uncertainties in the masses.
The black lines connect the states that we believe should be degenerate.  We use the optimal coupling $\ala=0.5$, with 
14 longitudinal basis functions, 7 transverse-magnitude basis functions, and 4 spin
basis functions, for a total of 392 basis functions.}
\end{figure}

\clearpage

\begin{figure}
\centerline{\epsffile{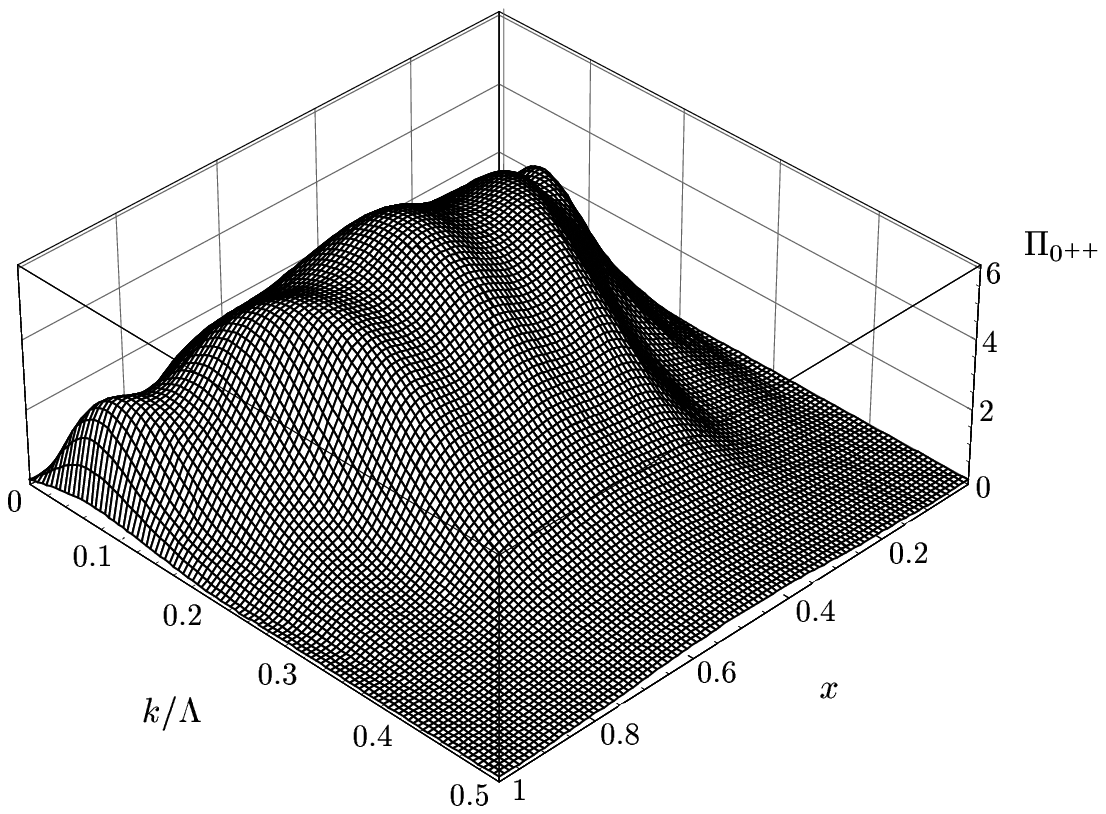}}
\caption{The 
probability density of the $0^{++}$ glueball.  We use the optimal coupling $\ala=0.5$, with 
14 longitudinal basis functions, 7 transverse-magnitude basis functions, and 4 spin
basis functions, for a total of 392 basis functions.}
\end{figure}

\clearpage

\begin{figure}
\centerline{\epsffile{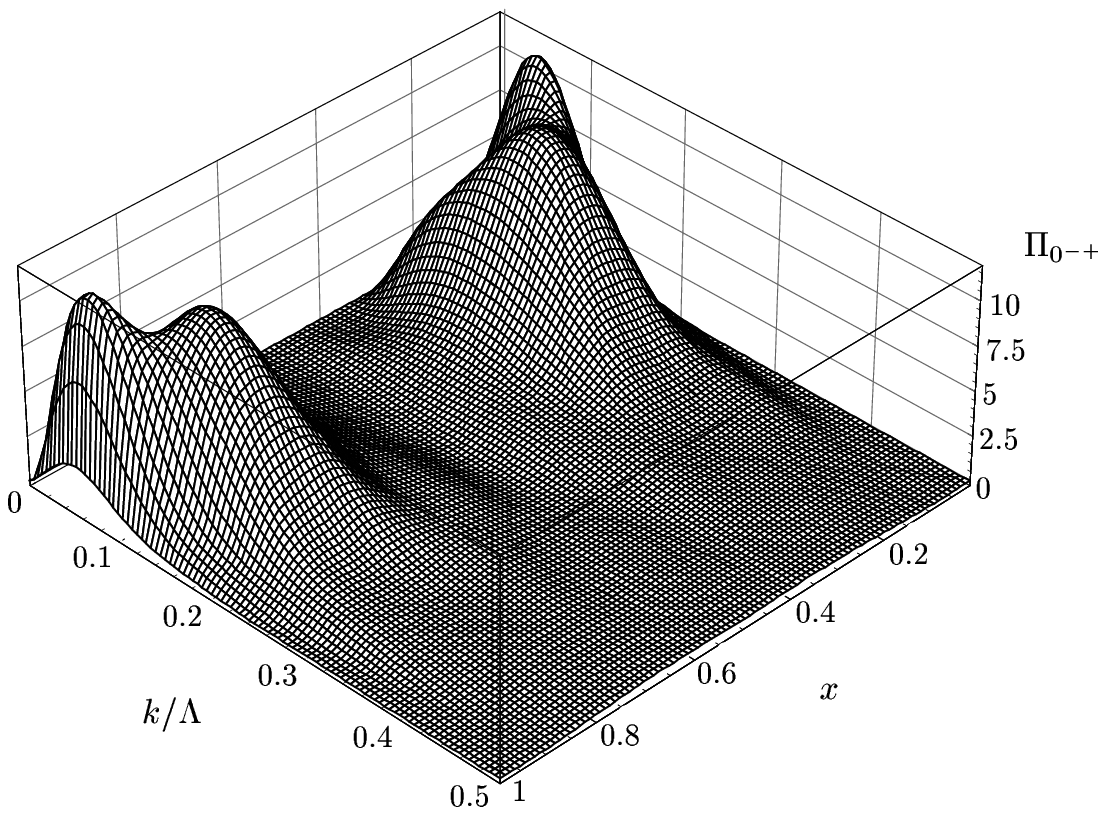}}
\caption{The 
probability density of the $0^{-+}$ glueball.  We use the optimal coupling $\ala=0.5$, with 
14 longitudinal basis functions, 7 transverse-magnitude basis functions, and 4 spin
basis functions, for a total of 392 basis functions.}
\end{figure}

\clearpage

\begin{figure}
\centerline{\epsffile{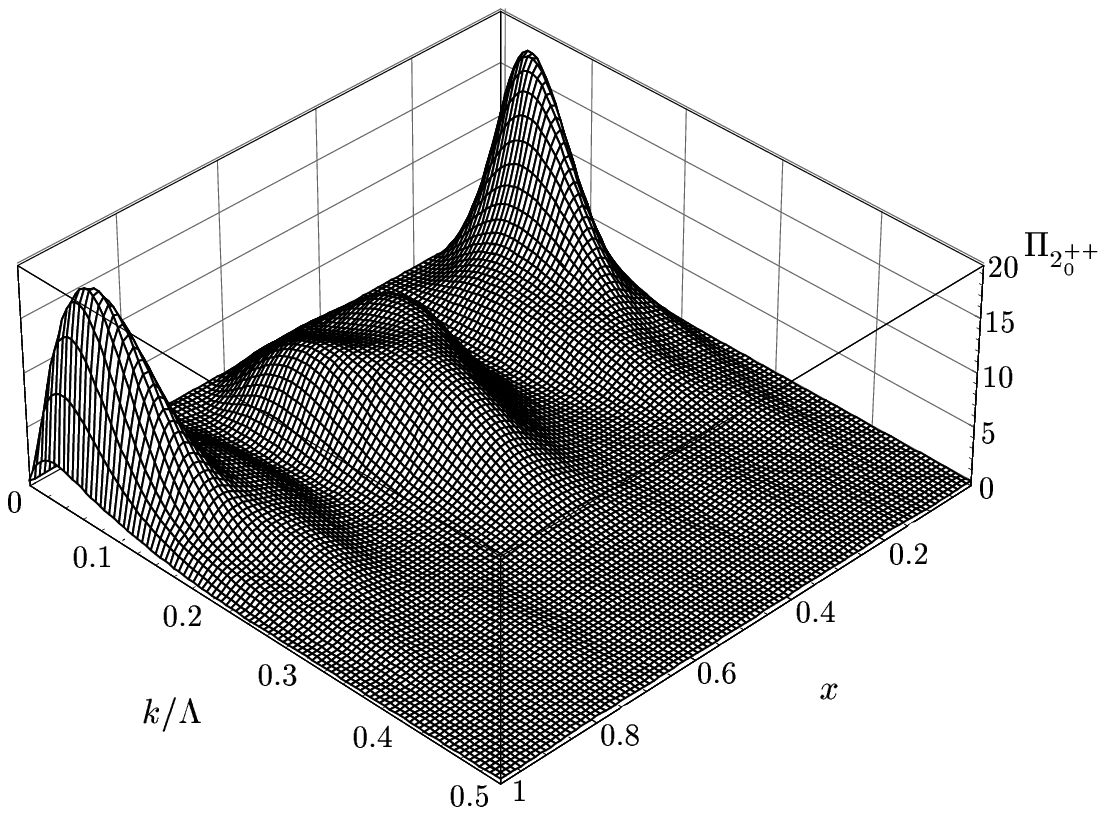}}
\caption{The 
probability density of the $2^{++}_0$ glueball.  We use the optimal coupling $\ala=0.5$, with 
14 longitudinal basis functions, 7 transverse-magnitude basis functions, and 4 spin
basis functions, for a total of 392 basis functions.}
\end{figure}

\clearpage

\begin{figure}
\centerline{\epsffile{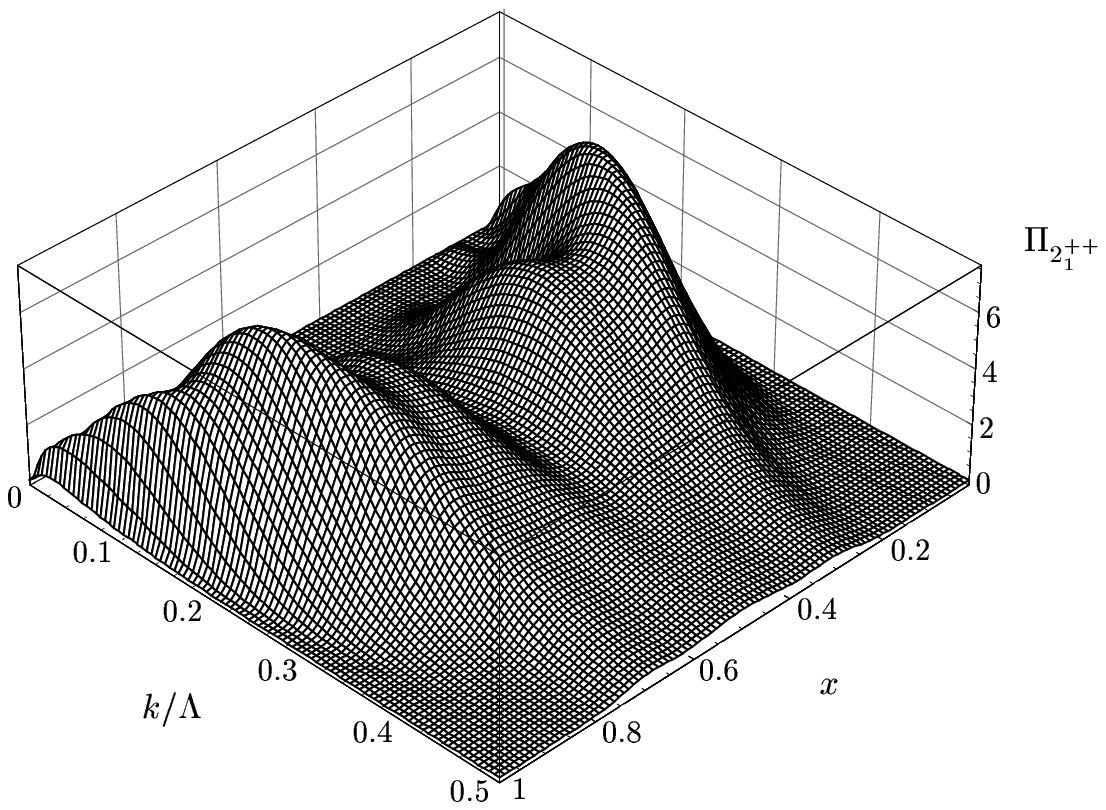}}
\caption{The 
probability density of the $2^{++}_1$ glueball.  We use the optimal coupling $\ala=0.5$, with 
14 longitudinal basis functions, 7 transverse-magnitude basis functions, and 4 spin
basis functions, for a total of 392 basis functions.}
\end{figure}

\clearpage

\begin{figure}
\centerline{\epsffile{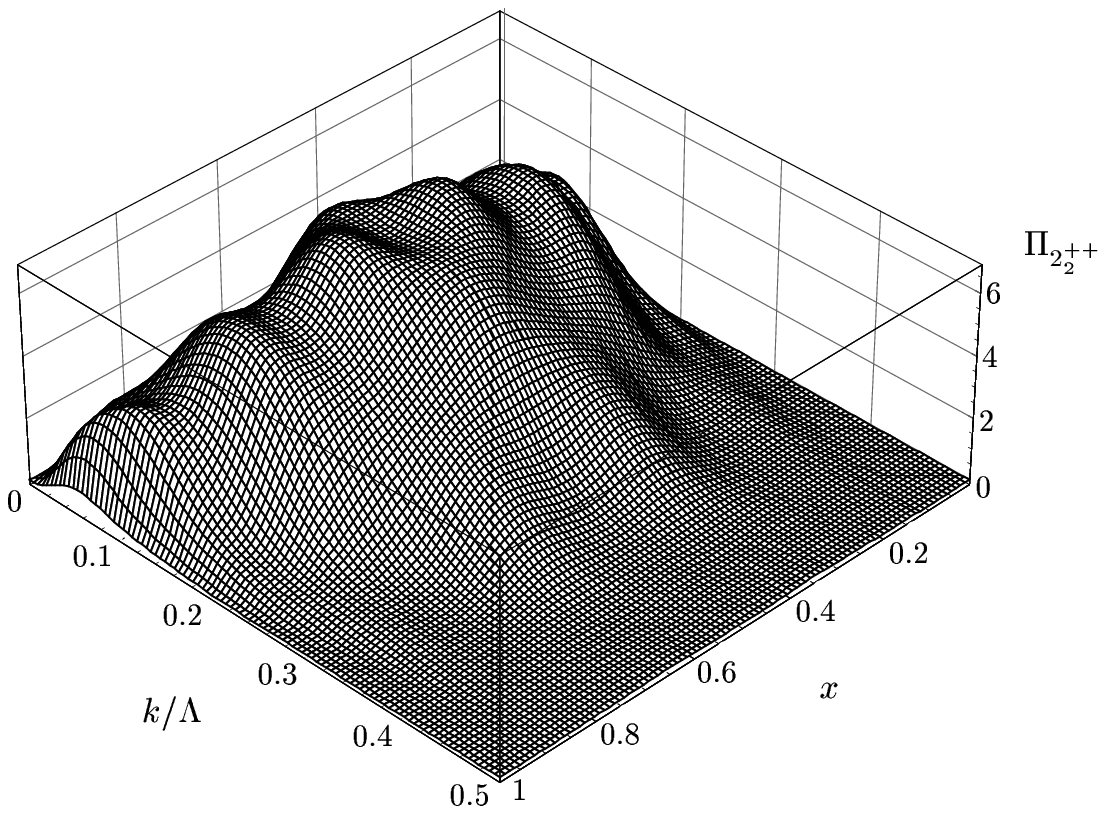}}
\caption{The 
probability density of the $2^{++}_2$ glueball.  We use the optimal coupling $\ala=0.5$, with 
14 longitudinal basis functions, 7 transverse-magnitude basis functions, and 4 spin
basis functions, for a total of 392 basis functions.}
\end{figure}

\clearpage

\begin{figure}
\centerline{\epsffile{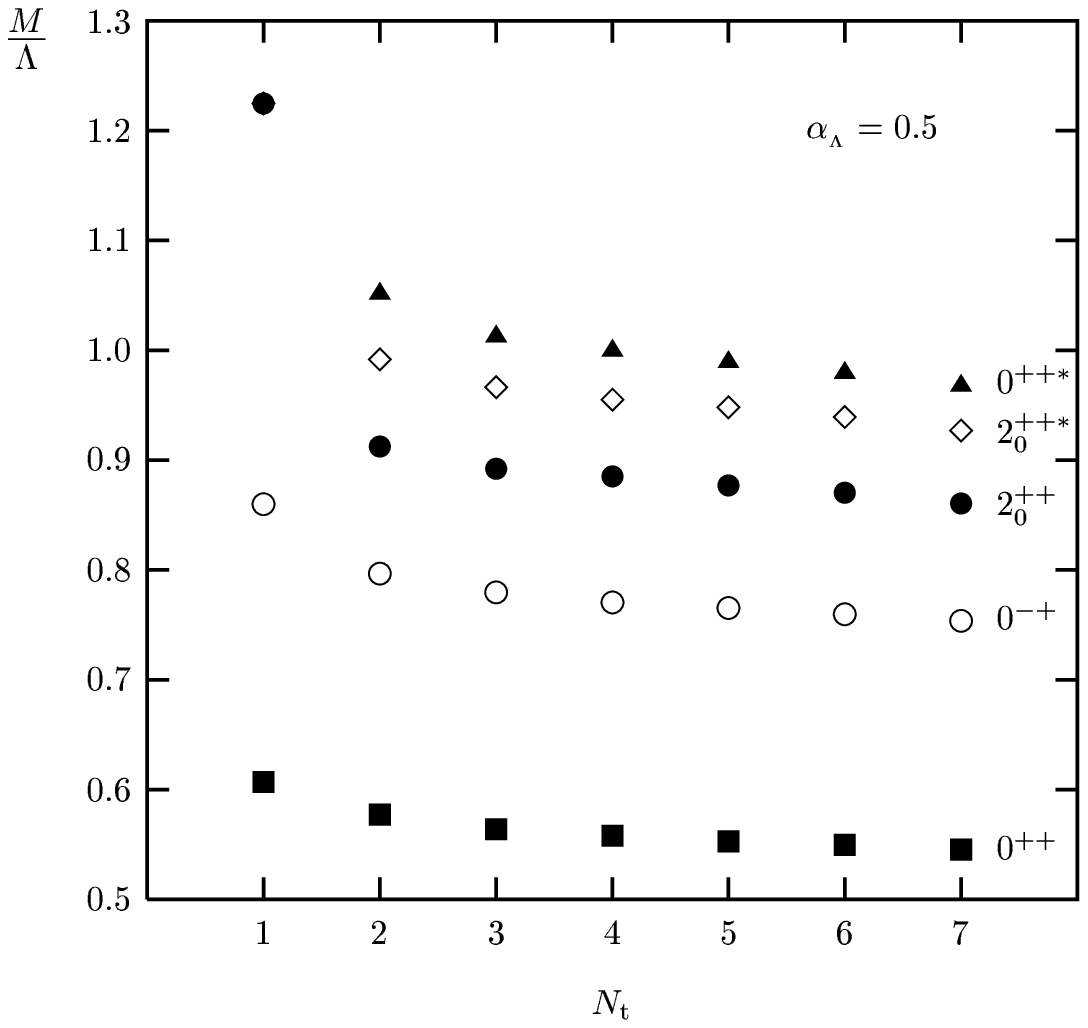}}
\caption{The 
masses of the five lightest glueballs with $j=0$, in units of the cutoff, as functions of the number of transverse-magnitude basis 
functions, $\Nt$.  We use the optimal coupling $\ala=0.5$, with
4 spin basis functions and $\Nl=2 \Nt$ longitudinal basis functions, for a total of $8 \Nt^2$ basis functions.}
\end{figure}

\clearpage

\begin{figure}
\centerline{\epsffile{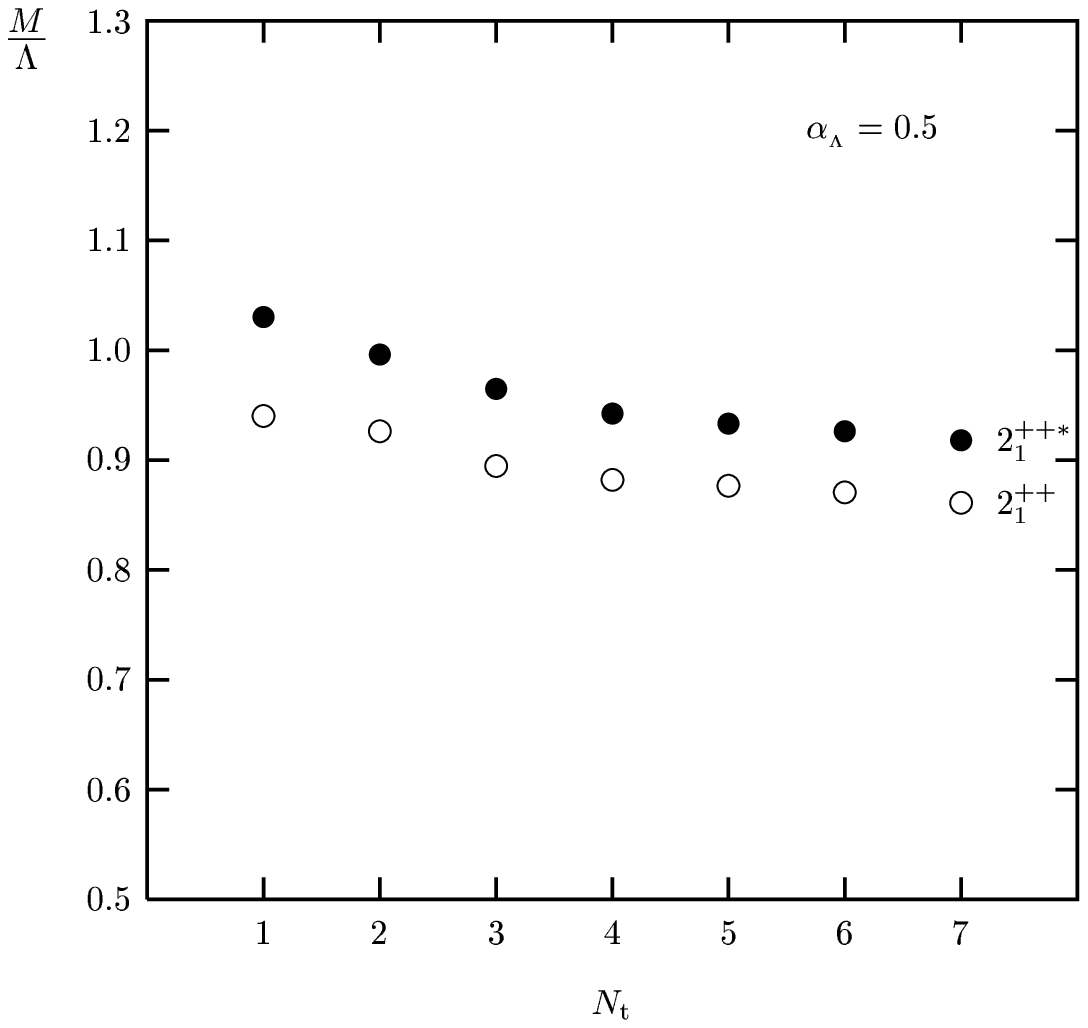}}
\caption{The masses of the two lightest glueballs 
with $j=1$, in units of the cutoff, 
as functions of the number of transverse-magnitude basis functions, $\Nt$.  We use the optimal coupling $\ala=0.5$, with
4 spin basis functions and $\Nl=2 \Nt$ longitudinal basis functions, for a total of $8 \Nt^2$ basis functions.}
\end{figure}

\clearpage

\begin{figure}
\centerline{\epsffile{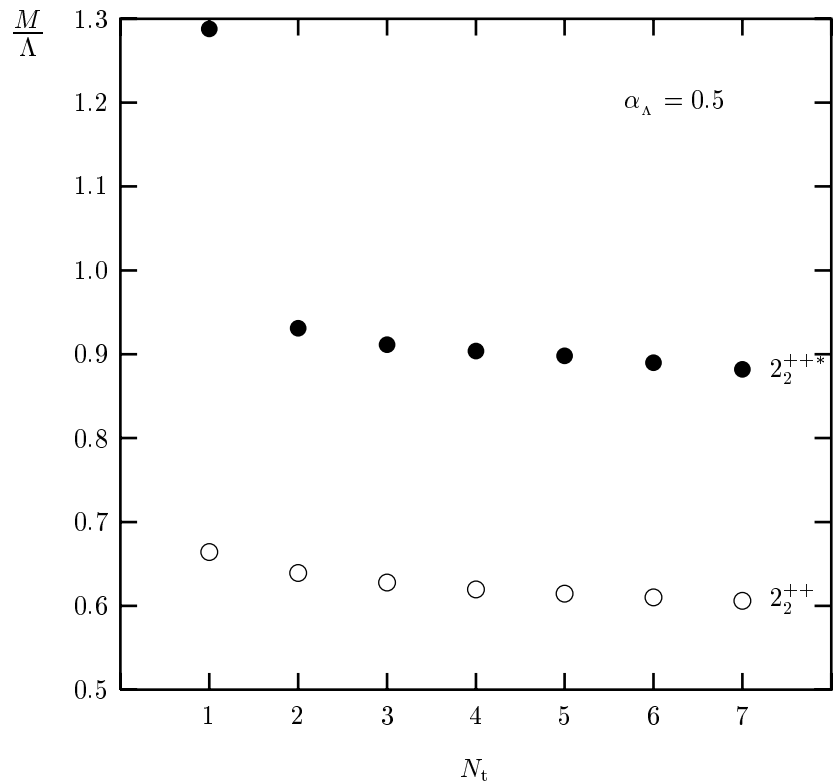}}
\caption{The masses of the two lightest 
glueballs with $j=2$, in units of the cutoff, 
as functions of the number of transverse-magnitude basis functions, $\Nt$.  We use the optimal coupling $\ala=0.5$, with
4 spin basis functions and $\Nl=2 \Nt$ longitudinal basis functions, for a total of $8 \Nt^2$ basis functions.}
\end{figure}

\clearpage

\section{Conclusion}
  
We have presented a formalism for pure-glue QCD that allows the physical states of the theory
to rapidly converge in a free-sector expansion.  In this approach, we force the free-state matrix elements of the
IMO to satisfy three conditions to make the desired expansion possible.  
First, the diagonal matrix elements of the IMO must be dominated by the free part of the 
IMO.  Second, the off-diagonal
matrix elements of the IMO must quickly decrease as the difference of the free masses of the states increases.
Third, the free mass of a free state must quickly increase as the number of particles in
the state increases.  

We assume that we can use perturbation theory to derive the operators of the theory, and if this is valid, 
then the first condition is automatically satisfied.  To satisfy the second condition, we place a smooth cutoff on the 
IMO.  We use LFFT so that
the effects of the vacuum are isolated in particles with zero longitudinal momentum, and we remove these particles from the
theory with the intent of replacing their physical effects with interactions.  This makes it reasonable to expect that
the third condition on the IMO is satisfied automatically due to the free-particle dispersion relation of LFFT.

The cutoff that we use violates a number of physical principles of light-front pure-glue QCD.  However, 
by requiring the IMO to produce
cutoff-independent physical quantities and by requiring it to respect the unviolated physical principles of the theory, 
we are able to derive recursion relations that uniquely determine the IMO to all orders in perturbation theory.

We have applied our method to the calculation of physical states and masses.
For this calculation, we approximated all physical states as two-gluon states.  We calculated the color parts of the states analytically,
and we expanded the states' momentum and spin degrees of freedom in terms of basis functions.  
We designed the states to be simultaneous eigenstates of
the IMO, the three-momentum operator, and the projection of the internal rotation generator onto the 3-axis.

Using our recursion relations for the IMO, we calculated to second order in perturbation theory 
the two-gluon to two-gluon matrix element of the IMO, which is 
required for the calculation of physical states.  We then used it to calculate the IMO matrix in terms of the basis functions.  
We showed that the infrared divergences in the matrix from exchanged gluons with infinitesimal longitudinal momentum 
cancel when treated properly.

In order to diagonalize the IMO matrix, we computed the five-dimensional integrals in the matrix elements using Monte Carlo methods.
We calculated the glueball spectrum for a range of couplings and found that we could not use more than about 400 basis functions without
the statistical errors becoming overwhelming.
We used the mass of our $0^{-+}$ glueball to compute the nonperturbative cutoff dependence of the coupling, and we 
analyzed the cutoff dependence of the spectrum.  We found that the cutoff that minimizes our errors is $\la/M_{0^{-+}}=1.33$.  The
corresponding coupling is $\ala=0.5$.  We presented the probability densities for some of our glueballs and found that our results 
for the spectrum compare
favorably with recent lattice data.  
The largest discrepancy seems to be the $2^{++}_2$ state, which is much too light.  Finally, we analyzed
the errors in our calculation from the various possible sources, and estimated the total uncertainty in our spectrum to be 33\%.

There are two main paths that we can take for future work with our approach.  
The first path is to further test our method with the theories that
we have considered so far.  Since the scalar theory that we considered in Ref. \cite{brent} is relatively simple, 
it would be interesting to use it compute the IMO to
higher orders in perturbation theory.  This would require us to solve the integral equation for the cutoff-independent
reduced interaction and could be used to further 
check our conjecture that our IMO leads to correct scattering amplitudes order-by-order in perturbation theory.  

In pure-glue QCD, we can further test our approach for calculating physical states 
by computing the IMO to higher orders in perturbation theory and by keeping more free sectors in the expansion of the states.
However, to keep more free sectors in the expansion of the states, we have to be able to calculate IMO matrices that have more degrees of freedom,
while controlling the statistical errors in the spectrum.  This means that we need a better algorithm for determining 
how accurately individual matrix elements of the IMO have to be computed in order to get a desired uncertainty in the spectrum.  We could also use
a better basis that requires fewer momentum functions to represent a wave function.
Overcoming these problems will be challenging, but it is important to test our method by studying the rate of convergence of the free-sector
expansion of states as a function of both the cutoff and the masses of the states. 

Another test of our method that we can do with pure-glue QCD is
to analyze the interaction that we have derived in this paper to test the conjecture that it is logarithmically confining.  It would also 
be interesting to analyze the long-range parts of higher-order interactions to see if the perturbative series for the interaction
is building towards a linearly confining potential.  Analyzing the long-range parts of higher-order interactions may be much
easier than computing these interactions in their entirety.

The second main path that we can take in the future is to extend our method to other theories and operators.
In order to compute quantities that can be compared with experiment, we wish to extend our method to full QCD\footnote{We are not thinking of
including QCD effects that require nonperturbative
renormalization or an infinite number of particles because these would require a 
new method rather than an extension of our current approach.}.  
This is complicated for two reasons.  First, there is additional algebraic and numerical complexity from the
the vertices involving quarks.  Second, quark masses complicate the method for determining the IMO because they increase the number
of reduced interactions that can be cutoff-independent \cite{roger}.  In addition, if large and small quark masses 
are considered simultaneously, then efficient numerical representation of the states and accurate calculation of the IMO's 
matrix elements become more difficult.  Masses also quickly enlarge the parameter space that must be explored to compare to 
experimental data.

We can also extend our method by applying it to the computation of operators other than
the IMO, such as the rotation generators, the parity operator, and currents.  The rotation generators and the parity operator are of particular interest
because they may aid in the classification of the physical states of a theory.

In summary, there are many avenues of research that must be explored, and some of them are quite complex.
However, all the improvements that we have discussed are necessary if we are to accurately represent the
physical states of quantum field theories as rapidly convergent expansions in free sectors.

\newpage
\begin{tabbing}
{\large \bf APPENDIX A:}$\;\;$\={\large \bf Conventions for Light-Front Pure-Glue QCD}
\end{tabbing}

The purpose of this appendix is to state our conventions.  With these conventions, any four-vector $a$ is written in the form

\bea
a = (a^+, a^-, \vec a_\perp) ,
\lll{four vec}
\eea

\noi
where in terms of equal-time vector components,

\bea
a^{\pm} = a^0 \pm a^3,
\eea

\noi
and

\bea
\vec a_\perp = \sum_{i=1}^2 a_\perp^i \hat e_i = \sum_{i=1}^2 a^i
\hat e_i,
\eea

\noi
where $\hat e_i$ is the unit vector pointing along the $i$-axis.
The inner product is

\bea
a \cdot b = \frac{1}{2} a^+ b^- + \frac{1}{2} a^- b^+ - \vec a_\perp
\!\cdot \vec b_\perp ,
\eea

\noi
and

\bea
\vec a_\perp^2 = \vec a_\perp \!\cdot \vec a_\perp .
\eea

A spacetime coordinate is a four-vector, and according to \reff{four vec}, it is
written

\bea
x = (x^+, x^-, \vec x_\perp) .
\eea

\noi
The time component is chosen to be $x^+$.  $x^-$ is referred to as the
longitudinal component, and $\vec x_\perp$ contains the transverse
components.

The gradient operator is treated just like any other four-vector.  Its
components are

\bea
\partial^{\pm} = 2 \frac{\partial}{\partial x^\mp} ,
\eea

\noi
and

\bea
\partial_\perp^i = \frac{-\partial}{\partial x_\perp^i} .
\eea

The canonical Lagrangian density for pure-glue QCD is

\bea
{\cal L} = - \frac{1}{4} F_{c \mu \nu} F^{\mu \nu}_c  ,
\eea

\noi
where

\bea
F_c^{\mu \nu} = \partial^\mu A_c^{\nu} - \partial^\nu A_c^{\mu} - g
A_{c_1}^{\mu} A_{c_2}^{\nu} f^{c_1 c_2 c} .
\eea

\noi
Greek indices are Lorentz indices, $c$'s are color indices, repeated
indices are summed over, and the $f$'s are the SU($N_\mathrm{c}$) structure constants.

We derive the canonical Hamiltonian from ${\cal L}$ by the following procedure:

\begin{enumerate}
\item We choose the light-cone gauge, $A^+_c=0$.
\item We derive the Euler-Lagrange equation that determines $A^-_c$ in terms of $\vec A_{\perp c}$.
\item Using the canonical procedure and treating the field classically (i.e. letting it and its derivatives
      commute), we derive the Hamiltonian in terms of $\vec A_{\perp c}$, dropping terms that are zero if the gluon
      field is zero at spacetime infinity.
\item We quantize the gluon field by expanding it in terms of free-particle creation and annihilation
      operators. (We define the field expansion and its inverse longitudinal derivative below.)
\item In each term in the Hamiltonian, we treat the creation and annihilation operators as if they
      commute and move all the creation operators to the left of all the annihilation operators.
      (This ``normal ordering'' drops the so-called ``self-inertias,'' as well as some constants.)
\item We drop the terms in the Hamiltonian that have no effect if there are no particles with $p^+=0$.

\end{enumerate}

We work in the Schr\"odinger representation, where operators are time-independent and states are time-dependent.  Thus we quantize the field by
defining it to be a superposition of solutions to the Klein-Gordon equation (since gluons are bosons), with the quantization surface $x^+=0$:

\bea
\vec A_{\perp c}(x^-,\vec x_\perp) &=& \int D_1 \delta_{c, c_1}
\left[ a_1
\vec \varepsilon_{\perp s_1} e^{-i p_1 \cdot x} + a_1^\dagger \vec
\varepsilon_{\perp s_1}^{\,*} e^{i p_1 \cdot x}
\right]\rule[-3mm]{.1mm}{7mm}_{\; x^+ = 0} , 
\eea

\noi
where

\bea
D_i = \sum_{c_i=1}^{N_\mathrm{c}} \sum_{s_i=-1,1} \frac{d^2 p_{i \perp} d p_i^+}{16
\pi^3 p_i^+} \theta(p_i^+ - \epsilon \p^+).
\eea

\noi
Here $s_i$ is the spin polarization of particle $i$, $\p$ is the four-momentum operator, $\epsilon$ is a positive
infinitesimal, and the gluon polarization vector is defined by

\bea
\vec \varepsilon_{\perp s} &=& \frac{-1}{\sqrt{2}}(s,i) .
\eea

\noi
$p^+$ and $\vec p_\perp$ are the momenta conjugate to $x^-$ and $\vec x_\perp$; so they are
referred to as the longitudinal and transverse momenta, respectively.  
The purpose of $\epsilon$ is to regulate divergent effects from exchanged gluons (either instantaneous or real) with 
infinitesimal longitudinal momentum.  We take $\epsilon \rightarrow 0$ before we calculate physical quantities (see Section 5).

In the process of deriving the canonical Hamiltonian, we need to take the inverse longitudinal derivative of the gluon field.  We do this
by moving the derivative inside the expansion:

\bea
\frac{1}{\partial^+} \vec A_{\perp c}(x^-,\vec x_\perp) &=& \int D_1 \delta_{c, c_1}
\left[ -\frac{1}{i p_1^+} a_1
\vec \varepsilon_{\perp s_1} e^{-i p_1 \cdot x} + \frac{1}{i p_1^+} a_1^\dagger \vec
\varepsilon_{\perp s_1}^{\,*} e^{i p_1 \cdot x}
\right]\rule[-4mm]{.1mm}{9.2mm}_{\; x^+ = 0} .
\eea

The expansion coefficients $a^\dagger_i$ and $a_i$ are identified as particle creation and annihilation operators.  
They follow the convention

\bea
a_i = a(p_i, s_i, c_i) ,
\eea

\noi
and have the commutation relations

\bea
[a_i , a_j^\dagger] &=& \delta_{i,j} \equiv 16 \pi^3 p_i^+ \delta^{(3)}(p_i -
p_j) \delta_{s_i, s_j} \delta_{c_i, c_j}
\eea

\noi
and

\bea
[a_i , a_j] = [a_i^\dagger , a_j^\dagger] = 0 ,
\eea

\noi
where

\bea
\delta^{(3)}(p_i - p_j) &=& \delta(p_i^+ - p_j^+) \delta^{(2)}(\vec p_{i
\perp} - \vec p_{j \perp}) .
\eea

Let ${\cal M}^2$ be the invariant-mass operator. 
The momentum conjugate to  $x^+$ is $p^-$; so the Hamiltonian is identified as $\p^-$ and it follows from

\bea
\p^2 = {\cal M}^2
\eea

\noi
that

\bea
\p^- = \frac{\vec \p_\perp^{\,2} + {\cal M}^2}{\p^+} .
\lll{ham}
\eea

The canonical Hamiltonian that results from the above procedure can be written as the sum
of a free part and an interacting part:

\bea
\p^- = \p^-_{\mathrm{free}} + v ,
\eea

\noi
where

\bea
\p^-_{\mathrm{free}} = \int D_1 \frac{\vec p_{1 \perp}^{\,2}}{p_1^+} a_1^\dagger a_1.
\eea

\noi
We give the interaction in terms of its ``modified'' matrix elements: 

\bea
v &=& g \frac{16 \pi^3}{2!} \int D_1 D_2 D_3 \, a_2^\dagger a_3^\dagger
a_1 \, \delta^{(3)}(p_1 - p_2 - p_3)
\lbrak g_2 g_3 | v | g_1 \rbrak \n
&+& g \frac{16 \pi^3}{2!} \int D_1 D_2 D_3 \, a_3^\dagger a_1 a_2 \,
\delta^{(3)}(p_1 + p_2 - p_3)
\lbrak g_3 | v | g_1 g_2 \rbrak \n
&+& g^2 \frac{16 \pi^3}{2! \, 2!} \int D_1 D_2 D_3 D_4 \, a_3^\dagger
a_4^\dagger a_1 a_2 \,
\delta^{(3)}(p_1 + p_2 - p_3 - p_4)  \sum_{i=1}^4
\lbrak g_3 g_4 | v | g_1 g_2 \rbrak_{\hspace{.005in} i} \n
&+& g^2 \frac{16 \pi^3}{3!} \int D_1 D_2 D_3 D_4 \, a_2^\dagger
a_3^\dagger a_4^\dagger a_1 \, \delta^{(3)}(p_1 - p_2 - p_3 - p_4)
\sum_{i=1}^4 \lbrak g_2 g_3 g_4 | v | g_1 \rbrak_{\hspace{.005in} i} \n
&+& g^2 \frac{16 \pi^3}{3!} \int D_1 D_2 D_3 D_4 \, a_4^\dagger a_1 a_2
a_3 \, \delta^{(3)}(p_1 + p_2 + p_3 - p_4)
\sum_{i=1}^4 \lbrak g_4 | v | g_1 g_2 g_3 \rbrak_{\hspace{.005in} i} ,
\eea

\noi
where a modified matrix element is defined by

\bea
\lbrak i | v | j \rbrak = \frac{< \! i | v | j \! >}{16 \pi^3 \delta^{(3)}(p_i - p_j)} \:\rule[-4mm]{.1mm}{10mm}_{\; g=1}  .
\eea

\noi
The modified matrix elements are

\bea
\lbrak g_2 g_3 | v | g_1 \rbrak &=& i f^{c_1 c_2 c_3} \left[
-\delta_{s_2, \bar s_3}
\vec \varepsilon_{\perp s_1} \! \cdot \left\{(\vec p_{2 \perp} -
\vec p_{3 \perp}) -
\frac{\vec p_{1 \perp}}{p_1^+} (p_2^+ - p_3^+) \right\}
\right. \n
&+& \left.  \delta_{s_1, s_3} \vec \varepsilon_{\perp s_2}^{\,*} \! \cdot
\left\{ (\vec p_{1 \perp} + \vec
p_{3\perp} ) -
\frac{\vec p_{2 \perp}}{p_2^+} (p_1^+ + p_3^+) \right\}  \right. \n
&+& \left.  \delta_{s_1, s_2} \vec \varepsilon_{\perp s_3}^{\, *}
\! \cdot \left\{ -(\vec p_{1 \perp} + \vec p_{2 \perp}) +
\frac{\vec p_{3 \perp}}{p_3^+} (p_1^+ + p_2^+) \right\} \right] , \n
\lbrak g_3 | v | g_1 g_2 \rbrak &=& - i f^{c_1 c_2 c_3} \left[
\delta_{s_2, s_3}
\vec \varepsilon_{\perp s_1} \! \cdot \left\{(\vec p_{3 \perp} +
\vec p_{2 \perp}) -
\frac{\vec p_{1 \perp}}{p_1^+} (p_3^+ + p_2^+) \right\}  \right. \n
&-& \left.  \delta_{s_1, \bar s_2} \vec \varepsilon_{\perp s_3}^{\, *}
\! \cdot\left\{ (\vec p_{1 \perp} -
\vec p_{2 \perp}) - \frac{\vec p_{3
\perp}}{p_3^+} (p_1^+ - p_2^+) \right\}  \right. \n
&+& \left.  \delta_{s_1, s_3} \vec \varepsilon_{\perp s_2} \! \cdot
\left\{ -(\vec p_{1 \perp} + \vec p_{3 \perp}) +
\frac{\vec p_{2 \perp}}{p_2^+} (p_1^+ + p_3^+) \right\}
\right] ,
\eea

\noi
and

\bea
\lbrak g_3 g_4 | v | g_1 g_2 \rbrak_{\hspace{.005in}1} &=& \left[
f^{c_1 c_3 c} f^{c_4 c_2 c} \left( \delta_{s_2, s_3}
\delta_{s_1, s_4}
- \delta_{s_3, \bar s_4} \delta_{s_1, \bar s_2} \right)  +  f^{c_1 c_4 c}
f^{c_3 c_2 c} \left( \delta_{s_2, s_4}
\delta_{s_1, s_3} - \delta_{s_3, \bar s_4}  \delta_{s_1, \bar s_2} \right)
\right. \n
&+& \left. f^{c_1 c_2 c} f^{c_3 c_4 c} \left( \delta_{s_2, s_4} \delta_{s_1,
s_3} - \delta_{s_1, s_4} \delta_{s_2,
s_3} \right) \right] , \n
\lbrak g_3 g_4 | v | g_1 g_2 \rbrak_{\hspace{.005in}2} &=& f^{c_1 c_3
c} f^{c_4 c_2 c} \delta_{s_1, s_3} \delta_{s_2, s_4}
\frac{p_1^+ + p_3^+}{p_1^+ - p_3^+} \frac{p_4^+ + p_2^+}{p_4^+ - p_2^+} ,
\n
\lbrak g_3 g_4 | v | g_1 g_2 \rbrak_{\hspace{.005in}3} &=& f^{c_1 c_4
c} f^{c_3 c_2 c} \delta_{s_1, s_4} \delta_{s_2, s_3}
\frac{p_1^+ + p_4^+}{p_1^+ - p_4^+} \frac{p_3^+ + p_2^+}{p_3^+ - p_2^+} ,
\n
\lbrak g_3 g_4 | v | g_1 g_2 \rbrak_{\hspace{.005in}4} &=& f^{c_1 c_2
c} f^{c_4 c_3 c} \delta_{s_1, \bar s_2} \delta_{s_3,
\bar s_4} \frac{p_1^+ - p_2^+}{p_1^+ + p_2^+} \frac{p_4^+ - p_3^+}{p_4^+ +
p_3^+} ,
\eea

\noi
and

\bea
\lbrak g_2 g_3 g_4 | v | g_1 \rbrak_{\hspace{.005in}1} &=& \left[
f^{c_1 c_3 c} f^{c_4 c_2 c} \left( -\delta_{\bar s_2, s_3}
\delta_{s_1, s_4}
+ \delta_{s_3, \bar s_4} \delta_{s_1, s_2} \right)  +  f^{c_1 c_4 c} f^{c_3
c_2 c} \left( -\delta_{\bar s_2, s_4}
\delta_{s_1, s_3} + \delta_{s_3, \bar s_4}  \delta_{s_1, s_2} \right) \right.
\n
&+& \left. f^{c_1 c_2 c} f^{c_3 c_4 c} \left( -\delta_{\bar s_2, s_4}
\delta_{s_1, s_3} + \delta_{s_1, s_4}
\delta_{\bar s_2, s_3}
\right) \right] , \n
\lbrak g_2 g_3 g_4 | v | g_1 \rbrak_{\hspace{.005in}2} &=& - f^{c_1 c_2
c} f^{c_4 c_3 c} \delta_{s_1, s_2} \delta_{\bar s_3,
s_4}
\frac{p_1^+ + p_2^+}{p_1^+ - p_2^+} \frac{p_4^+ - p_3^+}{p_4^+ + p_3^+} ,
\n
\lbrak g_2 g_3 g_4 | v | g_1 \rbrak_{\hspace{.005in}3} &=& - f^{c_1 c_3
c} f^{c_4 c_2 c} \delta_{s_1, s_3} \delta_{\bar s_2,
s_4}
\frac{p_1^+ + p_3^+}{p_1^+ - p_3^+} \frac{p_4^+ - p_2^+}{p_4^+ + p_2^+} ,
\n
\lbrak g_2 g_3 g_4 | v | g_1 \rbrak_{\hspace{.005in}4} &=& - f^{c_1 c_4
c} f^{c_2 c_3 c} \delta_{s_1, s_4} \delta_{\bar s_3,
s_2}
\frac{p_1^+ + p_4^+}{p_1^+ - p_4^+} \frac{p_2^+ - p_3^+}{p_2^+ + p_3^+} ,
\eea

\noi
and

\bea
\lbrak g_4 | v | g_1 g_2 g_3 \rbrak_{\hspace{.005in}1} &=& \left[
f^{c_1 c_3 c} f^{c_4 c_2 c}
\left( -\delta_{s_2, \bar s_3}  \delta_{s_1, s_4}
+ \delta_{s_3, s_4} \delta_{s_1, \bar s_2} \right)  +  \! f^{c_1 c_4 c}
f^{c_3 c_2 c} \left( -\delta_{s_2, s_4}
\delta_{s_1, \bar s_3} + \delta_{s_3, s_4}  \delta_{s_1, \bar s_2} \right)
\right. \n
&+& \left. f^{c_1 c_2 c} f^{c_3 c_4 c} \left( -\delta_{s_2, s_4}
\delta_{s_1, \bar s_3} + \delta_{s_1, s_4} \delta_{s_2,
\bar s_3}
\right) \right] , \n
\lbrak g_4 | v | g_1 g_2 g_3 \rbrak_{\hspace{.005in}2} &=& - f^{c_1 c_4
c} f^{c_3 c_2 c} \delta_{s_1, s_4}
\delta_{s_2, \bar s_3}  \frac{p_1^+ + p_4^+}{p_1^+ - p_4^+} \frac{p_3^+ -
p_2^+}{p_3^+ + p_2^+} , \n
\lbrak g_4 | v | g_1 g_2 g_3 \rbrak_{\hspace{.005in}3} &=& - f^{c_2 c_4
c} f^{c_3 c_1 c} \delta_{s_2, s_4}
\delta_{s_1, \bar s_3}  \frac{p_2^+ + p_4^+}{p_2^+ - p_4^+} \frac{p_3^+ -
p_1^+}{p_3^+ + p_1^+} , \n
\lbrak g_4 | v | g_1 g_2 g_3 \rbrak_{\hspace{.005in}4} &=& - f^{c_3 c_4
c} f^{c_1 c_2 c} \delta_{s_3, s_4}
\delta_{s_2, \bar s_1}  \frac{p_3^+ + p_4^+}{p_3^+ - p_4^+} \frac{p_1^+ -
p_2^+}{p_1^+ + p_2^+} ,
\eea

\noi
where $\bar s_n = - s_n$.

We work in the free basis, the basis of eigenstates of $\p^-_{\mathrm{free}}$.  They are given
by

\bea
\left| g_1 g_2 \cdots g_n\right> = a_1^\dagger a_2^\dagger \cdots
a_n^\dagger \left| 0 \right> ,
\eea

\noi
for any integer $n \ge 0$.  The associated eigenvalue equation is

\bea
\p^-_{\mathrm{free}} \left| g_1 g_2 \cdots g_n\right> = \sum_{i=1}^n p_i^- \left| g_1 g_2
\cdots
g_n\right> ,
\eea

\noi
where

\bea
p_i^- = \frac{\vec p_{i \perp}^{\,2}}{p_i^+}
\eea

\noi
(since $p_i^2=0$) and the sum is zero if $n=0$.

The noninteracting limit of \reff{ham} is

\bea
\p^-_{\mathrm{free}} = \frac{\vec \p_\perp^{\,2} + \Mf}{\p^+} ,
\eea

\noi
where $\Mf$ is the free invariant-mass operator.  It has the
eigenvalue equation

\bea
\Mf \left| g_1 g_2 \cdots g_n\right> = M^2 \left| g_1 g_2 \cdots
g_n\right> ,
\eea

\noi
where

\bea
M^2 = P^+ \sum_{i=1}^n p_i^- - \vec P_\perp^2 ,
\eea

\noi
and $P$ is the total momentum of the state.

Finally, in terms of the free states, the completeness relation is

\bea
{\bf 1} = \left| 0 \right> \left< 0 \right| + \int D_1 \left| g_1 \right>
\left< g_1 \right| +
\frac{1}{2!} \int D_1 D_2 \left| g_1 g_2 \right> \left< g_1 g_2 \right| +
\cdots .
\eea

\newpage

\begin{tabbing}
{\large \bf APPENDIX B:}$\;$\={\large \bf The Derivation of the Recursion Relations for the Reduced}\\
\>{\large \bf Interaction}
\end{tabbing}

\vspace{.1in}

This appendix is an extension of Appendix D of Ref. \cite{brent} to the case of pure-glue QCD.
In Subsection \ref{subsec: recursion}, we derived a constraint on the $\OR(\gla^r)$ reduced interaction for $r \ge 1$:

\bea
&&V^{(r)}_{\CD}(\la) - 
V^{(r)}_{\CD}(\la') = \delta V^{(r)} - \sum_{s=2}^{r-1}  B_{r-s,s}
V^{(r-s)}(\la) .
\lll{CD coupled D}
\eea

\noi
Since we already know the first-order reduced interaction [see \reff{O1}] and the cutoff-independent part of the
second-order reduced interaction [see \reff{O2}], we wish to use this equation to compute $V^{(r)}_\CD(\la)$ for 
$r \ge 2$ and $V^{(r)}_\CI$ for $r \ge 3$, in terms of lower-order reduced interactions.

\vspace{.20in}

\noi
{\large \bf B.1  \hspace{.05in} The Recursion Relation for the Cutoff-Dependent Part}
\vspace{.08in}

We begin by computing the recursion relation for the cutoff-dependent part of the reduced interaction.
Recall that momentum conservation implies that any matrix element 
of $V(\la)$ can be written as an expansion in unique products of
momentum-conserving delta functions.  This means that an arbitrary matrix element of 
\reff{CD coupled D} can be expanded in products of delta functions:

\bea
\sum_i \left<F \right| V^{(r)}_{\CD}(\la) \left| I \right>^{(i)} - 
\sum_i \left<F \right| V^{(r)}_{\CD}(\la') \left| I \right>^{(i)} = \sum_i \left[ \dVome{r} - \sum_{s=2}^{r-1}  B_{r-s,s}
\left<F \right| V^{(r-s)}(\la) \left| I \right>\right]^{(i)} \!\!\!\! ,
\eea

\noi
where the $(i)$ superscripts denote that we are considering the $i^{\hspace{.1mm}\mathrm{th}}$ 
product of delta functions that can occur 
in a delta-function expansion of $\left<F \right| V^{(r)}(\la) \left| I \right>$.  This
equation is equivalent to a set of equations, one for each possible product 
of delta functions:

\bea
\left<F \right| V^{(r)}_{\CD}(\la) \left| I \right>^{(i)} - 
\left<F \right| V^{(r)}_{\CD}(\la') \left| I \right>^{(i)} = \left[ \dVome{r} - \sum_{s=2}^{r-1}  B_{r-s,s}
\left<F \right| V^{(r-s)}(\la) \left| I \right>\right]^{(i)}  .
\lll{prod exp}
\eea

\noi
Cluster decomposition implies that we can write

\bea
\left<F \right| V^{(r)}_\CD(\la) \left| I \right>^{\,(i)} &=&
\left[ \prod_{j=1}^{N_{\delta}^{(i)}} \delta_j^{(i)} \right]
F^{(i)}_\CD(\{p_n\},\{s_n\},\{c_n\},\la),
\lll{gen delta prefactor}
\eea

\noi
where $\delta_j^{(i)}$ is the $j^\ths$ momentum-conserving delta function
in the $i^\ths$ product of
delta functions ($\delta_j^{(i)}$ includes a
longitudinal-momentum factor), $N_{\delta}^{(i)}$ is the number of delta
functions in the $i^\ths$ product, and
$F^{(i)}_\CD$ is a function of the cutoff and the quantum numbers of the
particles in the matrix element, but does not contain delta functions that fix momenta.
We define $N_{\mathrm{part}}$ to be the number of particles in state
$\left| I \right>$ plus the number of particles in
state $\left| F \right>$, and $n=1,2,3,\ldots,N_{\mathrm{part}}$.  The
momentum, spin polarization, and color of particle $n$
are given by $p_n$, $s_n$, and $c_n$.  We define $N_{\mathrm{int}}^{(i)}$
to be the number of particles in the matrix element that participate in an
interaction for the $i^\ths$
product of delta functions.  In order for the IMO to have the
dimensions $(\mathrm{mass})^2$,
$F^{(i)}_\CD$ must have the dimensions
$({\mathrm{mass}})^{4-N_{\mathrm{int}}^{(i)}}$.
Note that we are suppressing the dependence of the RHS of this equation on
$r$.

We have assumed that any matrix element of the IMO can be expanded in powers of transverse momenta, not
including the momentum-conserving delta functions; so

\bea
F^{(i)}_\CD(\{p_n\},\{s_n\},\{c_n\},\la) = \la^{4 -
N_{\mathrm{int}}^{(i)}} \sum_{\left\{ m_{nt} \right\}}
z^{\left\{m_{nt}
\right\}}_i
\Big( \{p_n^+\},\{s_n\},\{c_n\}\Big) \prod_{n=1}^{N_{\mathrm{part}}}
\prod_{t=1}^2 \left( \frac{ p_{n \perp}^t }{\la}
\right)^{m_{nt}} ,
\lll{F}
\eea

\noi
where $t$ denotes a component of
transverse momentum and  $m_{nt}$ is a non-negative integer index
associated with transverse-momentum component
$t$ of particle $n$.  The sum is over all values of each of the
$m_{nt}$'s,
subject to the constraint that

\bea
4 - N_{\mathrm{int}}^{(i)} - \sum_{n,t} m_{n t} \neq 0 ,
\lll{nonmarg cond}
\eea

\noi
which is necessary to avoid terms in the momentum expansion that are
cutoff-independent.
The $z^{\left\{m_{nt} \right\}}_i$'s are the coefficients for the momentum
expansion.
They depend on $i$ and the $m_{nt}$'s and are functions of the particles'
longitudinal momenta,
spin polarizations, and colors.

Since the RHS of \reff{prod exp} has the same product of delta functions as the LHS, we can write

\bea
\left[ \dVome{r} - \sum_{s=2}^{r-1}  B_{r-s,s}
\left<F \right| V^{(r-s)}(\la) \left| I \right>\right]^{(i)} = \left[
\prod_{j=1}^{N_{\delta}^{(i)}}
\delta_j^{(i)} \right]
G^{(i)}(\{p_n\},\{s_n\},\{c_n\},\la,\la'),
\lll{G prefactor}
\eea

\noi
where $G^{(i)}$ has dimensions $({\mathrm{mass}})^{4 - N_{\mathrm{int}}^{(i)}}$, and by inspection of the 
LHS of \reff{prod exp} and \reff{gen delta prefactor}, is a function of the quantum numbers 
of the particles, $\la$, and $\la'$.  Substitution of Eqs. (\ref{eq:G prefactor}) and (\ref{eq:gen delta
prefactor}) into \reff{prod exp} yields

\bea
F^{(i)}_\CD(\{p_n\},\{s_n\},\{c_n\},\la) -
F^{(i)}_\CD(\{p_n\},\{s_n\},\{c_n\},\la') =
G^{(i)}(\{p_n\},\{s_n\},\{c_n\},\la,\la') ,
\lll{eq}
\eea

\noi
where the momenta in this equation are constrained by the delta-function
conditions.

Since the LHS of \reff{eq} is the difference of a function of $\la$ and the
same function with $\la \rightarrow \la'$, $G^{(i)}$ must be as well.  Since the 
LHS of \reff{eq} can be expanded in powers of transverse momenta, $G^{(i)}$ 
must have the form

\bea
G^{(i)}(\{p_n\},\{s_n\},\{c_n\},\la,\la') &=& \sum_{\left\{ m_{nt}
\right\}} Z_i^{\left\{ m_{nt} \right\}}
\Big( \{p_n^+\},\{s_n\},\{c_n\}\Big) \n
&\times& \left[\la^{4 - N^{(i)}_{\mathrm{int}}} \prod_{n,t} \left( \frac{
p_{n \perp}^t
}{\la}
\right)^{m_{nt}} - \la'^{\, 4 - N^{(i)}_{\mathrm{int}}} \prod_{n,t} \left(
\frac{ p_{n \perp}^t }{\la'}
\right)^{m_{nt}} \right] ,
\lll{G}
\eea

\noi
where the sum is restricted by \reff{nonmarg cond}.

Substituting Eqs. (\ref{eq:G}) and (\ref{eq:F}) into \reff{eq}, we find that

\bea
&&\la^{4 - N^{(i)}_{\mathrm{int}}} \sum_{\left\{ m_{nt} \right\}}
z_i^{\left\{m_{nt}
\right\}}
\Big( \{p_n^+\},\{s_n\},\{c_n\}\Big) \prod_{n,t} \left( \frac{ p_{n
\perp}^t }{\la}
\right)^{m_{nt}} \n
&-& \la'^{\, 4 - N^{(i)}_{\mathrm{int}}} \sum_{\left\{ m_{nt} \right\}}
z_i^{\left\{m_{nt}
\right\}}
\Big( \{p_n^+\},\{s_n\},\{c_n\}\Big) \prod_{n,t} \left( \frac{ p_{n
\perp}^t
}{\la'}
\right)^{m_{nt}} \n
&=& \sum_{\left\{ m_{nt} \right\}} Z_i^{\left\{ m_{nt} \right\}}
\Big( \{p_n^+\},\{s_n\},\{c_n\}\Big) \left[\la^{4 -
N^{(i)}_{\mathrm{int}}} \prod_{n,t} \left( \frac{ p_{n \perp}^t
}{\la}
\right)^{m_{nt}} - \la'^{\, 4 - N^{(i)}_{\mathrm{int}}} \prod_{n,t} \left(
\frac{ p_{n \perp}^t }{\la'}
\right)^{m_{nt}}
\right] .
\eea

\noi
Matching powers of transverse momenta on both sides of this equation gives

\bea
&&z_i^{\left\{ m_{nt} \right\}}
\Big( \{p_n^+\},\{s_n\},\{c_n\}\Big) \left[ \la^{4 -
N^{(i)}_{\mathrm{int}} - \Sigma_{n,t} \, m_{nt}}  -
\la'^{\, 4 - N^{(i)}_{\mathrm{int}} - \Sigma_{n,t} \, m_{nt}} \right] \n
&=& Z_i^{\left\{ m_{nt} \right\}}
\Big( \{p_n^+\},\{s_n\},\{c_n\}\Big) \left[ \la^{4 -
N^{(i)}_{\mathrm{int}} - \Sigma_{n,t} \, m_{nt}}  -
\la'^{\, 4 - N^{(i)}_{\mathrm{int}} - \Sigma_{n,t} \, m_{nt}} \right] .
\lll{power match}
\eea

\noi
The factor in brackets cannot be zero because $\la \neq \la'$ and
\reff{nonmarg cond} holds.  Thus \reff{power match} implies that

\bea
z_i^{\left\{ m_{nt} \right\}} &=& Z_i^{\left\{ m_{nt} \right\}} .
\lll{cc exp sol}
\eea

\noi
Then Eqs. (\ref{eq:F}), (\ref{eq:G}), and (\ref{eq:cc exp sol}) imply that

\bea
F^{(i)}_\CD(\{p_n\},\{s_n\},\{c_n\},\la) =
G^{(i)}(\{p_n\},\{s_n\},\{c_n\},\la,\la') \, \rule[-1.7mm]{.1mm}{4.4mm}_{\; \la \; \mathrm{terms}} \,\,\, ,
\lll{cc F}
\eea

\noi
where ``$\la \; \mathrm{terms}$" means that $G^{(i)}$ is to be expanded in powers of transverse
momenta and only the terms in the expansion that are proportional to powers or inverse powers of $\la$ contribute.  From
Eqs. (\ref{eq:gen delta prefactor}), (\ref{eq:G prefactor}), and (\ref{eq:cc F}),

\bea
\left<F \right| V^{(r)}_{\CD}(\la) \left| I \right>^{\,(i)} = \left[ \dVome{r} - \sum_{s=2}^{r-1}  B_{r-s,s}
\left<F \right| V^{(r-s)}(\la) \left| I \right>\right]^{(i)}_\lz ,
\eea

\noi
where it is understood that the momentum-conserving delta functions are ignored for the purposes of
transverse-momentum expansions.
Since a matrix element is the sum of the contributions to it from different products of delta functions, both 
sides of this equation can be summed over $i$ to obtain

\bea
\left<F \right| V^{(r)}_{\CD}(\la) \left| I \right> = \left[ \dVome{r} - \sum_{s=2}^{r-1}  B_{r-s,s}
\left<F \right| V^{(r-s)}(\la) \left| I \right>\right]_\lz .
\lll{cc soln D}
\eea

\noi
This equation tells us how to calculate the cutoff-dependent part of the $\OR(\gla^r)$ reduced interaction 
in terms of lower-order contributions.  

\vspace{.20in}

\noi
{\large \bf B.2  \hspace{.05in} The Recursion Relation for the Cutoff-Independent Part}

\vspace{.08in}

Since we have specified $V^{(1)}$ and $V^{(2)}_\CI$, we need to determine
$V^{(r)}_\CI$ for $r \ge 3$.
It is useful to first consider which contributions to $V^{(r)}(\la)$ can
be cutoff-independent.

A matrix element of the cutoff-independent part of $V^{(r)}(\la)$ can be
expanded in products of delta
functions and in powers of transverse momenta just as was done for the
cutoff-dependent part.  Thus we can write

\bea
\left<F \right| V^{(r)}_\CI \left| I \right> = \sum_i \left<F \right|
V^{(r)}_\CI \left| I \right>^{\,(i)} ,
\eea

\noi
where

\bea
\left<F \right| V^{(r)}_\CI \left| I \right>^{\,(i)} &=&
\left[ \prod_{j=1}^{N_{\delta}^{(i)}} \delta_j^{(i)} \right]
F^{(i)}_\CI(\{p_n\},\{s_n\},\{c_n\})
\lll{CI prefactor}
\eea

\noi
and

\bea
F^{(i)}_\CI(\{p_n\},\{s_n\},\{c_n\}) = \la^{4 - N_{\mathrm{int}}^{(i)}}
\sum_{\left\{ m_{nt} \right\}}
w^{\left\{m_{nt}
\right\}}_i
\Big( \{p_n^+\},\{s_n\},\{c_n\}\Big) \prod_{n=1}^{N_{\mathrm{part}}}
\prod_{t=1}^2 \left( \frac{ p_{n \perp}^t }{\la}
\right)^{m_{nt}} .
\lll{FCI}
\eea

\noi
The sum is over all values of each of the $m_{n t}$'s, subject to the
constraint that

\bea
4 - N_{\mathrm{int}}^{(i)} - \sum_{n,t} m_{n t} = 0 ,
\lll{marg cond}
\eea

\noi
which ensures that all the terms in the expansion of $F^{(i)}_\CI$ are
cutoff-independent.

\reff{marg cond} places constraints on the possible cutoff-independent
contributions to the reduced interaction.   Any contribution to a matrix
element of $V^{(r)}(\la)$ has an $N^{(i)}_{\mathrm{int}} \ge 2$, but
\reff{marg cond} can only hold if $N_{\mathrm{int}}^{(i)} \le 4$.

Suppose that
$N_{\mathrm{int}}^{(i)}=2$.  In this case, \reff{marg cond} implies that
$F^{(i)}_\CI$ is quadratic in transverse momenta.  Due to
approximate cluster decomposition, only interacting particles'
transverse momenta can appear in $F^{(i)}_\CI$.  So any contribution
to $F^{(i)}_\CI$ can depend on the transverse momenta of two
interacting particles.  Thus
$F^{(i)}_\CI$ can be written
as a sum of terms, where each term corresponds to a distinct pair of
interacting particles.  The momentum dependence of each term in
$F^{(i)}_\CI$ is limited to dependence on the momenta of the interacting
particles and the total longitudinal momentum:

\bea
F^{(i)}_\CI(\{p_n\},\{s_n\},\{c_n\}) &=& \sum_m F^{(i,m)}_\CI(k_m, k_m',
P^+,\{s_n\},\{c_n\}) \n
&=& \sum_m F^{(i,m)}_\CI(k_m, P^+,\{s_n\},\{c_n\}) ,
\eea

\noi
where $k_m$ and $k_m'$ are the momenta for the initial and final
particles in the $m^\ths$ interacting pair,  and where we have used the fact
that for $N_{\mathrm{int}}^{(i)}=2$, momentum conservation implies that
$k_m=k_m'$.   $F^{(i,m)}_\CI$ must be quadratic in $\vec k_{m \perp}$ or
it must be zero.
The matrix elements of the IMO are boost-invariant,  as
is the delta-function product in
\reff{CI prefactor}.  This means that $F^{(i)}_\CI$ must be boost-invariant, but it
cannot be if $F^{(i,m)}_\CI$ is quadratic in $\vec k_{m \perp}$; so $F^{(i,m)}_\CI$ must be zero.
Thus the reduced interaction does not contain any cutoff-independent two-point interactions.

Note that two-point interactions are self-energies, and they change the
particle dispersion relation.
If they change the dispersion relation such that the coefficients of the
free relation
become modified by interactions, then this can be viewed as
renormalization of the field operators, i.e. wave-function
renormalization.
This effect is absent unless either $F^{(i)}_\CI$ or $F^{(i)}_\CD$ can be
quadratic in
transverse momenta for $N^{(i)}_{\mathrm{int}}=2$.  We have just shown
that boost
invariance prevents this for
$F^{(i)}_\CI$, and according to \reff{nonmarg cond}, $F^{(i)}_\CD$ cannot be
quadratic in
transverse momenta for $N^{(i)}_{\mathrm{int}}=2$; so there
is no wave-function renormalization at any order in $\gla$ in our approach.

According to \reff{marg cond}, if $N_{\mathrm{int}}^{(i)}=3$, then $F^{(i)}_\CI$
has to be linear in transverse momenta, and
if $N_{\mathrm{int}}^{(i)}=4$, then $F^{(i)}_\CI$ has to be independent of all
transverse momenta.  According to assumptions that we made in Subsection \ref{subsubsec: rep of toi}, 
if $r$ is odd then
$V^{(r)}_\CI$ has no $N_{\mathrm{int}}^{(i)}=4$ part, and if $r$ is even then
$V^{(r)}_\CI$ has no $N_{\mathrm{int}}^{(i)}=3$ part.  This means that if $r$ is odd, then
$V_\CI^{(r)}$ can contain only three-point interactions that are linear in transverse momenta,
and if $r$ is even, then $V_\CI^{(r)}$ can contain only four-point interactions that are independent of all
transverse momenta.

To calculate $V^{(r)}_\CI$, we consider \reff{prod exp} with $r \rightarrow r+2$:

\bea
\left<F \right| V^{(r+2)}_\CD(\la) \left| I \right>^{(i)} -
\left<F \right| V^{(r+2)}_\CD(\la') \left| I \right>^{(i)} = \left[
\dVome{r+2} - \sum_{s=2}^{r+1}  B_{r+2-s,s}
\left<F \right| V^{(r+2-s)}(\la) \left| I \right>\right]^{(i)}  \hspace{-3mm}.
\lll{marg exp}
\eea

\noi
In the remainder of this appendix, we assume that $r$ is odd.  Then we need to consider only 
$N_{\mathrm{int}}^{(i)}=3$ initially.  We expand
\reff{marg exp} in powers of transverse momenta and keep only the linear term:

\bea
0 = \left[
\dVome{r+2} - \sum_{s=2}^{r+1}  B_{r+2-s,s}
\left<F \right| V^{(r+2-s)}(\la) \left| I
\right>\right]^{(i)}_{\vec p_\perp^{\, 1} \; \mathrm{term}}  ,
\eea

\noi
where we have used the fact that $\left<F \right| V^{(r+2)}_\CD(\la) \left| I \right>^{(i)}$ has no 
linear part when $N^{(i)}_\mathrm{int}=3$ [see \reff{nonmarg cond}].
We move the first term in the sum on the RHS to the LHS:

\bea
B_{r,2}
\left<F \right| V^{(r)}_\CI \left| I \right>^{(i)} = \left[
\dVome{r+2} - \sum_{s=3}^{r+1}  B_{r+2-s,s}
\left<F \right| V^{(r+2-s)}(\la) \left| I
\right>\right]^{(i)}_{\vec p_\perp^{\, 1} \; \mathrm{term}}  .
\eea

\noi
Now we can sum over all values of $i$ corresponding to three-point interactions:

\bea
\left<F \right| V^{(r)}_\CI \left| I \right> &=& \frac{1}{B_{r,2}} \left[ \dVome{r+2} -
\sum_{s=3}^{r+1}  B_{r+2-s,s}
\left<F \right| V^{(r+2-s)}(\la) \left| I \right>\right]^\mathrm{3-point}_{\vec p_\perp^{\, 1} \; \mathrm{term}} .
\lll{CI cc soln B1}
\eea

To use this equation, we also need to use \reff{marg exp} to solve for $V^{(r+1)}_\CI$.  Since $r$ is odd,
$r+1$ is even.  Thus $V^{(r+1)}_\CI$ will contain only transverse-momentum-independent four-point interactions.
Using steps analogous to those that led to \reff{CI cc soln B1}, we find that

\bea
\left<F \right| V^{(r+1)}_\CI \left| I \right> &=& \frac{1}{B_{r+1,2}} \left[ \dVome{r+3} -
\sum_{s=3}^{r+2}  B_{r+3-s,s}
\left<F \right| V^{(r+3-s)}(\la) \left| I \right>\right]^\mathrm{4-point}_{\vec p_\perp^{\, 0} \; \mathrm{term}} .
\lll{CI cc soln B2}
\eea

\noi
To use these equations, the right-hand sides have to be expanded in powers of transverse momenta.  Only 
three-point interactions that are linear in transverse momenta contribute to $V^{(r)}_\CI$, and only 
four-point interactions that are independent of all transverse momenta contribute to $V^{(r+1)}_\CI$.  

These equations are coupled integral equations\footnote{It is very difficult to prove that integral equations of this type have a unique
solution; so we simply assume that it is true in this case.} because both $V^{(r)}_\CI$ and $V^{(r+1)}_\CI$ appear
on the RHS of \reff{CI cc soln B1} inside integrals in $\delta V^{(r+2)}$, and $V^{(r+1)}_\CI$ appears on the
RHS of \reff{CI cc soln B2} inside integrals in $\delta V^{(r+3)}$.  It would seem that $V^{(r+2)}_\CI$ also appears
on the RHS of \reff{CI cc soln B2} inside integrals in $\delta V^{(r+3)}$, but $V^{(r+2)}_\CI$
cannot couple to $V^{(1)}$ to produce a transverse-momentum-independent four-point contribution to $\delta V^{(r+3)}$.  
This is because the cutoff function
$T_2^{(\la,\la')}$ vanishes when the intermediate state is massless and all external transverse momenta are zero.  This means that since we
specified $V^{(1)}$ and $V^{(2)}_\CI$ in Subsection \ref{subsubsec: rep of toi}, we can use Eqs. (\ref{eq:CI cc soln B1}) and
(\ref{eq:CI cc soln B2}) to solve for $V^{(3)}_\CI$ and $V^{(4)}_\CI$ simultaneously, and $V^{(5)}_\CI$ and $V^{(6)}_\CI$ simultaneously,
and so on.  Note that before we can use these equations to solve for $V^{(r)}_\CI$ and $V^{(r+1)}_\CI$ simultaneously, we must
first use \reff{cc soln D} both to compute $V^{(r)}_\CD(\la)$ in terms of lower-order interactions and to express $V^{(r+1)}_\CD(\la)$
in terms of lower-order interactions and $V^{(r)}(\la)$.

Before concluding this appendix, we would like to deduce a bit more about the 
relationship of $\gla$ to $\glap$.
According to \reff{running 1} and the surrounding discussion, 
this relationship is determined by the matrix element
$\left< g_2 g_3 \right| \delta V \left| g_1 \right>$, which can be expanded in powers
of $\glap$:

\bea
\left< g_2 g_3 \right| \delta V \left| g_1 \right> = \sum_{t=3}^\infty \glap^t 
\left< g_2 g_3 \right| \delta V^{(t)} \left| g_1 \right> .
\lll{last run}
\eea

\noi
Recall that $\delta V^{(t)}$ is built from products of $V^{(r)}(\la')$'s.   This implies that $\delta V^{(t)}$
can change particle number by 1 only if $t$ is odd, and thus \reff{last run} implies that the coupling runs at
odd orders; i.e. $C_s$ is zero if $s$ is even [see \reff{scale dep}].

\newpage

\begin{tabbing}
{\large \bf APPENDIX C:}$\;$\={\large \bf Technical Issues in the Numerical Calculation of Matrix}\\
\>{\large \bf Elements}
\end{tabbing}

\vspace{.1in}

In this appendix, we discuss some of the technical issues involved in the numerical calculation of the matrix
elements $\ME$.  In Section C.1 we discuss how we put the matrix elements into a form that is amenable to numerical calculation.
In Section C.2 we briefly cover three topics: we show how the glueball state with $\jj_3^\mathrm{R}$ eigenvalue $-j$ can be
written in terms of the glueball state with $\jj_3^\mathrm{R}$ eigenvalue $j$, we list a few tricks that allow us to 
reduce the number of matrix elements that we must compute, and
we present our method for estimating how numerical uncertainties in the matrix elements translate into uncertainties 
in the spectrum.

\vspace{.20in}

\noi
{\large \bf C.1  \hspace{.05in} Preparation of Integrals for Monte Carlo}

\vspace{.08in}

There are two types of contributions to $\ME$: finite sums and five-dimensional integrals.  We use Mathematica \cite{math}
to evaluate the finite sums to as many digits as we wish.  To evaluate the integrals, we combine them into
one integral and use the VEGAS Monte Carlo routine \cite{vegas}.  It takes a bit of work to put the
integral into a form that will converge.

There are two main difficulties with 
getting the integral to converge.  The first difficulty is that $\ME\INBF$ looks divergent: as $x' \rightarrow x$,
the sum in \reff{inbf} diverges.  This divergence is misleading because it actually contributes nothing to the integral 
(assuming that we calculate the integral carefully; see the discussion below).  
Left unchecked, this false divergence prevents the integral from converging with VEGAS.  To rectify
the problem, we want to subtract the false divergence from the integrand.  Since it integrates to zero, this is allowed.

The second main difficulty with getting the integral to converge is roundoff error.  Even after we subtract the false
divergence from the integrand, the integrand peaks around $x'=x$, and in this region there are large cancellations in some of the quantities that we
have defined.  To prevent these cancellations from causing roundoff error, we rewrite these quantities so that the
cancellations are explicit, before we turn to numerics.

\vspace{.10in}

\noi
{\bf C.1.1  \hspace{.05in} Subtraction of the False Divergence}

\vspace{.04in}

We begin by defining a set of variables that is natural for dealing with the false divergence.  We define

\bea
\eta = x-x' .
\eea

\noi
We change variables from $\kp$ and $\kpp$ to the dimensionless transverse variables $\rp$ and $\Wp$:

\bea
\rp &=& \frac{d}{2} (\kp + \kpp) ,\n
\Wp &=& \frac{d}{2} \frac{\kp - \kpp}{\sqrt{\eta}} .
\eea

\noi
We define the angle between $\rp$ and $\Wp$ to be $\beta$:

\bea
\rp \cdot \Wp = r w \cos \beta.
\eea

\noi
We also define

\bea
r_\pm &=& \sqrt{r^2 + \eta w^2 \pm 2 r w \sqrt{\eta} \cos \beta} ,
\eea

\noi
and then we can derive a host of useful relations:

\bea
\kp &=& \frac{\rp + \sqrt{\eta} \Wp}{d} ,\n
\kpp &=& \frac{\rp - \sqrt{\eta} \Wp}{d} , \n
k &=& \frac{r_+}{d} ,\n
k' &=& \frac{r_-}{d} , \n
\kp \cdot \kpp &=& \frac{1}{d^2} (r^2 - \eta w^2) ,\n
\cos \gamma &=& \frac{r^2 - \eta w^2}{r_+ r_-} ,\n
\sin \gamma &=& \frac{-2 r w \sqrt{\eta}}{r_+ r_-} \sin \beta ,
\lll{relations}
\eea

\noi
and

\bea
dk \BIT dk' \BIT d\gamma \BIT k \BIT k' \BIT \theta(k) \BIT \theta(k') \BIT \theta(\gamma) \BIT \theta(2 \pi - \gamma) &=& 
\frac{4 \eta}{d^4} \BIT dr \BIT dw \BIT d\beta \BIT r \BIT w \BIT \theta(r) \BIT \theta(w) \BIT \theta(\beta)
\BIT \theta(2 \pi - \beta) .
\eea

In terms of these variables, $\ME\INBF$ takes the form

\bea
\ME\INBF &=& -\frac{N_\mathrm{c} \gla^2}{2 \pi^3 d^2} \int D' \log\eta \BIT W_{q,q'} \T_{t'}(r_-) 
\T_t(r_+) e^{-(\la d)^{-4} (\bar \Delta_{FK}^2 + \bar \Delta_{IK}^2)} \tim \sum_{i=1}^5 E'_i 
\prod_{m=1; \; m \neq i}^5 E_m ,
\lll{new var}
\eea

\noi
where

\bea
D' = dx \BIT dx' \BIT dr \BIT dw \BIT d\beta \BIT r \BIT w \BIT \eta \BIT \theta(x) \BIT \theta(1-x) \BIT \theta(x') \BIT \theta(x-x') \BIT \theta(r) 
\BIT \theta(w) \BIT \theta(\beta) \BIT \theta(2 \pi - \beta) \bar L^{(e)}_l(x) .
\eea

\noi
To avoid roundoff error, we have defined simplified dimensionless versions of the differences of 
the free masses and the derivatives of these differences:

\bea
\bar \Delta_{FI} &=& \frac{r^2 \eta (1-x-x') + w^2 \eta^2 (1-x-x') - 2 w r \sqrt{\eta} (x[1-x]+x'[1-x']) \cos \beta}
{x (1-x) x' (1-x')} ,\n
\bar \Delta_{FK} &=& - \frac{r^2 \eta + w^2 (1-x+1-x')^2 + 2 r w \sqrt{\eta} (1-x+1-x') \cos \beta}{(1-x) (1-x')},\n
\bar \Delta_{IK} &=& - \frac{r^2 \eta + w^2 (x+x')^2 - 2 r w \sqrt{\eta} (x+x') \cos \beta}{x x'},\n
\bar \Delta_{FK}' &=& \frac{r_-^2}{(1-x')^2} - 4 \frac{w^2}{\eta},\n
\bar \Delta_{IK}' &=& \frac{r_-^2}{\xpsq} - 4 \frac{w^2}{\eta}.
\eea

\noi
\reff{new var} is now dimensionless, except for the factor in front of the integral, which is proportional to $1/d^2$.

As $x' \rightarrow x$ ($\eta \rightarrow 0$), the contributions to the integrand from the first and second terms in the sum diverge.
In the limit $x' \rightarrow x$, the contribution to the integral from these terms can be written\footnote{We do not replace all the
occurrences of $x'$ with $x$ because doing so hampers the convergence of the integral.}

\bea
&& -\frac{N_{\mathrm{c}} \gla^2}{2 \pi^3 d^2} \delta_{q,q'} \int D' \log\eta \BIT \T_{t'}(r)
\T_t(r) e^{-32 (\la d)^{-4} w^4} \left[ \frac{1}{\eta^2} - 
64 (\la d)^{-4} \frac{w^4}{\eta^2} \right] \bar L_{l'}^{(e)}(x') (x+x') \tim (1-x+1-x') .
\eea

\noi
An examination of this integral reveals a problem: the transverse integrals are zero and the longitudinal integrals are infinite.
To solve this problem, we consider 
what would have happened if we had not yet taken $\epsilon \rightarrow 0$.  In this case, 
the transverse integrals would be zero and the longitudinal integrals would be finite.  Thus the $\epsilon \rightarrow 0$ limit of this
integral would be zero.  This means that this integral is actually zero, and we can subtract it from the full integral
in \reff{new var}:

\bea
&& \ME\INBF = -\frac{N_{\mathrm{c}} \gla^2}{2 \pi^3 d^2} \int D' \log\eta \tim 
\Bigg[ \BIT W_{q,q'} \T_{t'}(r_-)
\T_t(r_+) e^{-(\la d)^{-4} (\bar \Delta_{FK}^2 + \bar \Delta_{IK}^2)} \sum_{i=1}^5 E'_i 
\prod_{m=1; \; m \neq i}^5 E_m \n
&-& \delta_{q,q'} \T_{t'}(r) 
\T_t(r) e^{-32 (\la d)^{-4} w^4} \left( \frac{1}{\eta^2} - 
64 (\la d)^{-4} \frac{w^4}{\eta^2} \right) \bar L_{l'}^{(e)}(x') (x+x') (1-x+1-x') \Bigg].
\eea

\noi
Once we have performed this subtraction, there is no ambiguity about the value of the full integral, and it converges when computed numerically.

\vspace{.10in}

\noi
{\bf C.1.2  \hspace{.05in} Combination of the Integrals}

\vspace{.04in}

We now combine all the five-dimensional integrals into one integral:

\bea
\ME_{5-\mathrm{D}} = - \frac{N_\mathrm{c} \gla^2}{2 \pi^3 d^2} \int D' \left[ I_\CON + I\EXF + I_\INX + I\INBF \right] ,
\lll{semifinal}
\eea

\noi
where

\bea
I_\CON &=& \left[ \delta_{j,-2} \delta_{q',2} \delta_{q,2} + \delta_{j, 2}
\delta_{q', 1} \delta_{q, 1} - \delta_{j,0} \delta_{q',3} \delta_{q,3} \right]
\bar L^{(e)}_{l'}(x') \T_{t'}(r_-) \T_t(r_+) e^{-(\la d)^{-4} \bar \Delta_{FI}^2} , \n
I\EXF &=& \bar L^{(e)}_{l'}(x')  \T_{t'}(r_-) \T_t(r_+)
e^{-(\la d)^{-4} \bar \Delta_{FI}^2} \frac{1}{\eta} \left( \frac{1}{\bar \Delta_{FK}} +
\frac{1}{\bar \Delta_{IK}} \right) \left( 1 - e^{-2 (\la d)^{-4} \bar \Delta_{FK} \bar \Delta_{IK}} \right) \tim
\left[ \frac{r_+^2}{x(1-x)} S_{q,q'}^{(1)} + \frac{r_-^2}{x'(1-x')} S_{q,q'}^{(2)} + \frac{r_+ r_-}{x(1-x)x'(1-x')} S_{q,q'}^{(3)} \right] ,\n
I_\INX &=& W_{q,q'} \bar L^{(e)}_{l'}(x') \T_{t'}(r_-) \T_t(r_+)
e^{-(\la d)^{-4} \bar \Delta_{FI}^2} \frac{\left( 1 - e^{-2 (\la d)^{-4} \bar \Delta_{FK} \bar \Delta_{IK}}
\right)}{x (1-x) x' (1-x')} \frac{1}{\eta} \frac{1}{\bar \Delta_{FK} \bar \Delta_{IK}} \tim \Big[ 
-\eta^5 w^4 - 2 \eta^4 w^2 (r^2 + 2 w^2 [1-2x]) - \eta^3 (r^4+4r^2 w^2 [1-2x] + 6 w^4 [1-2x]^2) \n
&-& 4 \cos \beta \sqrt{\eta} r w ([-1+\eta^2] r^2 + \eta w^2 [1-x-x']^2) (1-x-x') + 
8 r^2 w^2 x (1-x) \n
&-& 4 \cos^2 \!\beta \, r^2 w^2 (\eta^4 + 2 \eta^3 [1 - 2 x] + 4 x [1-x] - 4 \eta^2 x [1-x] - 2 \eta [1 - 2 x]) \n
&-& 4 \eta^2 w^2 (w^2 [1-2x]^3 + r^2 [1-2 x (1-x) ]) \n
&+& 2 \eta (r^4 - 2 r^2 w^2 [1-2x] - 4 w^4 x [-1+3 x-4x^2+2x^3]) \Big] ,\n
I\INBF &=& \log\eta \bigg[ W_{q,q'} \T_{t'}(r_-)
\T_t(r_+) e^{-(\la d)^{-4} (\bar \Delta_{FK}^2 + \bar \Delta_{IK}^2)} \sum_{i=1}^5 E'_i 
\prod_{m=1; \; m \neq i}^5 E_m \n
&-& \delta_{q,q'} \T_{t'}(r) \T_t(r) e^{-32 (\la d)^{-4} w^4} \left( \frac{1}{\eta^2} - 
64 (\la d)^{-4} \frac{w^4}{\eta^2} \right) \bar L_{l'}^{(e)}(x') (x+x') \tim (1-x+1-x') \bigg] .
\eea

\noi
We have rewritten the integrand of $\ME_\INX$ to eliminate large roundoff errors, at the expense of making it more complicated. 
To further avoid roundoff errors, we rewrite a few of the $S_{q,q'}^{(i)}$'s:

\bea
S_{3,3}^{(1)} &=& \cos (j \gamma) (2 \eta^2 + 2 \eta (1-2x) + (1-2x)^2) ,\n
S_{3,4}^{(3)} &=& - \sin \gamma \sin(j \gamma) (2 \eta^2 [1-2x] + \eta [1 - 6x + 6x^2] - 2 x [1 - 3x + 2x^2]) ,      \n
S_{4,4}^{(1)} &=& \cos (j \gamma) (1-2x) (1 + 2 \eta - 2 x) .
\eea

\noi
Note that to compute some of the trigonometric functions that appear in this integral [such as $\cos(j\gamma)$] in terms
of the integration variables, it is necessary
to use recursion relations that define these functions in terms of $\cos\gamma$ and $\sin\gamma$ so that we can use
\reff{relations}.

The integral in \reff{semifinal} converges slowly.  This is because it is strongly peaked when $x' \simeq x$, even though we have
subtracted the false divergence.  To spread out this region, we change variables from $x'$ to $p$ where

\bea
x' = x \left(1-e^{-p} \right).
\eea

\noi
Now the integral is

\bea
\ME_{5-\mathrm{D}} &=& - \frac{N_\mathrm{c} \gla^2}{2 \pi^3 d^2} \int_0^1 dx \int_0^\infty dp \int_0^\infty dr \int_0^\infty dw \int_0^{2 \pi} d\beta 
\BIT r \BIT w \BIT \eta^2 \bar L_l^{(e)}(x) \tim \left[ I_\CON + I\EXF + I_\INX + I\INBF \right] .
\lll{semifinal2}
\eea

As a final step, we note that VEGAS requires the region of integration to be finite.  Thus we change variables from
$p$, $r$, and $w$ to $y_p$, $y_r$, and $y_w$:

\bea
p &=& \frac{2}{1+y_p} - 1, \n
r &=& \frac{2}{1+y_r} - 1, \n
w &=& \frac{2}{1+y_w} - 1,
\eea

\noi
and then the final expression for the contribution to the matrix elements from the five-dimensional integral is 

\bea
&& \ME_{5-\mathrm{D}} = - \frac{4 N_\mathrm{c} \gla^2}{\pi^3 d^2} \int_0^1 dx \int_{-1}^1 d y_p \int_{-1}^1 d y_r \int_{-1}^1 d y_w 
\int_0^{2 \pi} d\beta 
\BIT r \BIT w \BIT \frac{1}{(1+y_p)^2} \tim \frac{1}{(1+y_r)^2} \frac{1}{(1+y_w)^2} \eta^2 \bar L_l^{(e)}(x) 
\left[ I_\CON + I\EXF + I_\INX + I\INBF \right] .
\lll{final}
\eea

\noi
The integral converges nicely in this form.

\vspace{.20in}

\noi
{\large \bf C.2  \hspace{.05in} Miscellaneous Issues}

\vspace{.08in}

\vspace{.10in}

\noi
{\bf C.2.1  \hspace{.05in} Rotational Symmetry: $\bbox{j \rightarrow -j}$}

\vspace{.04in}

To compute the glueball spectrum, we compute the matrix $\ME$ and diagonalize it for each value of $j$ separately.
For a given $|j|$, the matrices 
with $j=|j|$ and $j=-|j|$ are simply related, and we can use this fact to avoid computing and diagonalizing both of them.
By inspection, we determine that

\bea
\left< q',l',t',-j \right| \M \left| q,l,t,-j \right> = \sum_{q'',q'''} \left< q''',l',t',j \right| \M \left| q'',l,t,j \right> f(q,q'')f(q',q''') ,
\eea

\noi
where

\bea
f(q,q') = \delta_{q,1} \delta_{q',2} + \delta_{q,2} \delta_{q',1} + \delta_{q,3} \delta_{q',3} - \delta_{q,4} \delta_{q',4} .
\eea

\noi
This simply means that the basis $\left| q,l,t,-j \right>$ is related to the basis $\left| q,l,t,j \right>$ by swapping the
states $\left| 1,l,t,j \right>$ and $\left| 2,l,t,j \right>$, and changing the sign of $\left| 4,l,t,j \right>$.  Renaming basis states and
changing their phases has no
effect on the eigenvalues of the matrix; so $\left| \Psi^{(-j)n}(P) \right>$ has the same mass as
$\left| \Psi^{jn}(P) \right>$.  It also means that since

\bea
\left| \Psi^{jn}(P) \right> &=& \sum_{qlt} R^{jn}_{qlt} \left| q,l,t,j \right> \n
&=& \sum_{lt} \left[ R^{jn}_{1lt} \left| 1,l,t,j \right> + 
R^{jn}_{2lt} \left| 2,l,t,j \right> + R^{jn}_{3lt} \left| 3,l,t,j \right> + R^{jn}_{4lt} \left| 4,l,t,j \right> \right],
\eea

\noi
we have

\bea
\left| \Psi^{(-j)n}(P) \right> &=& \sum_{lt} \left[ R^{jn}_{1lt} \left| 2,l,t,j \right> + 
R^{jn}_{2lt} \left| 1,l,t,j \right> + R^{jn}_{3lt} \left| 3,l,t,j \right> - R^{jn}_{4lt} \left| 4,l,t,j \right> \right] \n
&=& \sum_{q'qlt} R^{jn}_{qlt} f(q,q') \left| q',l,t,j \right> .
\eea

\noi
Thus by diagonalizing $\left< q',l',t',j \right| \M \left| q,l,t,j \right>$, we also obtain the eigenvalues and eigenstates of
$\left< q',l',t',-j \right| \M \left| q,l,t,-j \right>$.

\vspace{.10in}

\noi
{\bf C.2.2  \hspace{.05in} Reducing the Number of Matrix Elements to Compute}

\vspace{.04in}

There are a few facts that allow us to reduce the number of matrix elements that we have to compute.  First, because of
gluon-exchange symmetry,
the basis state $\left| q,l,t,j \right>$ is zero if $l+j$ is even and $q=4$, or if $l+j$ is odd and $q \neq 4$.
Second, $\M$ is Hermitian, and its matrix elements in this basis are real; so it is symmetric in this basis.
Third, by inspection, we see that

\bea
\left< 1,l',t',j \right| \M \left| 2,l,t,j \right> = 0.
\eea

\noi
Finally, there are some redundancies and additional zeros in the matrix when $j=0$:

\bea
\left< 2,l',t',0 \right| \M \left| 2,l,t,0 \right> &=& \left< 1,l',t',0 \right| \M \left| 1,l,t,0 \right>, \n
\left< 1,l',t',0 \right| \M \left| 3,l,t,0 \right> &=& \left< 2,l',t',0 \right| \M \left| 3,l,t,0 \right>, \n
\left< 1,l',t',0 \right| \M \left| 4,l,t,0 \right> &=& -\left< 2,l',t',0 \right| \M \left| 4,l,t,0 \right>, \n
\left< 3,l',t',0 \right| \M \left| 4,l,t,0 \right> &=& 0.
\eea

\vspace{.10in}

\noi
{\bf C.2.2  \hspace{.05in} Estimating Uncertainties in the Spectrum}

\vspace{.04in}

When we use the VEGAS Monte Carlo routine to compute the matrix elements, the results for the matrix elements have statistical uncertainties.
In order to control the resulting uncertainties in the spectrum, we would like to have a method that allows us
to estimate how accurately we must calculate any given matrix element in order for the spectrum to have a desired
uncertainty.  

Suppose
a diagonal matrix element $\left< q,l,t,j \right| \M \left| q,l,t,j \right>$ is given by

\bea
\left< q,l,t,j \right| \M \left| q,l,t,j \right> = Z \pm \delta ,
\eea

\noi
where $Z$ is the Monte Carlo estimate of the matrix element and $\delta$ is the associated absolute uncertainty.  Using Mathematica 
with test matrices,
it is straightforward to convince oneself that if $\delta$ is small, then it will yield a relative uncertainty 
$e_{_{M_n^2}} \sim \delta/Z$ in the eigenvalues of the matrix\footnote{In the development of this method, we are guided by the principles of 
quantum-mechanical perturbation theory, 
although we cannot legitimately use perturbation theory to analyze the uncertainties.}.  This translates to a relative uncertainty of 
$e_{_{M_n}} \sim \delta/(2 Z)$ in the masses.

Estimating the uncertainty in the spectrum due to uncertainties in off-diagonal matrix elements is more difficult.  Using Mathematica with
test matrices, the simplest method that we have found that is reasonably reliable is to use a type of degenerate perturbation theory.  When we have
an off-diagonal matrix element given by

\bea
\left< q',l',t',j \right| \M \left| q,l,t,j \right> = Z \pm \delta ,
\eea

\noi
we diagonalize the two matrices

\bea
\left(
\begin{array}{cc}
\left< q,l,t,j \right| \M \left| q,l,t,j \right> & Z+\delta \\
Z+\delta & \left< q',l',t',j \right| \M \left| q',l',t',j \right>
\end{array}
\right) ,
\eea

\noi
and

\bea
\left(
\begin{array}{cc}
\left< q,l,t,j \right| \M \left| q,l,t,j \right> & Z-\delta \\
Z-\delta & \left< q',l',t',j \right| \M \left| q',l',t',j \right>
\end{array}
\right) ,
\eea

\noi
and we compare their eigenvalues to the eigenvalues of the matrix

\bea
\left(
\begin{array}{cc}
\left< q,l,t,j \right| \M \left| q,l,t,j \right> & Z \\
Z & \left< q',l',t',j \right| \M \left| q',l',t',j \right>
\end{array}
\right) .
\eea

\noi
We then define $e_{_{M_n^2}}$ to be the largest relative deviation that we have found in the eigenvalues,
and we estimate the resulting relative uncertainty in the mass spectrum to be $e_{_{M_n}} \sim e_{_{M_n^2}}/2$.
This estimate tends to work well unless there are too many diagonal matrix elements that are nearly degenerate
with either $\left< q,l,t,j \right| \M \left| q,l,t,j \right>$ or $\left< q',l',t',j \right| \M \left| q',l',t',j \right>$.

To achieve a relative uncertainty of $\OR(\varepsilon)$ in the glueball masses, we require $e_{_{M_n}} < \varepsilon$
for each matrix element.  This method tends to work reasonably well.  The main difficulty is that we have neglected to consider
the highly nonlinear couplings between the uncertainties in different matrix elements.  For this reason, as we increase the size of the matrix,
the error in our estimate of the uncertainty eventually becomes critical.  At this point, the spectrum that we get when
we diagonalize the matrix becomes completely unreliable.  (The evidence of the breakdown is sudden contamination of the low-lying wave functions
with high-order components.)  It should be possible to develop more sophisticated methods
of estimating uncertainties to suppress this problem.

\newpage
\noi
{\large \bf Acknowledgments}

We would like to thank Roger D. Kylin, Richard J. Furnstahl, and John Hiller for useful discussions.  This work has been partially supported by 
National Science Foundation grant PHY-9800964.


\end{document}